%% file: main.tex
\documentclass[a4paper,11pt]{article}

\usepackage{booktabs}
\usepackage{jheppub}
\usepackage{lineno}
\usepackage{overpic}
\usepackage{physics}
\usepackage{float}
\usepackage{orcidlink}
\usepackage[separate-uncertainty=true]{siunitx}[=v2] 

\newcommand{\Llh}{{\cal L}}

\graphicspath{{figures/}}
\input{belle2-symbols}

\title{Measurement of $\CP$ asymmetries and branching-fraction ratios for
	$B^\pm \to DK^\pm$ and $D\pi^\pm$ with
  $D\to \KS K^\pm\pi^\mp$ using Belle and Belle~II data}

\input{authors.tex}
\emailAdd{xiaodong.shi@kek.jp, coll-publications@belle2.org}

\abstract{

    We measure $\CP$ asymmetries and branching-fraction ratios for
    $B^\pm \to DK^\pm$ and $D\pi^\pm$ decays with
    $D\to \KS K^\pm\pi^\mp$, where $\PD$ is a superposition of
    $\Dz$ and $\Dzb$. We use the full data set of the Belle
    experiment, containing \num{772e6}~$\BBbar$ pairs, and data from
    the Belle~II experiment, containing \num{387e6}~$\BBbar$ pairs,
    both collected in electron-positron collisions at the $\Upsilon(4S)$ resonance. 
		Our results provide model-independent information on
		the unitarity triangle angle $\phi_3$.
    
}

\begin{document} 
\begin{flushright}
Belle II Preprint 2023-010 \\
KEK Preprint 2023-8
\end{flushright}

\maketitle

\flushbottom

\section{Introduction}


In the Standard Model, $\CP$ violation is described by a single irreducible complex phase of the Cabibbo-Kobayashi-Maskawa (CKM) quark-mixing matrix~\cite{Cabibbo:1963yz,Kobayashi:1973fv}.
The CKM matrix is unitary, so $V_{ud}^{}V_{ub}^*+ V_{cd}^{}V_{cb}^*+V_{td}^{}V_{tb}^*=0$, where $V_{ij}$ is the CKM matrix element coupling quark flavour $i$ to quark flavour $j$.
This condition, represented by a triangle in the complex plane, provides a promising way to verify the Standard Model by testing the closure of the triangle.
The interior angle $\phi_3$~(also known as $\gamma$), defined as $\arg(-V_{ud}^{}V_{ub}^*/V_{cd}^{}V_{cb}^*)$,
is independent of top-quark couplings and a benchmark of the Standard Model.
Through the interference of $b\to c\bar{u}s$ and $b\to u\bar{c}s$ transition amplitudes 
in tree-level $b$-hadron decays, where non-Standard-Model effects are negligible~\cite{Brod:2014bfa},
$\phi_3$ can be studied in a theoretically reliable way~\cite{Brod:2013sga}.
An improved direct measurement of $\phi_3$ would provide a key input for 
Standard Model predictions, which can be compared with predictions of other measured observables
that are sensitive to new particles and interactions.

The world average of $\phi_3$ measurements, $(65.9^{+3.3}_{-3.5})\si{\degree}$~\cite{ParticleDataGroup:2022pth}, in which the LHCb experiment contributes most,
is dominated by studies of the interference of
$\Bpm\to\Dz\Kpm$ and $\Bpm\to\Dzb\Kpm$ decays in which the
$\Dz$ and $\Dzb$ decay to a common final
state~\cite{BaBar:2013caj,Belle:2021efh,LHCb:2022awq}. 
Grossman, Ligeti, and Soffer~(GLS) proposed a method to measure $\phi_3$ with singly 
Cabibbo-suppressed decays of $D$ mesons, $\PD \to \KS\Kpm\pimp$~\cite{Grossman:2002aq}, 
where $\PD$ is a superposition of $\Dz$ and $\Dzb$ mesons.
Experimentally, one measures seven observables to access $\phi_3$, 
including four $\CP$ asymmetries and three branching-fraction ratios in 
$\Bpm \to \PD h^\pm$ decays, where $h$ is a pion or kaon. 
From these, the $\phi_3$-related information is extracted without model dependent uncertainties from the amplitude model of $D$ decay.
However, information about the $\PD \to \KS\Kpm\pimp$ dynamics is necessary.
The CLEO experiment measured this information using all such $\PD$ decays and also using only 
decays in which the $\KS\pimp$ pair has a mass within \SI{100}{MeV/\it{c}^2} of the known
$\Kstar(892)^\mp$ mass~\cite{CLEO:2012obf}.\footnote{The $\Kstar(892)^{\pm}$ is henceforth referred to as $\Kstarpm$.} In that region, 
the interference of $\Bpm \to \Dz h^\pm$ and $\Bpm \to \Dzb h^\pm$ decays is expected to be enhanced 
due to the large coherence factor in $\PD\to\KS\Kpm\pimp$ decays, 
possibly leading to a more precise determination of $\phi_3$.
The principal experimental challenge is extracting the signal given the small branching fractions of these channels.

The LHCb collaboration reported the most precise GLS measurement to date~\cite{LHCb:2020vut}. 
In this paper, we present a similar measurement 
for $\Bpm\to\PD\Kpm$ and $\Bpm\to\PD\pipm$ using \num{772e6}~$\BBbar$ pairs
collected by the Belle experiment~\cite{Belle:2000cnh} and \num{387e6}~$\BBbar$ pairs
collected by the Belle~II experiment~\cite{Belle-II:2010dht}, produced in electron-positron
collisions at the $\Upsilon(4S)$ resonance.\footnote{Here, $\PB$ indicates either a $\Bp$ or $\Bz$.}
We fit to the distributions of two signal-discriminating observables, simultaneously in both data sets and all channels to extract the seven GLS observables. 
An additional measurement restricted to a $\Kstarpm$-enriched region of 
the $D$ meson phase space is also reported, as well as results based on the Belle data only.

\section{Formalism}
\label{sec:formalism}

We categorise $\Bpm \to \PD h^\pm$ followed by $\PD \to \KS\Kpm\pimp$ 
as same-sign~(SS) or opposite-sign~(OS) decays according to 
the charge of the $\Kpm$ produced by the $\PD$ meson relative to the charge
of the $\Bpm$ meson.  The four $\CP$ asymmetries are
\begin{equation}
  \mathcal{A}_m^{{\it Dh}} \equiv \frac{N_m^{{\it D h^-}} -~N_m^{{\it D h^+}}}{N_m^{{\it D h^-}} +~N_m^{{\it D h^+}}}
  \qq{with}
  h = \Ppi, \PK,
\end{equation}
where $N_m^{{\it D h^\pm}}$ is the number of $\Bpm \to \PD h^\pm$ decays,
and $m$ denotes the decay type, which is either SS or OS.
The three branching-fraction ratios are
\begin{equation}
  \mathcal{R}_m^{{\it DK/D\pi}} \equiv \frac{N_m^{{\it D K^-}} +~N_m^{{\it D K^+}}}{N_m^{{\it D \pi^-}} +~N_m^{{\it D \pi^+}}}
\end{equation}
and
\begin{equation}
  \mathcal{R}_{\text{SS}/\text{OS}}^{{\it D \pi}} \equiv \frac{N_{\text{SS}}^{{\it D \pi^-}} +~N_{\text{SS}}^{{\it D \pi^+}}}{N_{\text{OS}}^{{\it D \pi^-}} +~N_{\text{OS}}^{{\it D \pi^+}}}.
\end{equation}

The relations between the seven observables and $\phi_3$ are given by the following equations:
\begin{equation}
\begin{split}
\mathcal{A}^{DK}_{\text{SS}} & = \frac{2r^{DK}_Br_D\kappa_D\sin(\delta^{DK}_B-\delta_D)\sin\phi_3}{1+(r^{DK}_B)^2r^2_D+2r^{DK}_Br_D\kappa_D\cos(\delta^{DK}_B-\delta_D)\cos\phi_3},  \\
\mathcal{A}^{DK}_{\text{OS}} & = \frac{2r^{DK}_Br_D\kappa_D\sin(\delta^{DK}_B+\delta_D)\sin\phi_3}{(r^{DK}_B)^2+r^2_D+2r^{DK}_Br_D\kappa_D\cos(\delta^{DK}_B+\delta_D)\cos\phi_3}, \\ 
\mathcal{A}^{D\pi}_{\text{SS}} & = \frac{2r^{D\pi}_Br_D\kappa_D\sin(\delta^{D\pi}_B-\delta_D)\sin\phi_3}{1+(r^{D\pi}_B)^2r^2_D+2r^{D\pi}_Br_D\kappa_D\cos(\delta^{D\pi}_B-\delta_D)\cos\phi_3},  \\
\mathcal{A}^{D\pi}_{\text{OS}} & = \frac{2r^{D\pi}_Br_D\kappa_D\sin(\delta^{D\pi}_B+\delta_D)\sin\phi_3}{(r^{D\pi}_B)^2+r^2_D+2r^{D\pi}_Br_D\kappa_D\cos(\delta^{D\pi}_B+\delta_D)\cos\phi_3}. \\ 
\mathcal{R}^{DK/D\pi}_{\text{SS}} & =R\frac{1+(r^{DK}_B)^2r^2_D+2r^{DK}_Br_D\kappa_D\cos(\delta^{DK}_B-\delta_D)\cos\phi_3}{1+(r^{D\pi}_B)^2r^2_D+2r^{D\pi}_Br_D\kappa_D\cos(\delta^{D\pi}_B-\delta_D)\cos\phi_3}, \\
\mathcal{R}^{DK/D\pi}_{\text{OS}} & =R\frac{(r^{DK}_B)^2+r^2_D+2r^{DK}_Br_D\kappa_D\cos(\delta^{DK}_B+\delta_D)\cos\phi_3}{(r^{D\pi}_B)^2+r^2_D+2r^{D\pi}_Br_D\kappa_D\cos(\delta^{D\pi}_B+\delta_D)\cos\phi_3}, \\
\mathcal{R}^{D\pi}_{\text{SS/OS}} & =\frac{1+(r^{D\pi}_B)^2r^2_D+2r^{D\pi}_Br_D\kappa_D\cos(\delta^{D\pi}_B-\delta_D)\cos\phi_3}{(r^{D\pi}_B)^2+r^2_D+2r^{D\pi}_Br_D\kappa_D\cos(\delta^{D\pi}_B+\delta_D)\cos\phi_3},
\end{split}
\end{equation}
where $r^{DK}_B(r^{D\pi}_B)$ is the ratio of the magnitudes of the suppressed-to-favoured amplitudes for the $B^+\to DK^+(D\pi^+)$ decay, $\delta^{DK}_B(\delta^{D\pi}_B)$ is the relative strong-phase difference between those amplitudes, $r_D$ and $\delta_D$ are the amplitude ratio and strong-phase difference, respectively, between $D^0\to \KS K^-\pi^+$ and $D^0\to \KS K^+\pi^-$ decays, $\kappa_D$ is the coherence factor of these $D$ decays~\cite{CLEO:2012obf}, and $R$ is the ratio between $B^+\to\Dzb\Kp$ and $B^+\to\Dzb\pip$ branching fractions.

\section{Belle and Belle~II detectors}
\label{sec:det}

The Belle detector~\cite{Belle:2000cnh, Belle:2012iwr} was a
large-solid-angle magnetic spectrometer at the KEKB
accelerator~\cite{Kurokawa:2001nw, Abe:2013kxa}, which collided
\SI{8}{GeV} electrons with \SI{3.5}{GeV} positrons. The subdetectors
of Belle most relevant for our study are the silicon vertex
detector and the central drift chamber for
charged-particle tracking and ionization-energy loss measurement and the aerogel
threshold Cherenkov counters and time-of-flight scintillation
counters for charged particle identification~(PID). They were
situated in a uniform axial magnetic field of \SI{1.5}{T}.

The Belle~II detector~\cite{Belle-II:2010dht} is an upgrade of the
Belle detector at the SuperKEKB accelerator~\cite{Akai:2018mbz}, which
collides \SI{7}{GeV} electrons with \SI{4}{GeV} positrons. 
Innermost is a tracking system, including
two layers of silicon pixel sensors, four
layers of silicon strip detectors, 
and the central drift chamber.
Only 15\% of the azimuthal angle is covered by the second layer of the pixel 
detector for the Belle~II data used in this paper.
Outside the drift chamber, the time-of-propagation 
and aerogel ring-imaging Cherenkov subdetectors 
cover the barrel and forward endcap regions, respectively.
Outside these subdetectors are the electromagnetic calorimeter and 
a solenoid, which provides a uniform \SI{1.5}{T} magnetic field.
A $\KL$ and muon detector is installed in the iron flux-return yoke of 
the solenoid.

\section{Simulation}

We use simulated samples to identify sources of background, optimise
selection criteria, calculate selection efficiencies, and distinguish fit
models. We generate $\epem \to \Upsilon(4S) \to \BBbar$ events, 
and simulate particle decays with
\textsc{EvtGen}~\cite{Lange:2001uf}; we generate continuum
$\epem \to \qqbar$ where $\quark$ is an $\uquark, \dquark,
\cquark,$ or $\squark$ quark with \textsc{Pythia}~\cite{Sjostrand:2007gs} for
Belle and \textsc{KKMC}~\cite{Jadach:1999vf} and \textsc{Pythia} for Belle~II; we simulate
final-state radiation with \textsc{Photos}~\cite{Barberio:1993qi}; we 
simulate detector response using \textsc{Geant3}~\cite{Brun:1987ma}
for Belle and \textsc{Geant4}~\cite{GEANT4:2002zbu} for Belle~II. We
model our signal processes using both nonresonant
$\PD\to\KS\Kpm\pimp$ and $\PD\to\PK^{*\mp}\Kpm$ decays. In the Belle
simulation, beam backgrounds are taken into account by overlaying random trigger data. In the Belle~II simulation, they
are accounted for by simulating the Touschek effect~\cite{Bernardini:1963dyf},
beam-gas scattering, and luminosity-dependent backgrounds from Bhabha
scattering and two-photon quantum-electrodynamic
processes~\cite{Lewis:2018ayu, Natochii:2023thp}.

\section{Event selection}
\label{sec:evtsel}

We reconstruct events using the Belle~II analysis software
for both Belle and Belle~II data~\cite{Gelb:2018agf,Kuhr:2018lps,basf2-zenodo}.
All events are required to pass the online selection criteria based on either total 
energy deposition in the electromagnetic calorimeter or the number of charged-particle tracks in the central drift chamber.
The efficiency of the online selection is found to be close to 100\%.

Tracks originating from $\PK^\pm$ and $\Ppi^\pm$ are selected by
requiring that each have a distance of closest approach to the
$\epem$ interaction point smaller than \SI{1.0}{cm}
in the longitudinal direction (parallel to the $\ep$ beam at
Belle and the principal axis of the magnet at Belle~II) and smaller than
\SI{0.2}{cm} in the transverse plane. We identify the species of each
charged hadron using $\Llh(K/\pi) = \Llh(K) / [\Llh(K) + \Llh(\pi)]$,
where $\Llh(h)$ is the likelihood for hadron $h$ to have produced the relevant track based on
information from the aerogel threshold Cherenkov counters, time-of-flight scintillation counters, and the central drift chamber for Belle and all subdetectors for
Belle~II. 

We identify an $h^\pm$ as a kaon if $\Llh(K/\pi) > 0.6$. 
To improve signal efficiency, no PID requirement is applied to the pion candidate from $D$ decay.
Only prompt pion candidates, which are produced directly from $B^\pm$ decay, are required to satisfy $\Llh(K/\pi) < 0.6$.
In Belle~II data, we also restrict the polar angle of the prompt $h^\pm$ in the laboratory frame to be within the acceptance of the PID detectors.
The kaon-identification efficiency depends on the
particle momentum and is in the range of 86\%--90\%. The rate to misidentify a pion as a
kaon is in the range of 3\%--9\%.

We reconstruct $\KS$ mesons via their decays to $\pip\pim$. Each
candidate $\KS$ is formed from a pair of oppositely-charged particles
with no PID requirements, constrained to come from a common vertex.
The resulting dipion mass must be in the range of \SIrange{487}{508}{MeV/}$c^2$, which corresponds to
$\pm3\sigma$ around the known $\Kz$
mass~\cite{ParticleDataGroup:2022pth}, with $\sigma$ being the mass
resolution. To improve purity, we reject combinatorial background based on the output
of a neural network for Belle ~\cite{Belle:2018xst} and a boosted decision
tree~(BDT)~\cite{Keck:2017gsv} for Belle~II. 
For the latter one, 15 input variables are selected including 
kinematic quantities and the number of hits in the vertex detector 
associated to the $\pipm$ tracks. The most discriminating variables
are the angle between the directions of the $\KS$ momentum and the 
decay position seen from the beam interaction point in the laboratory frame
and the flight length of the $\KS$ normalised by its uncertainty.

Each neutral $\PD$ candidate, reconstructed from $\KS$, $\Kpm$, and
\pimp candidates, must have a mass in the range of \SIrange{1.85}{1.88}{GeV}/$c^2$,
which corresponds to $\pm3\sigma$ around the known $\Dz$
mass~\cite{ParticleDataGroup:2022pth}, where $\sigma$ is the typical $D$ mass resolution.

Each $\Bpm$ candidate is reconstructed from $\PD$ and prompt $h^\pm$
candidates. To suppress continuum background, we
require that the beam-constrained mass,
\begin{equation}
  M_{\text{bc}} \equiv \sqrt{s/4 - |\vec{p}_{\PB}c|^2}/c^2,
\end{equation}
exceed \SI{5.27}{GeV}/$c^2$, where $s$ is the squared collision 
energy and $\vec{p}_{\PB}$ is the $\PB$ momentum, both defined in the
$\epem$ centre-of-mass~(c.m.) frame. We also require
$|\Delta E| < \SI{0.15}{GeV}$, where
$\Delta E \equiv E_{\PB} - \sqrt{s}/2$
and $E_{\PB}$ is the $\PB$ energy in the c.m.\ frame.

The remaining background comes mostly from continuum events,
which are topologically distinguishable from $\BBbar$ events. Since
the momentum of a $\PB$ is only \SI{333}{MeV}/$c$ in the c.m.\ frame, the
final-state particles of a $\BBbar$ event are almost isotropically
distributed in the c.m.\ frame. The final-state particles of continuum
events form mainly back-to-back jets. We discriminate between $\BBbar$ and
continuum events using modified Fox-Wolfram
moments~\cite{Fox:1978vu, Belle:2003fgr}, the thrust of the $\PB$ decay
products~\cite{Farhi:1977sg}, the angle between the axis of this thrust and that of the
particles in the rest of the event~(ROE), the polar angle of the $\PB$ 
momentum, the distance between the $\PB$ decay vertex and that of the
ROE in the longitudinal direction, and the output of a
$\PB$-flavour-tagging algorithm~\cite{Belle:2004uxp,Belle-II:2021zvj}\footnote{For Belle II, a category-based algorithm is adopted.}. 
All frame-dependent
quantities are calculated in the c.m.\ frame. 
Simulated samples show that these variables have correlations of 4\% or smaller with $\Delta E$,
which is used in the signal-extraction fit.
We train BDT classifiers with these variables separately for Belle and
Belle~II.  The classifier output, $C$, is distributed between zero and
one. We require $C > 0.3$, which rejects 64.3\% of
continuum background candidates and retains 95.3\% of signal
candidates for Belle and rejects 63.0\% of continuum background
candidates and retains 97.2\% of signal candidates for Belle~II.

We suppress $\Bpm$ candidates in which the $\PD$ comes from a $\Dstarpm$ decay
by reconstructing possible $\Dstarpm$ candidates from
the $\PD$ and a charged particle~(assumed to be a pion) from the ROE and vetoing
any $\Bpm$ candidate whose $\PD$ forms a $\Dstarpm$ with mass difference
$M(\KS\Kp\pim\pipm) - M(\KS\Kp\pim)$ in the range of \SIrange{143}{148}{MeV/}$c^2$.

Remaining background candidates come from continuum events, cross-feed background from signal
events in which the prompt $\Kpm$ is misidentified as a $\pipm$ or the
reverse, or from other $\PB$ decays such as $\Bpm \to \Dstar h^\pm$ or
$\Bpm \to \PD \Kstarpm$. 
Sidebands in $D$ mass, \SIrange{1.73}{1.85}{GeV/}$c^2$ and \SIrange{1.88}{1.94}{GeV/}$c^2$,
show no significant backgrounds from 
\Bpm decays without intermediate $D$ mesons.

On average, 1.02 $\Bpm$ candidates are reconstructed per event. In
events with multiple candidates, we keep only the one with the
smallest $\chi^2$ calculated from the measured and known masses of the
\PD, $M_{\text{bc}}$, the known $\PB$ mass, and the resolutions on
both measured masses. Simulated samples show that we
select the correct candidate in 78\% of such events.

\section{Signal extraction}
\label{sec:signal}

To determine the $\CP$ asymmetries and branching-fraction ratios, we fit to
the two-dimensional distributions of $\Delta E$ and $C'$,
a transformation of $C$ that is uniformly distributed between 
zero and one for signal and peaks at zero for continuum background
~\cite{Belle:2021efh}. We model their distributions independently since they
 have negligible correlations according to simulation.
We perform an unbinned extended maximum-likelihood fit 
simultaneously in sixteen subsets of the data formed by the Cartesian
product of the two charges of the $\Bpm$, the two relative charges of
the $\Kpm$ from $\PD$, the two species of the prompt $h^\pm$, and the two
experiments. The fit function accounts for contributions from the signal decays,
continuum background, cross-feed background, and other $\BBbar$
backgrounds.
We perform the fit separately for the full $D$ phase space and for the $\Kstarpm$ region. 

For the signal component, we model $\Delta E$ as a sum of two Gaussian
functions and an asymmetric Gaussian function with all parameters
fixed to values determined from simulated samples, except for the common
mean of all three Gaussian functions and a common multiplier for all their
widths, which account for differences in resolution between the experiment
and simulation. We model $C'$ as uniformly distributed.

For the continuum component, we model $\Delta E$ as a straight line
and $C'$ as the sum of two exponential functions. All parameters but
the slope of the line and the rate parameter for the steeper
exponential are fixed to values determined from simulated samples.

For the cross-feed component, we model $\Delta E$ identically to
signal, but with its own parameters, and $C'$ as a straight line, with
independent parameters for the ${\it D K}$ and ${\it D \pi}$ data and all
parameters fixed to values determined from simulated samples.

For the $\BBbar$-background component, we model $\Delta E$ as a sum
of two exponential functions and $C'$ as a straight line, with
independent parameters for ${\it D K}$ and ${\it D \pi}$ and all parameters
except for the rate parameter of the steeper exponential function
fixed to values determined from simulated samples.

For each data subset $i$, the total number of events $n_{\text{tot},i}^{Dh^\pm}$ 
with no PID requirement are related to the observed numbers of signal events 
$n^{{\it D h^\pm}}_{\text{sig},i}$, cross-feed background yields 
$n^{{\it D h^\pm}}_{\text{bkg},i}$, and the PID efficiencies $\epsilon_{h^\pm}$ as follows:
\begin{alignat}{1}
  n^{{\it D K^\pm}}_{\text{sig},i}  & = \epsilon_{\PK^\pm}      n^{{\it D K^\pm}}_{\text{tot},i},\\
  n^{{\it D \pi^\pm}}_{\text{sig},i} & = \epsilon_{\Ppi^\pm}     n^{{\it D \pi^\pm}}_{\text{tot},i},\\
  n^{{\it D K^\pm}}_{\text{bkg},i}  & = (1-\epsilon_{\Ppi^\pm}) n^{{\it D \pi^\pm}}_{\text{tot},i},\\
  n^{{\it D \pi^\pm}}_{\text{bkg},i} & = (1-\epsilon_{\PK^\pm})  n^{{\it D K^\pm}}_{\text{tot},i}.
\end{alignat}
In the fit, the $n^{{\it D K^\pm}}_{\text{tot},i}$ and $n^{{\it D \pi^\pm}}_{\text{tot},i}$ yields
are expressed in terms of the $\CP$ asymmetries and branching-fraction ratios, the sum of all
$n^{{\it D \pi}}_{\text{tot},i}$, and the ratio $\delta$ of the
efficiency for detecting $\Bpm \to {\it D K^\pm}$ over that for $\Bpm \to
{\it D \pi^\pm}$ decays. The PID efficiencies and $\delta$ are fixed
from simulated samples and corrected for discrepancies between
experiment and simulation that depend on particle momentum, direction,
charge, and species. Those corrections range between 0.960 and 0.984, 
with typical uncertainties of 0.008 to 0.009, 
as estimated using the control channels 
$\Dstarp\to \Dz(\to\Km\pip)\pip$, and $\KS\to\pip\pim$ for $\Kpm$ and $\pipm$.
Table~\ref{tab:PIDEff} lists the post-correction
efficiencies and $\delta$ with uncertainties including those on the
corrections. We validate the fit strategy using simplified simulated experiments and find unbiased estimates and Gaussian uncertainties for all parameters of interest. 

\begin{table}[!t]
  \centering
  \caption{PID and tracking efficiencies.}
  \label{tab:PIDEff}

  \vspace\baselineskip
  
  \sisetup{round-mode=places,round-precision=3}
  
  \begin{tabular}{lll}
    \hline
    \hline
    
    & {Belle}
    & {Belle~II}
    \\

    \hline
    
    $\epsilon_{\Kp}$
    & \num{0.8446} $\pm$ \num{0.0086}
    & \num{0.8201} $\pm$ \num{0.0045}
    \\

    $\epsilon_{\Km}$
    & \num{0.8526} $\pm$ \num{0.0080}
    & \num{0.8208} $\pm$ \num{0.0043}
    \\

    $\epsilon_{\pi^+}$
    & \num{0.9262} $\pm$ \num{0.0078}
    & \num{0.9287} $\pm$ \num{0.0030}
    \\

    $\epsilon_{\pim}$
    & \num{0.9304} $\pm$ \num{0.0085}
    & \num{0.9182} $\pm$ \num{0.0032}
    \\
    
    $\delta$
    & \num{0.973} $\pm$ \num{0.005}
    & \num{0.972} $\pm$ \num{0.004}
    \\

    \hline
    \hline

  \end{tabular}
\end{table}

The fits determine the sums of all $n^{{\it D \pi}}_{\text{tot},i}$ yields in the
full $\PD$ phase space to be \num{2209 +- 59} for Belle and \num{1210 +-
  39} for Belle II and in the region of the $\Kstarpm$ to be \num{1337 +-
  42} for Belle and \num{732 +- 30} for Belle~II. 
The sum of all $n^{{\it D K}}_{\text{tot},i}$ yields is calculated using the 
fit results, and is \num{238 +- 21} for Belle and \num{131 +-
  12} for Belle II for the full $\PD$ phase space and 
	\num{126 +- 15} for Belle and \num{69 +- 9} for Belle~II in the region of the $\Kstarpm$. 
Figures~\ref{fig:PHSPB1DK}--\ref{fig:PHSPB2DPi} show the data
and fit results for the full $\PD$ phase space and
figures~\ref{fig:KstarKB1DK}--\ref{fig:KstarKB2DPi} show the same distributions for
the $\Kstarpm$ region.  We enhance the signal in the plots by displaying  
$\Delta E$ distributions for events with $C' > 0.6$ and $C'$ distributions for
events with $|\Delta E| < \SI{0.05}{GeV}$.
In the $\Delta E$ distributions of the ${\it DK}$ data subset, the cross-feed component 
from ${\it D\pi}$ is visible as a peak at higher $\Delta E$ values.

\newcommand{\ResultFigureCaption}[3]{Distributions of $C'$ and $\Delta E$ for $\Bpm \to \PD #1$ candidates reconstructed in the #3 data for the #2, with the fit results overlaid. The SS or OS indicates the type of the signal decay chain, same-sign or opposite-sign, respectively.}

\newcommand{\FullResultsFigureCaption}[2]{\ResultFigureCaption{#1}{full $\PD$ phase space}{#2}}
\newcommand{\KstarResultsFigureCaption}[2]{\ResultFigureCaption{#1}{$\Kstarpm$ region}{#2}}

\begin{figure}[H]
  \centering

  \begin{overpic}[width=0.49\textwidth]{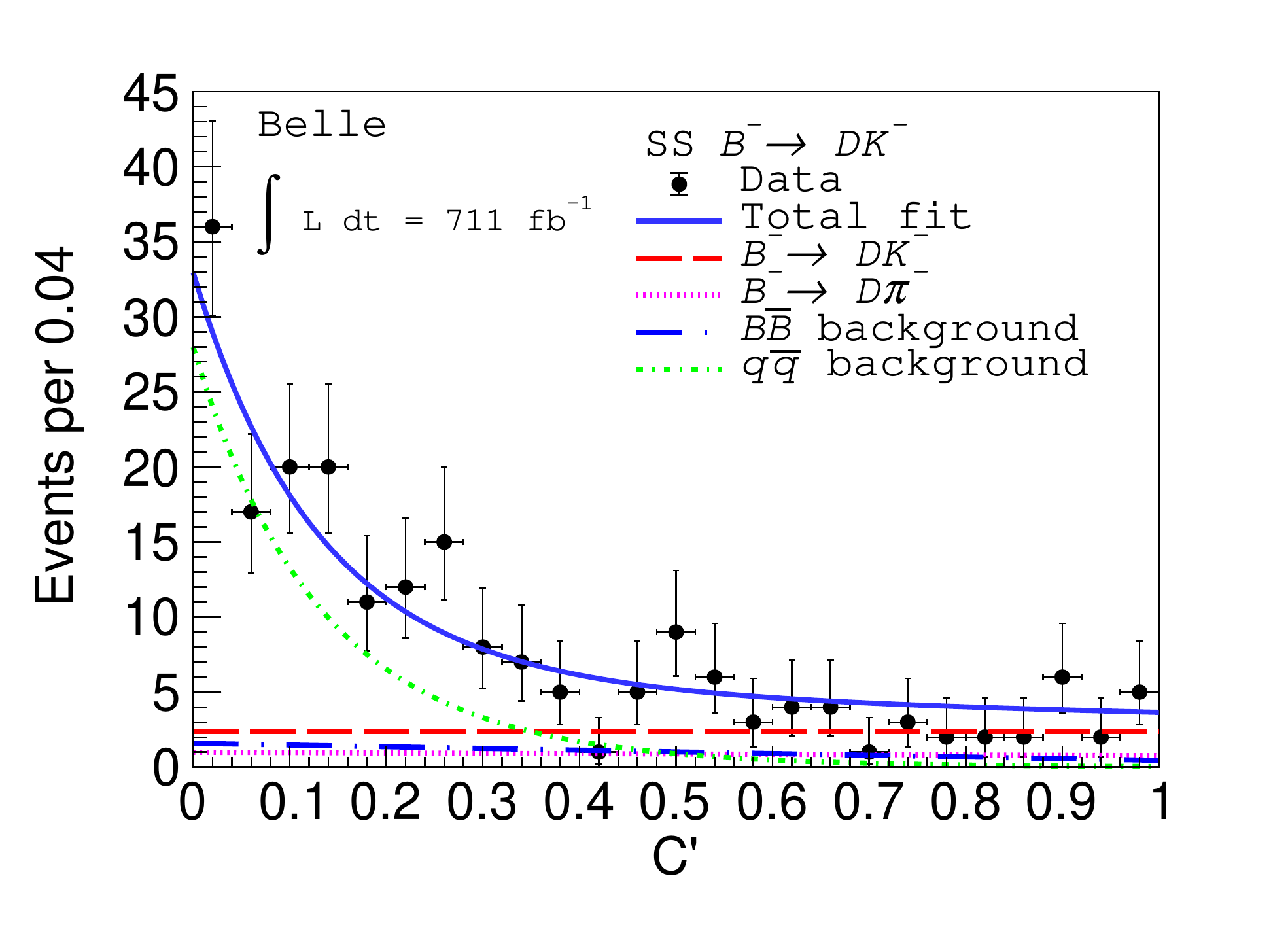}\put(40,55){}\end{overpic}
  \begin{overpic}[width=0.49\textwidth]{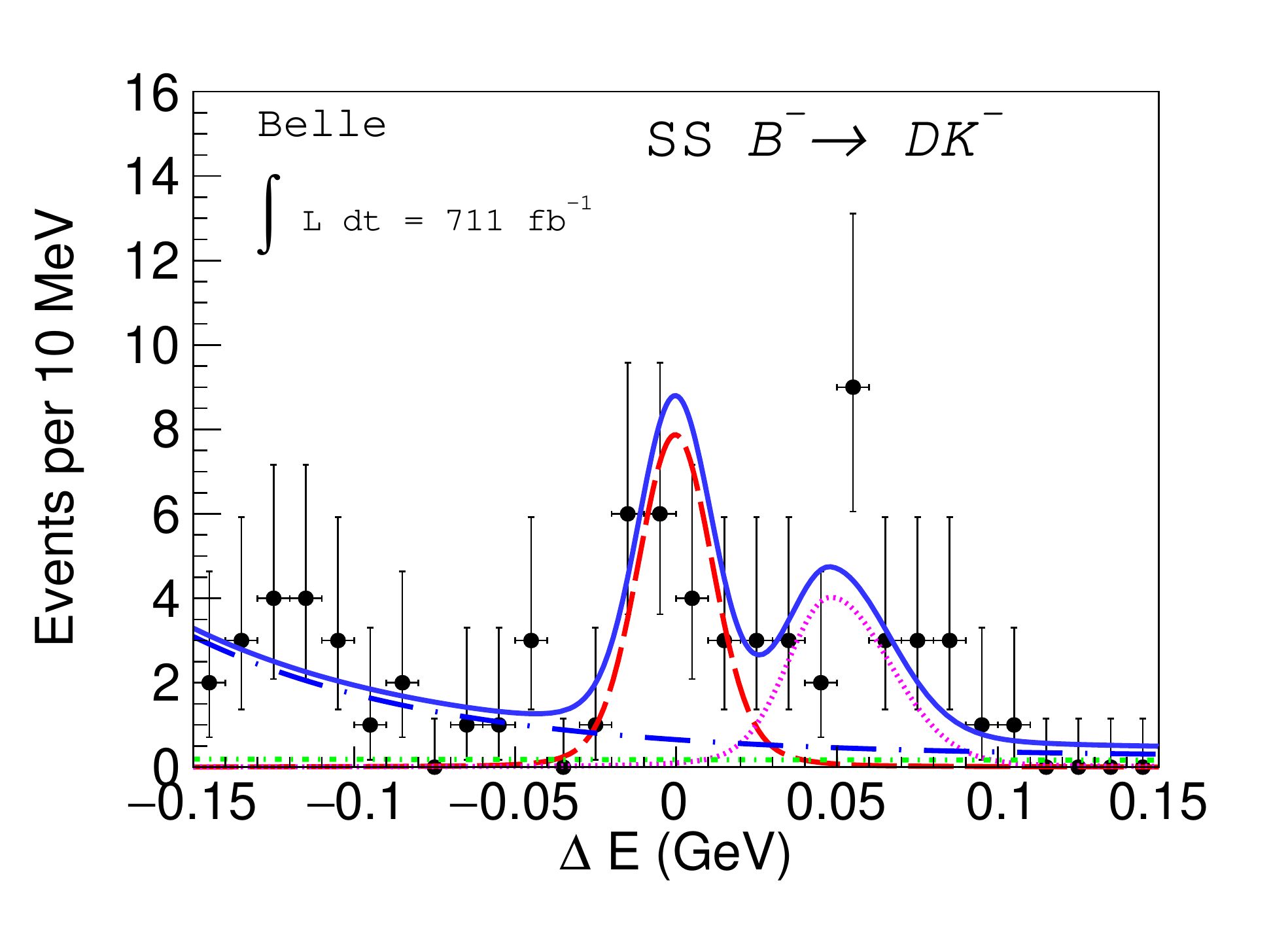}\put(40,55){}\end{overpic}
  \begin{overpic}[width=0.49\textwidth]{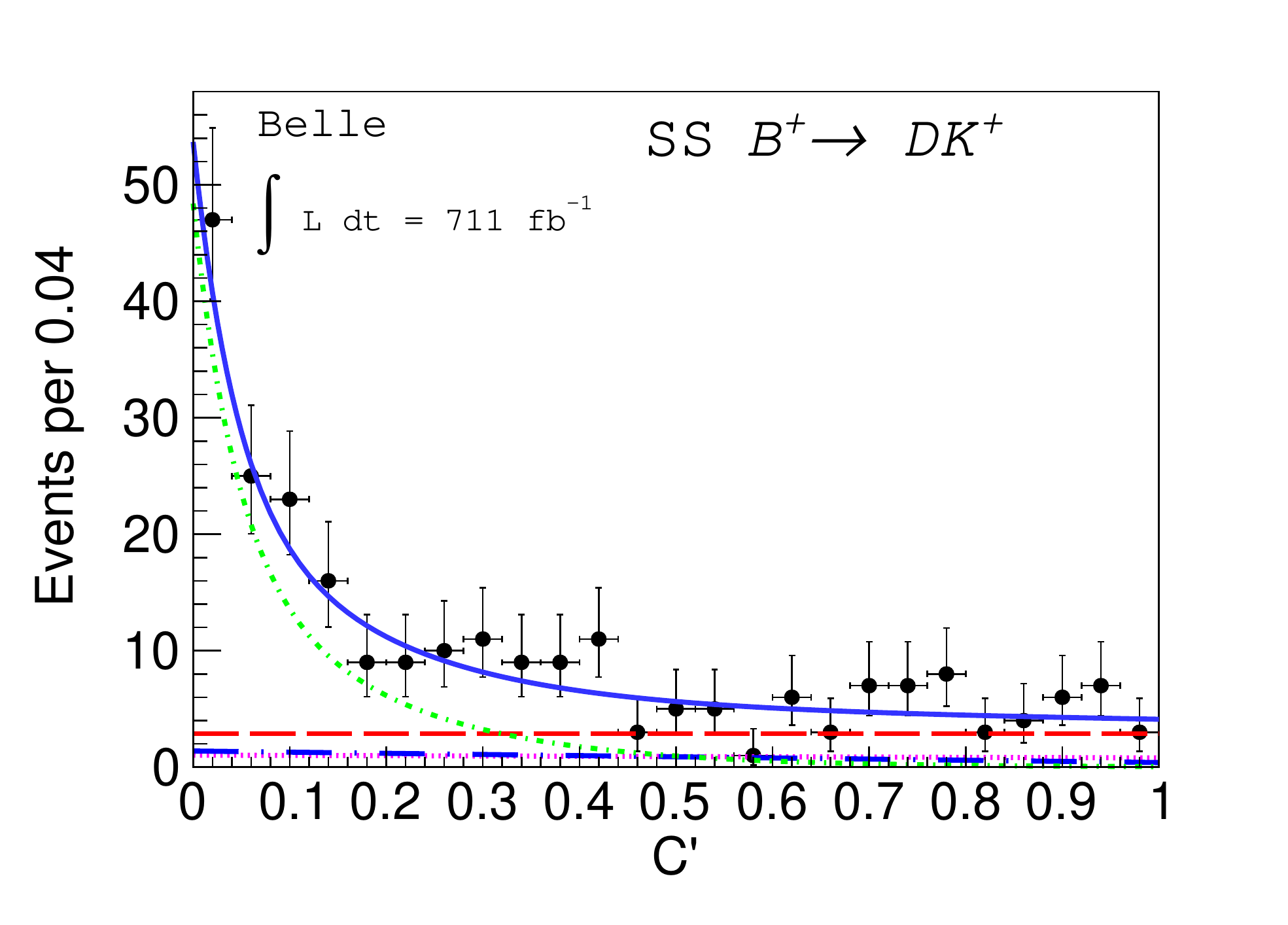}\put(40,55){}\end{overpic}
  \begin{overpic}[width=0.49\textwidth]{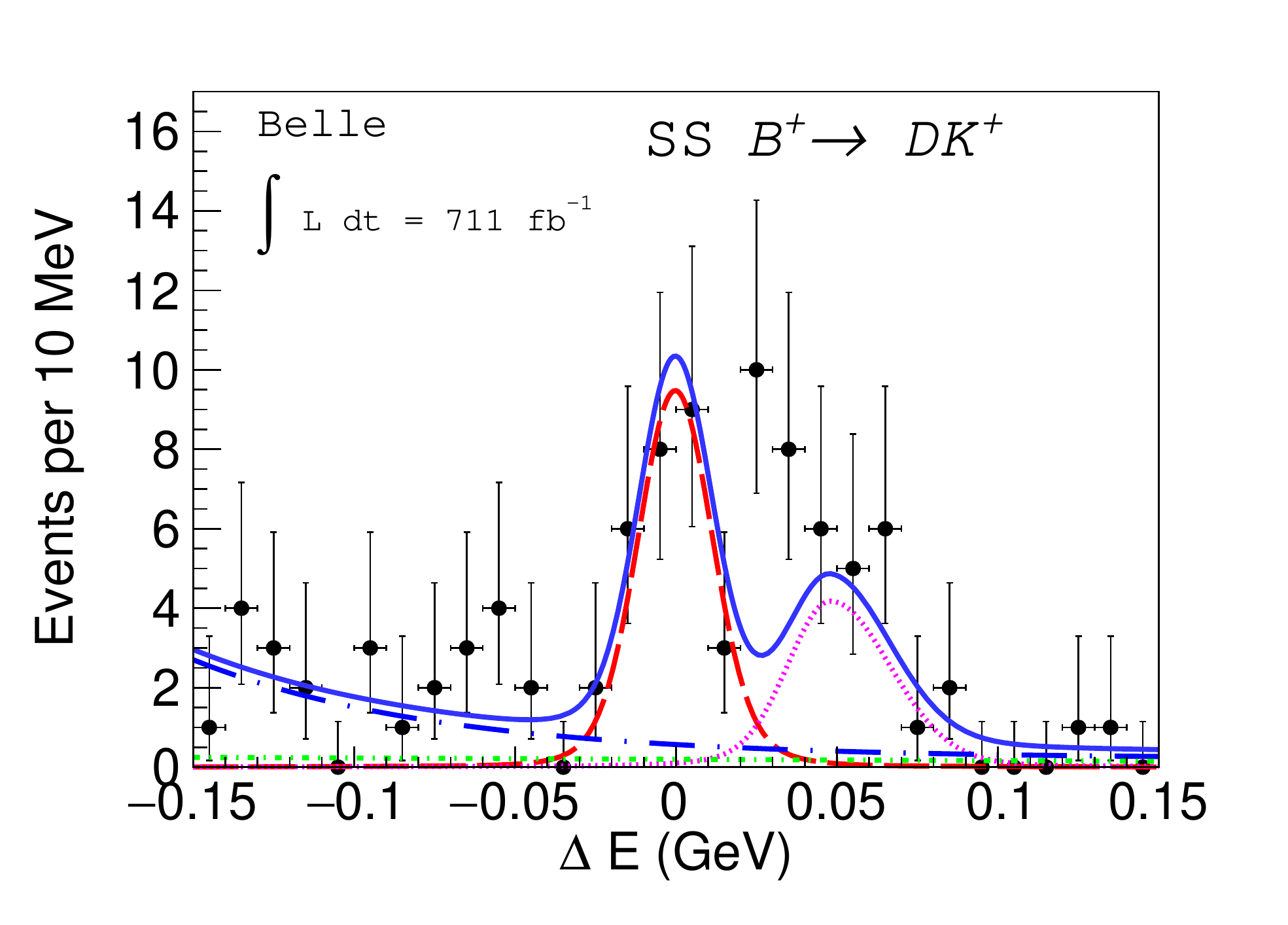}\put(40,55){}\end{overpic}
  \begin{overpic}[width=0.49\textwidth]{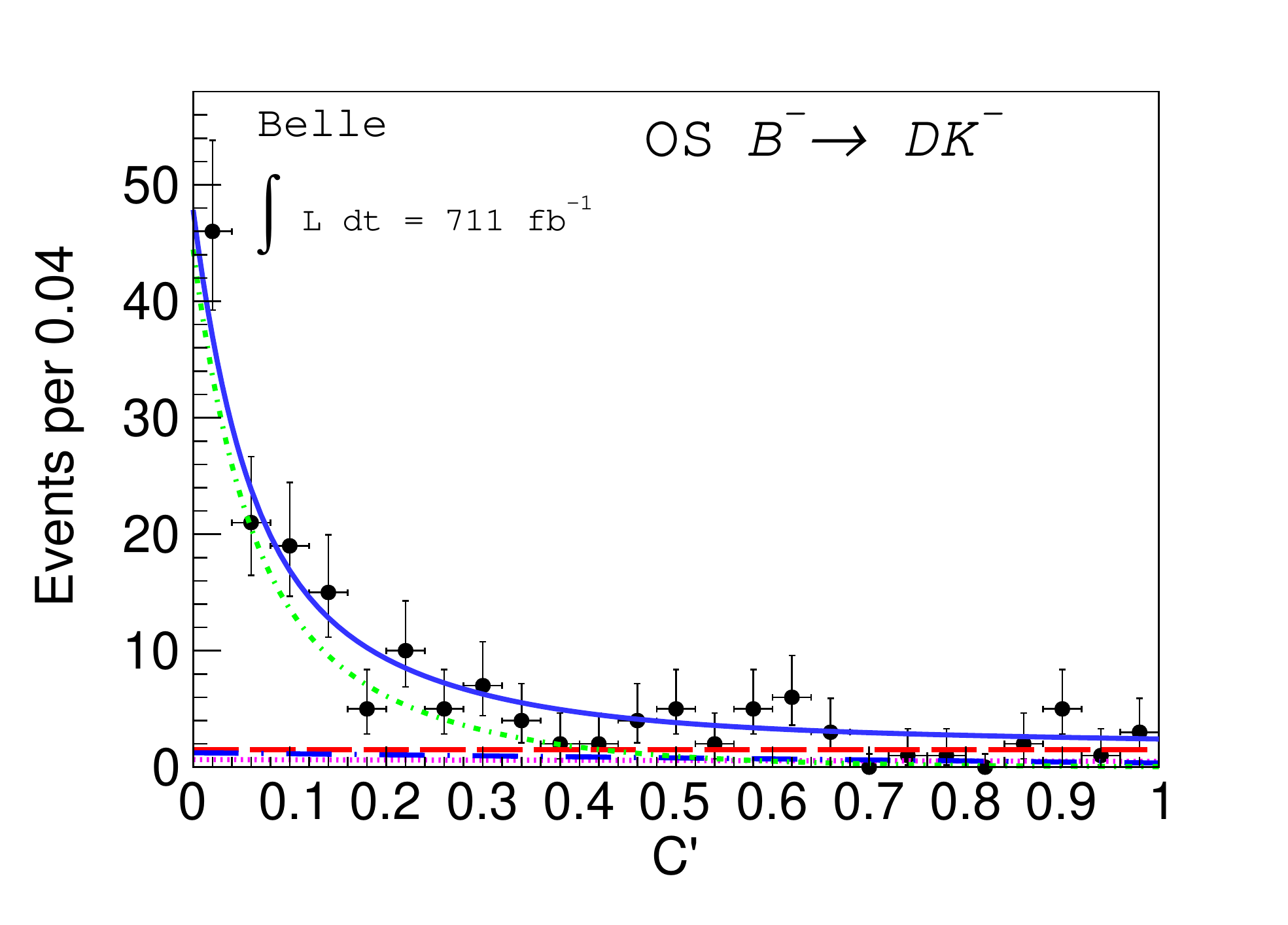}\put(40,55){}\end{overpic}
  \begin{overpic}[width=0.49\textwidth]{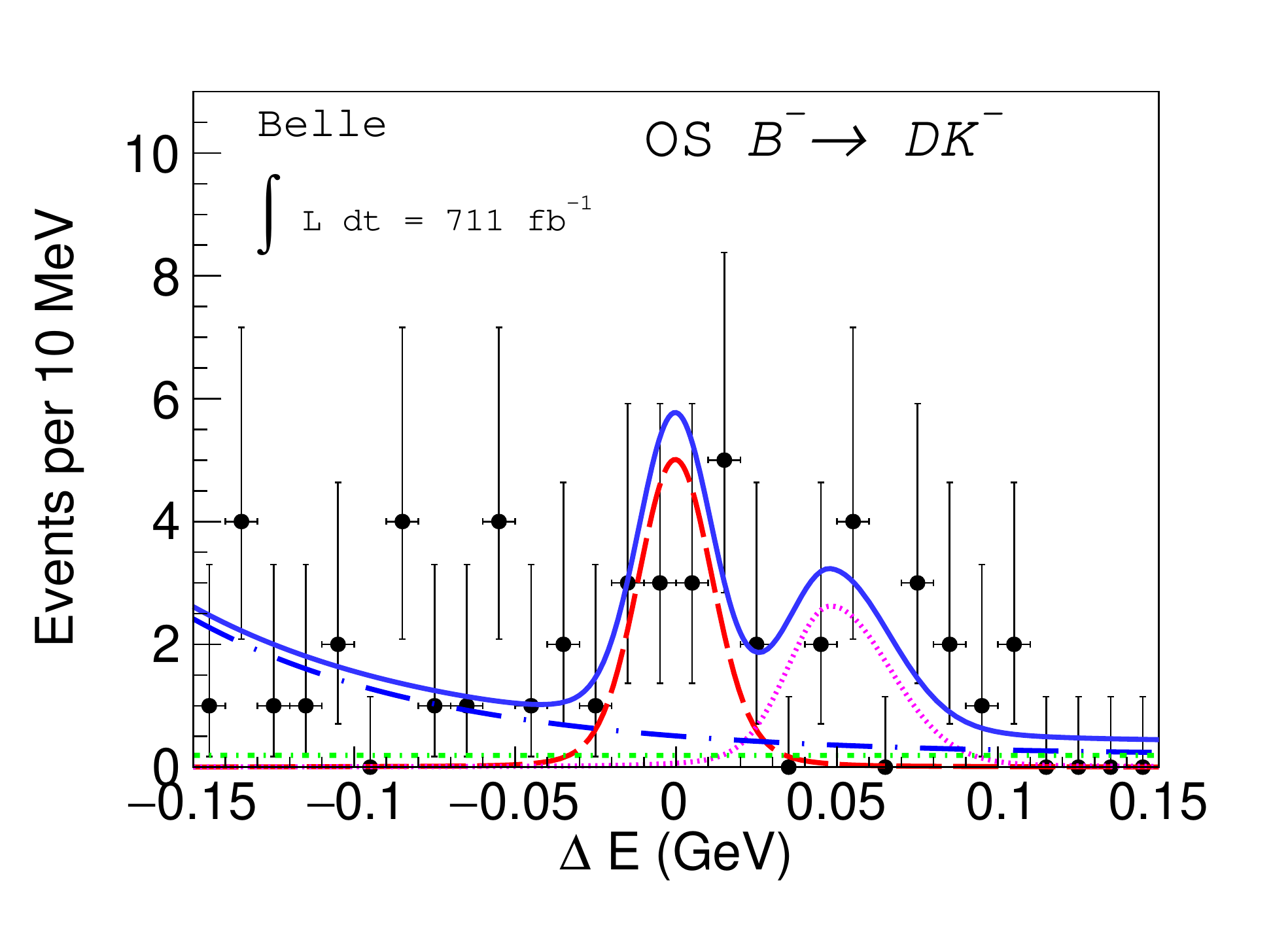}\put(40,55){}\end{overpic}
  \begin{overpic}[width=0.49\textwidth]{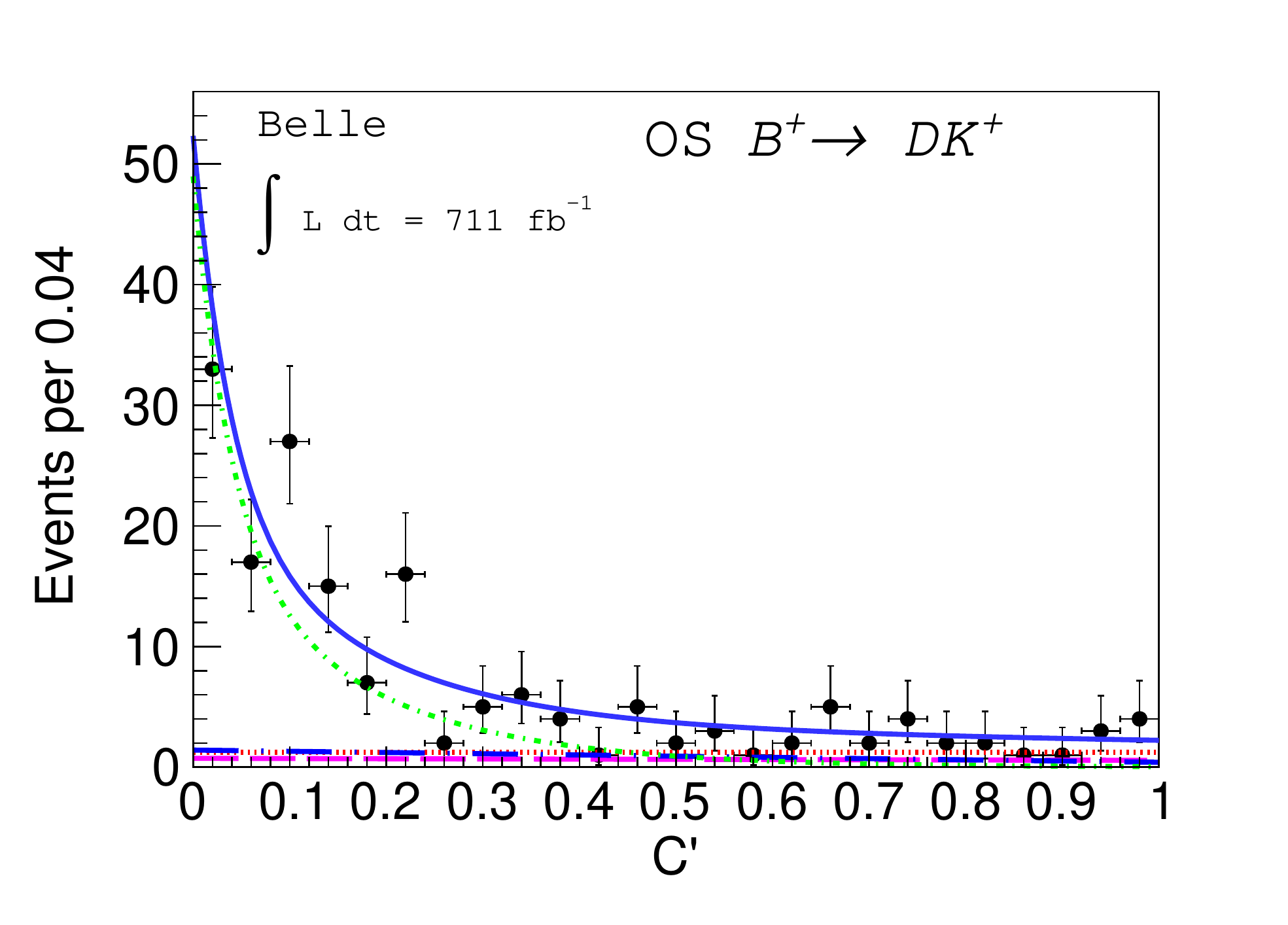}\put(40,55){}\end{overpic}
  \begin{overpic}[width=0.49\textwidth]{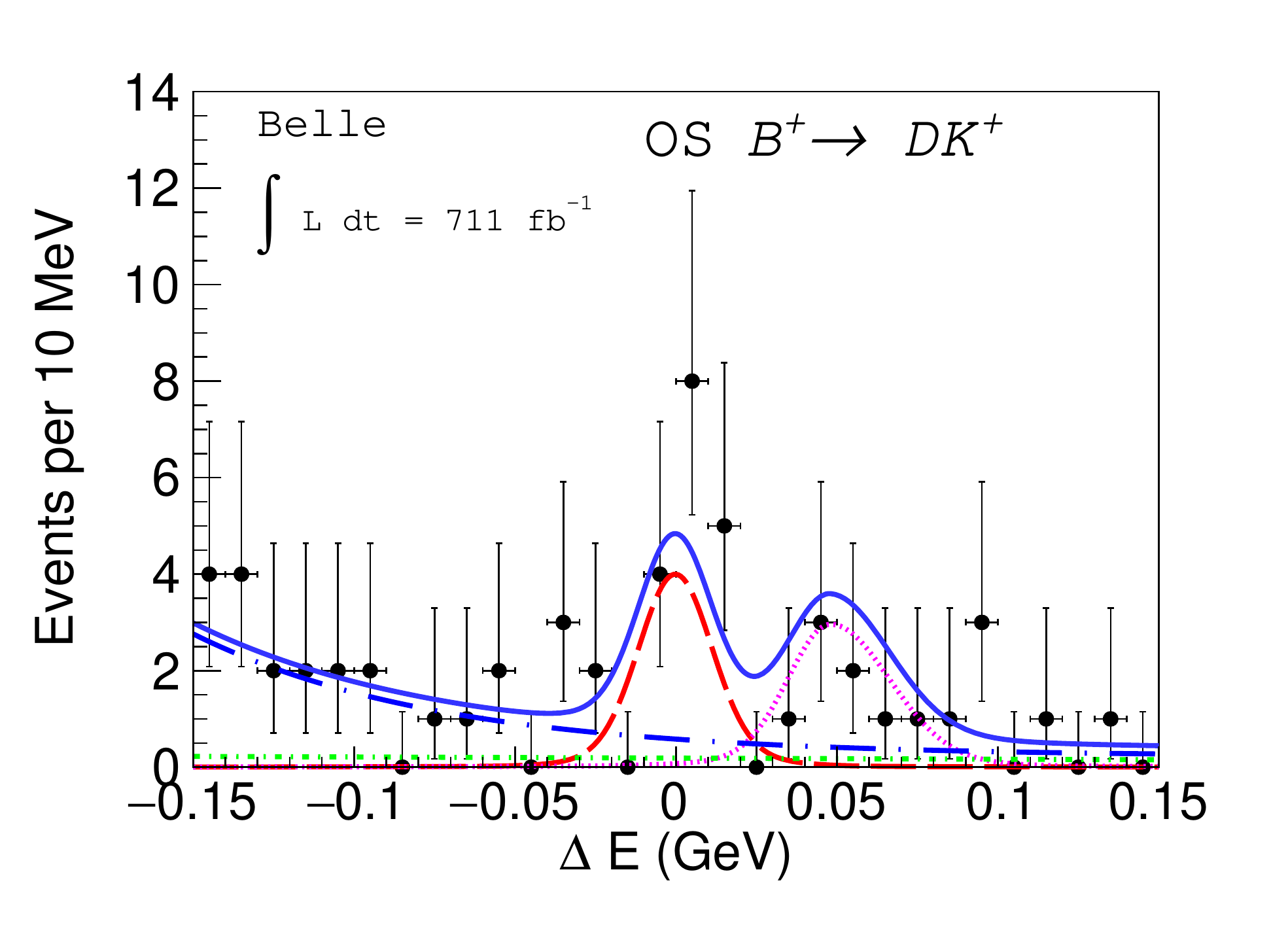}\put(40,55){}\end{overpic}

  \caption{\FullResultsFigureCaption{\Kpm}{Belle}}
  \label{fig:PHSPB1DK}
\end{figure}

\begin{figure}[H]
  \centering
  
  \begin{overpic}[width=0.49\textwidth]{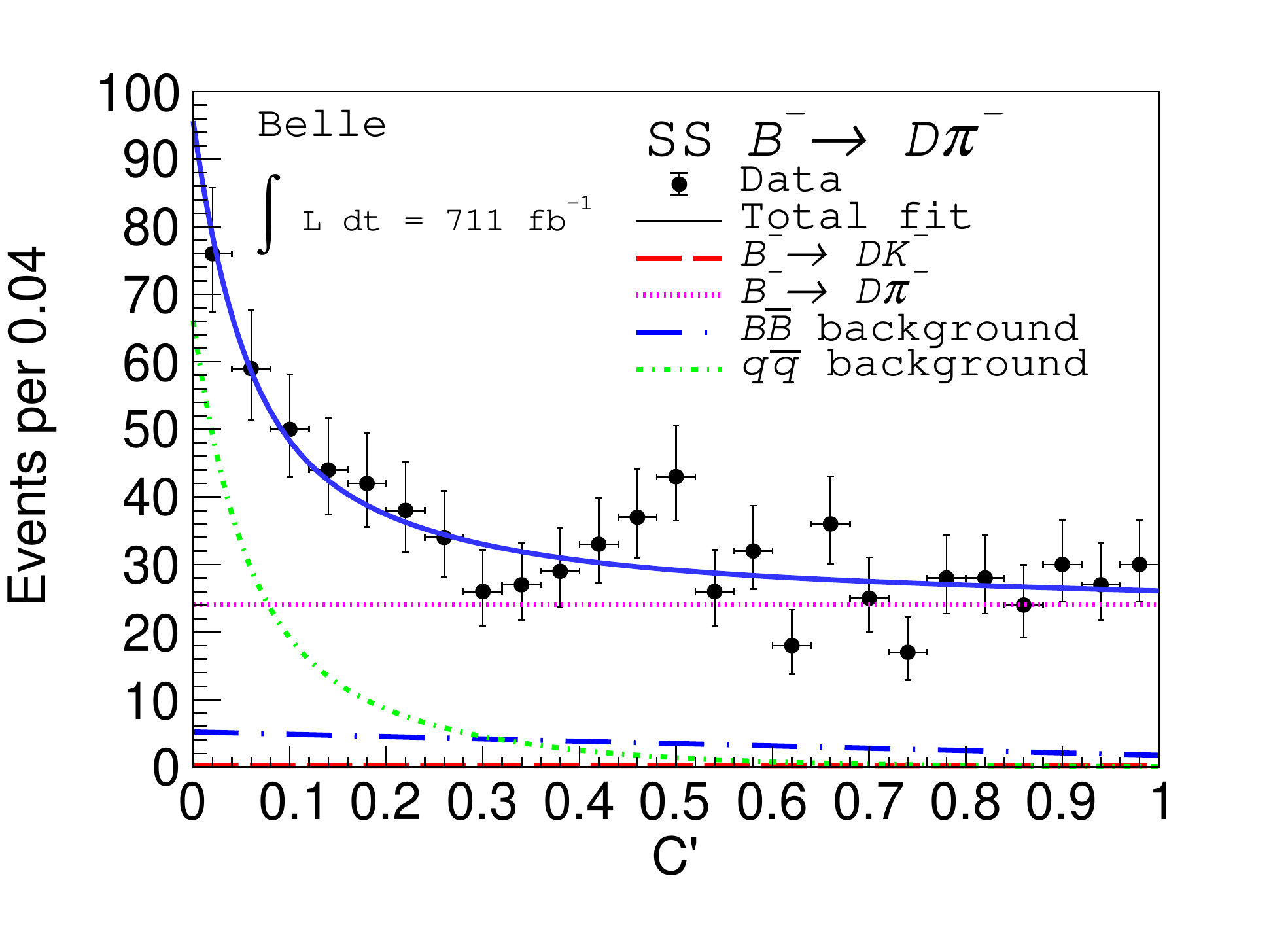}\put(40,55){}\end{overpic}
  \begin{overpic}[width=0.49\textwidth]{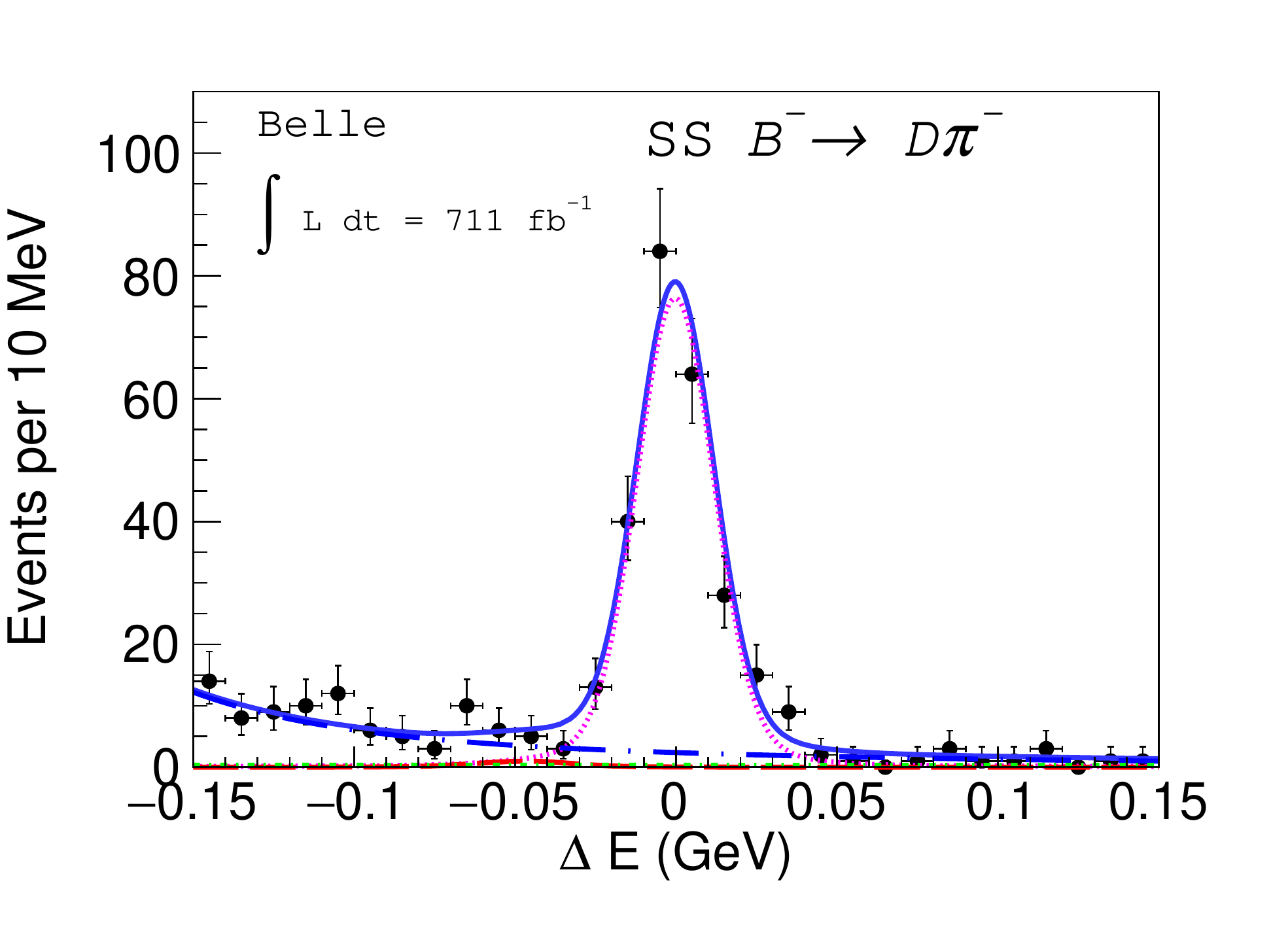}\put(40,55){}\end{overpic}
  \begin{overpic}[width=0.49\textwidth]{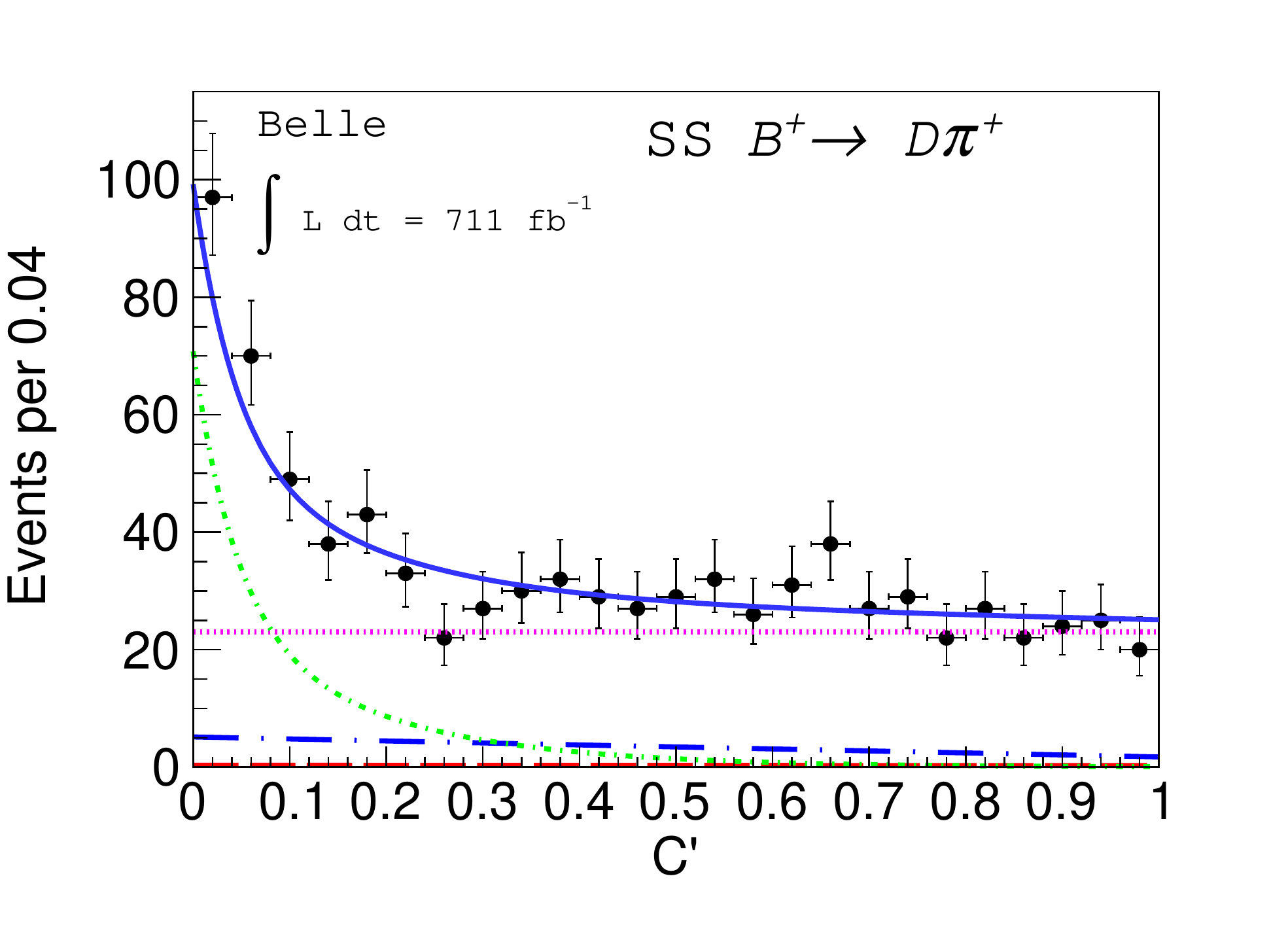}\put(40,55){}\end{overpic}
  \begin{overpic}[width=0.49\textwidth]{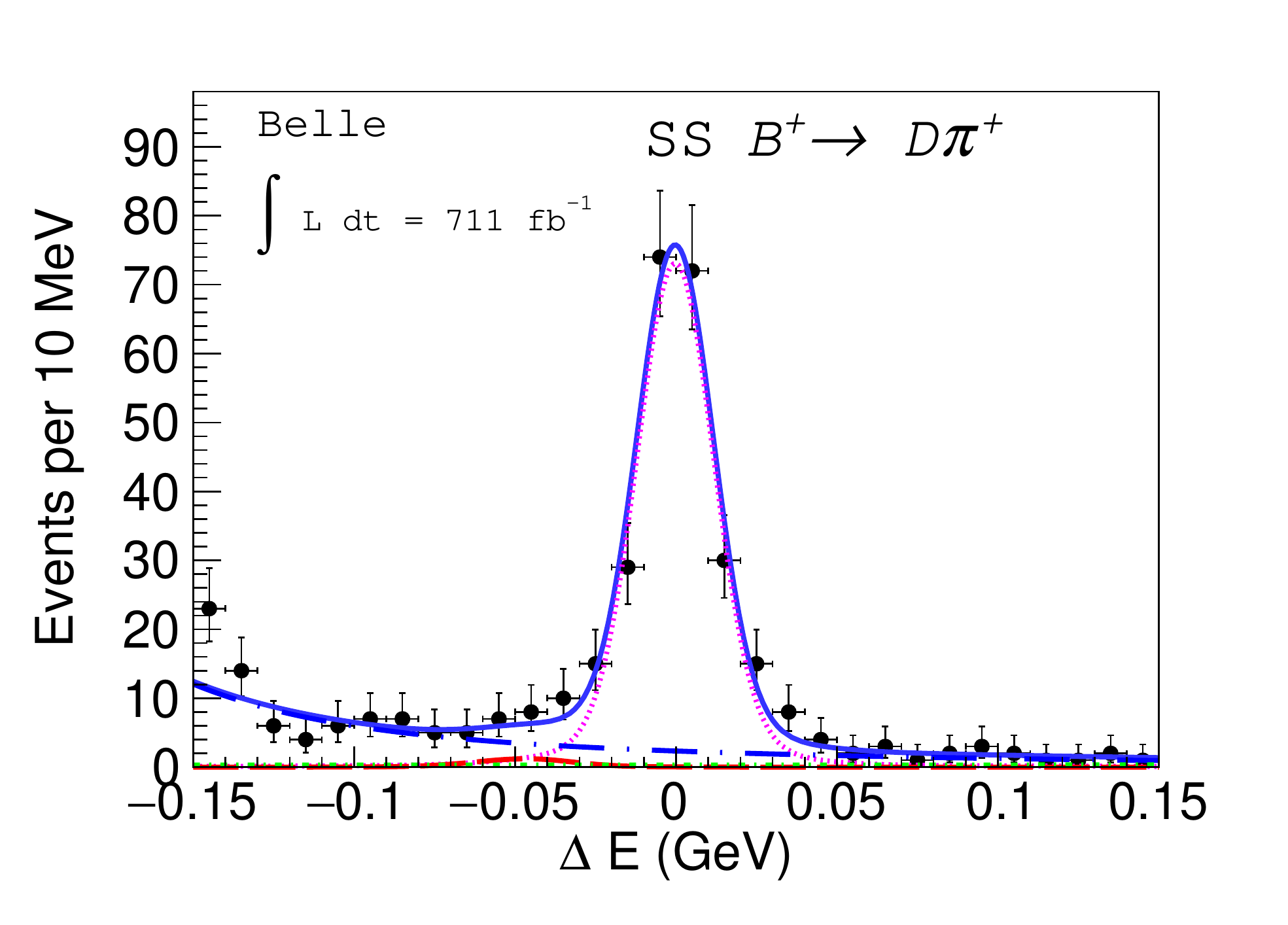}\put(40,55){}\end{overpic}
  \begin{overpic}[width=0.49\textwidth]{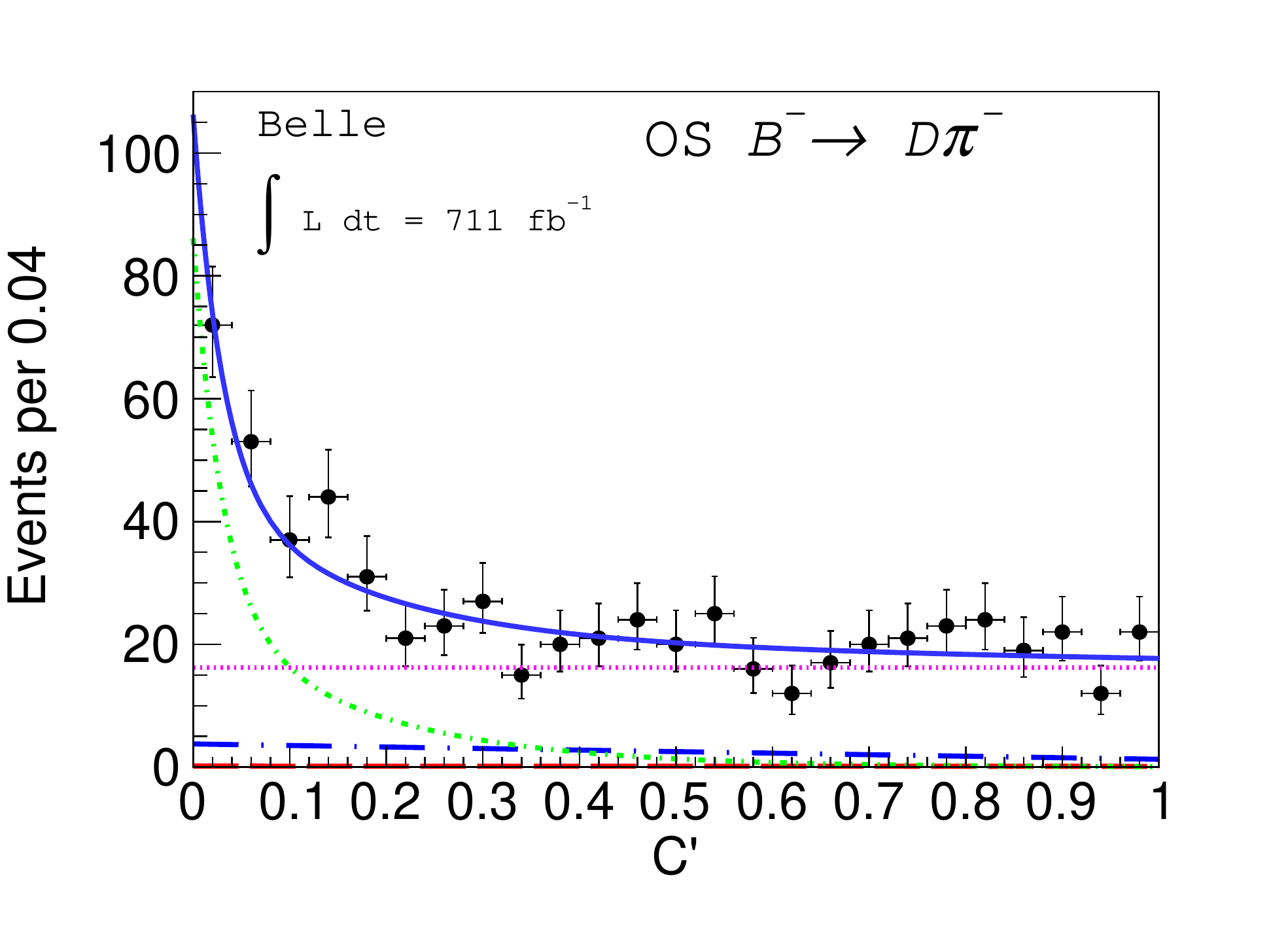}\put(40,55){}\end{overpic}
  \begin{overpic}[width=0.49\textwidth]{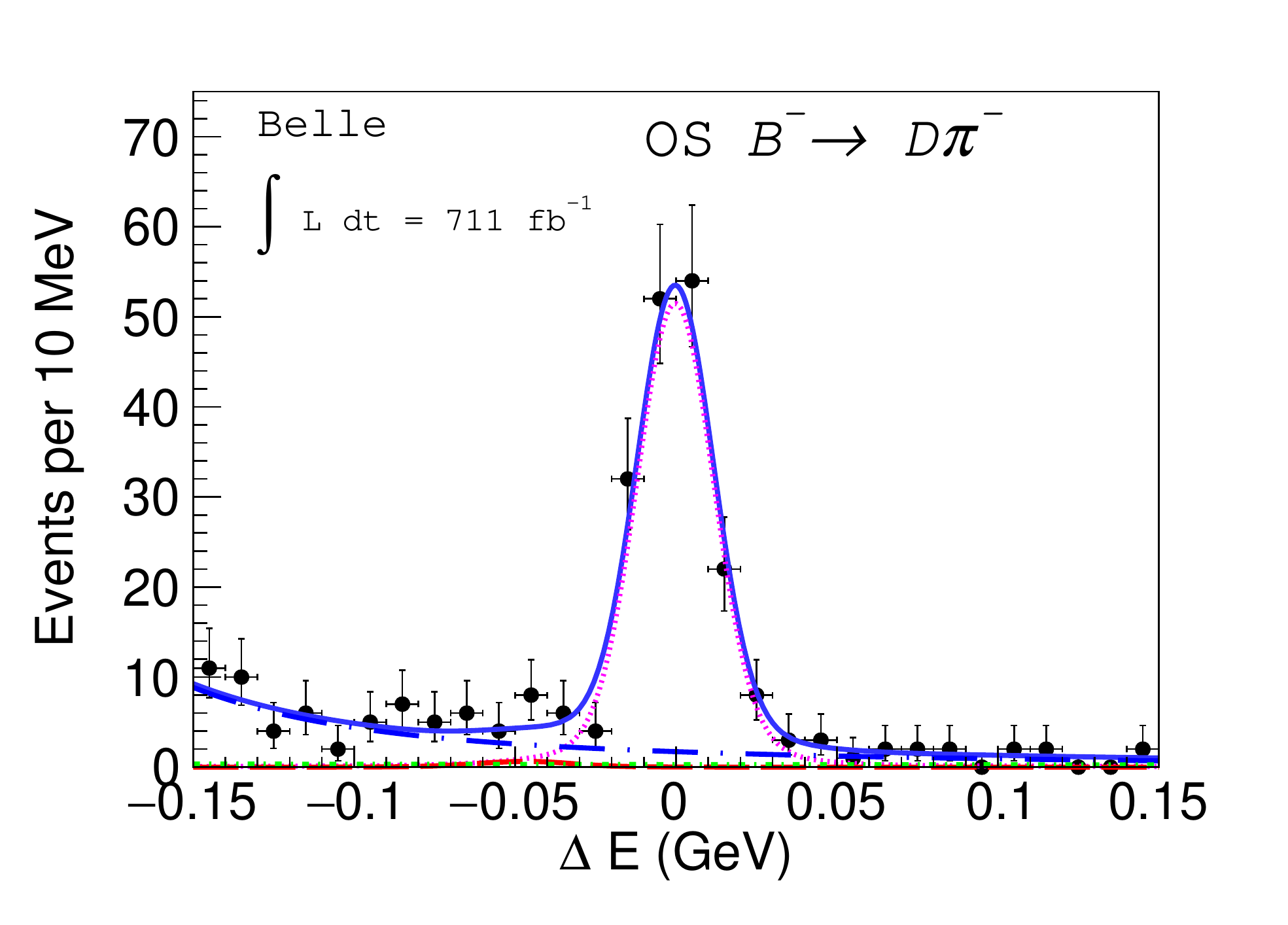}\put(40,55){}\end{overpic}
  \begin{overpic}[width=0.49\textwidth]{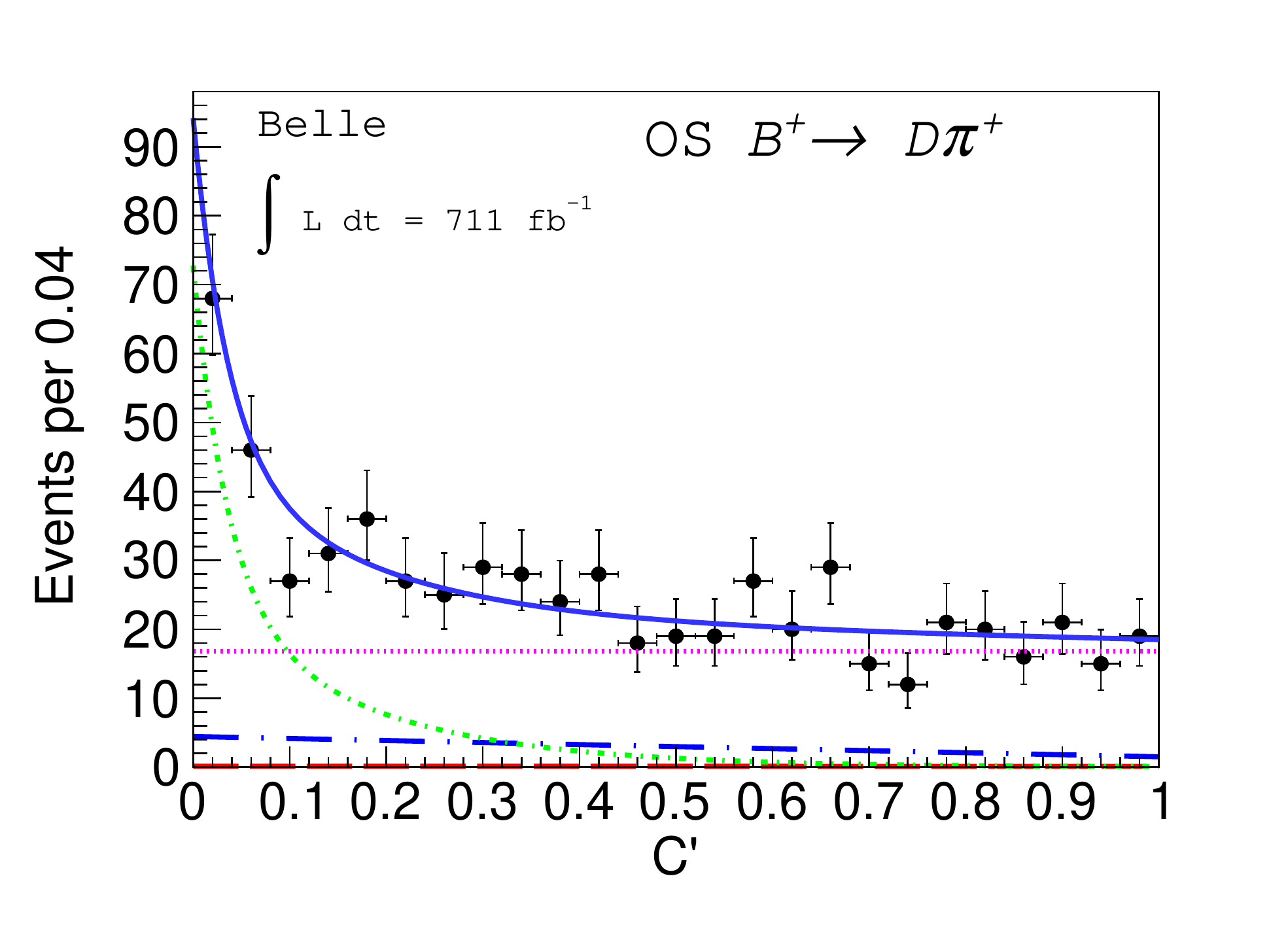}\put(40,55){}\end{overpic}
  \begin{overpic}[width=0.49\textwidth]{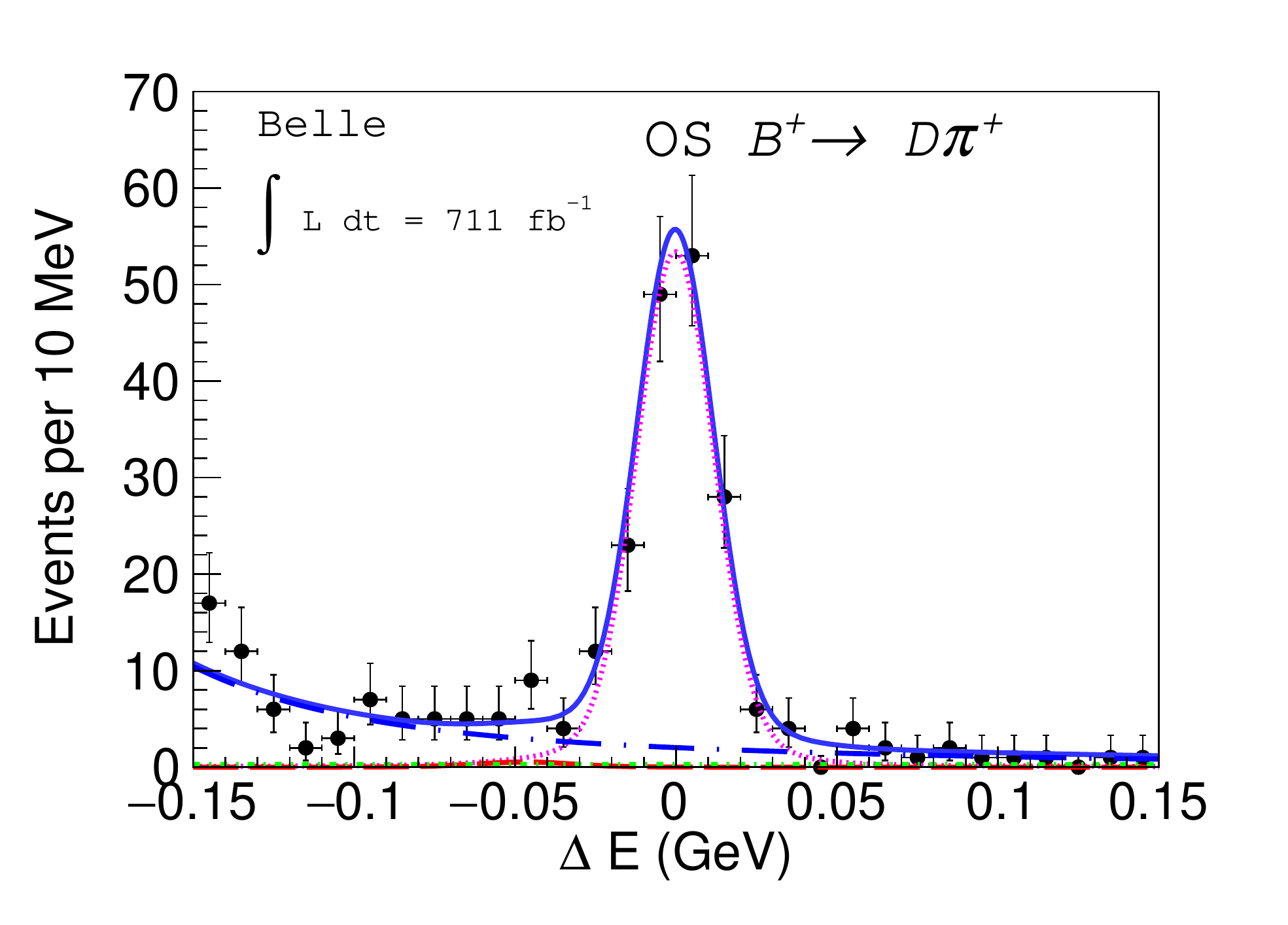}\put(40,55){}\end{overpic}

  \caption{\FullResultsFigureCaption{\pipm}{Belle}}
  \label{fig:PHSPB1DPi}
\end{figure}

\begin{figure}[H]
  \centering

  \begin{overpic}[width=0.49\textwidth]{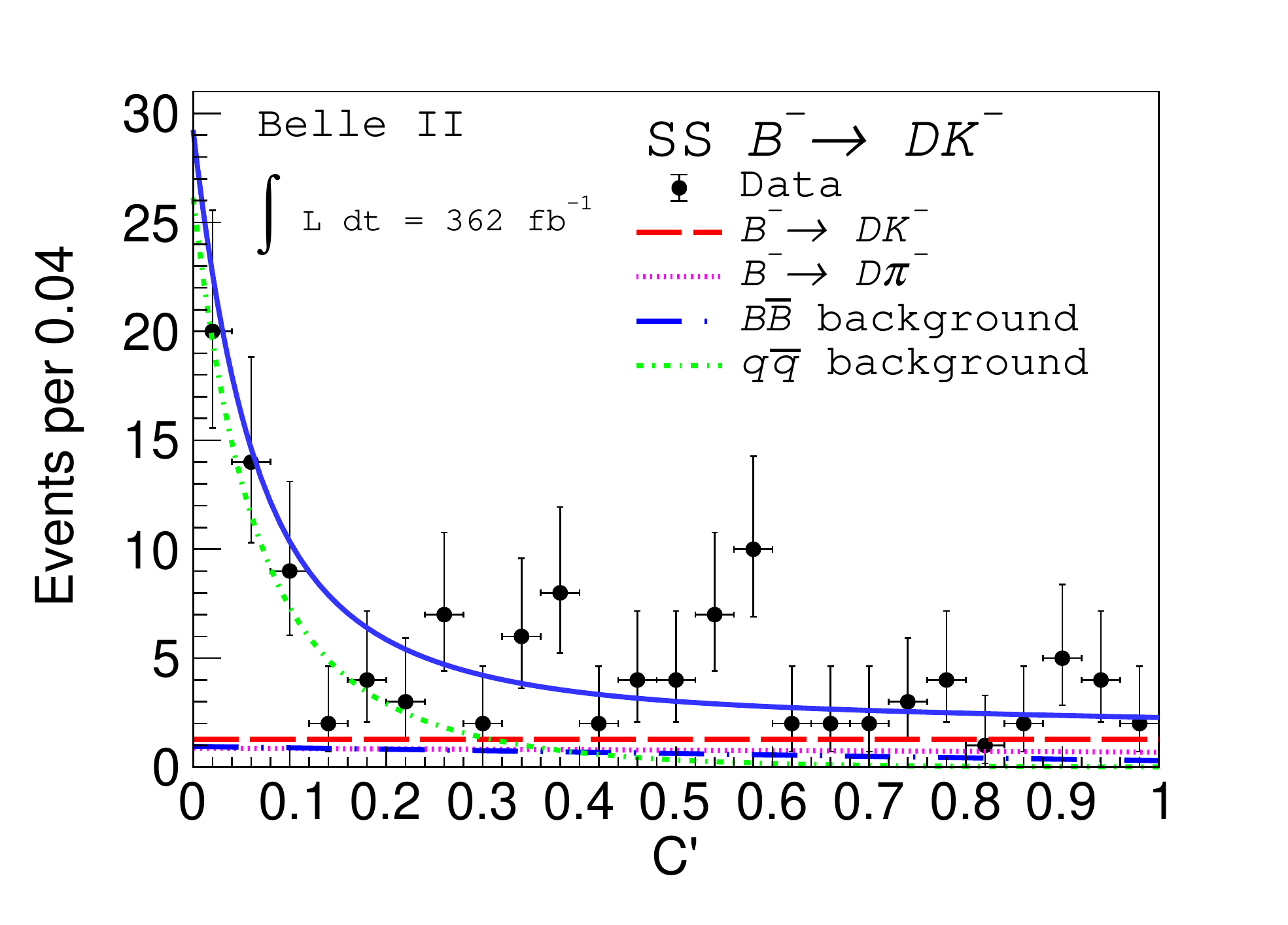}\put(40,55){}\end{overpic}
  \begin{overpic}[width=0.49\textwidth]{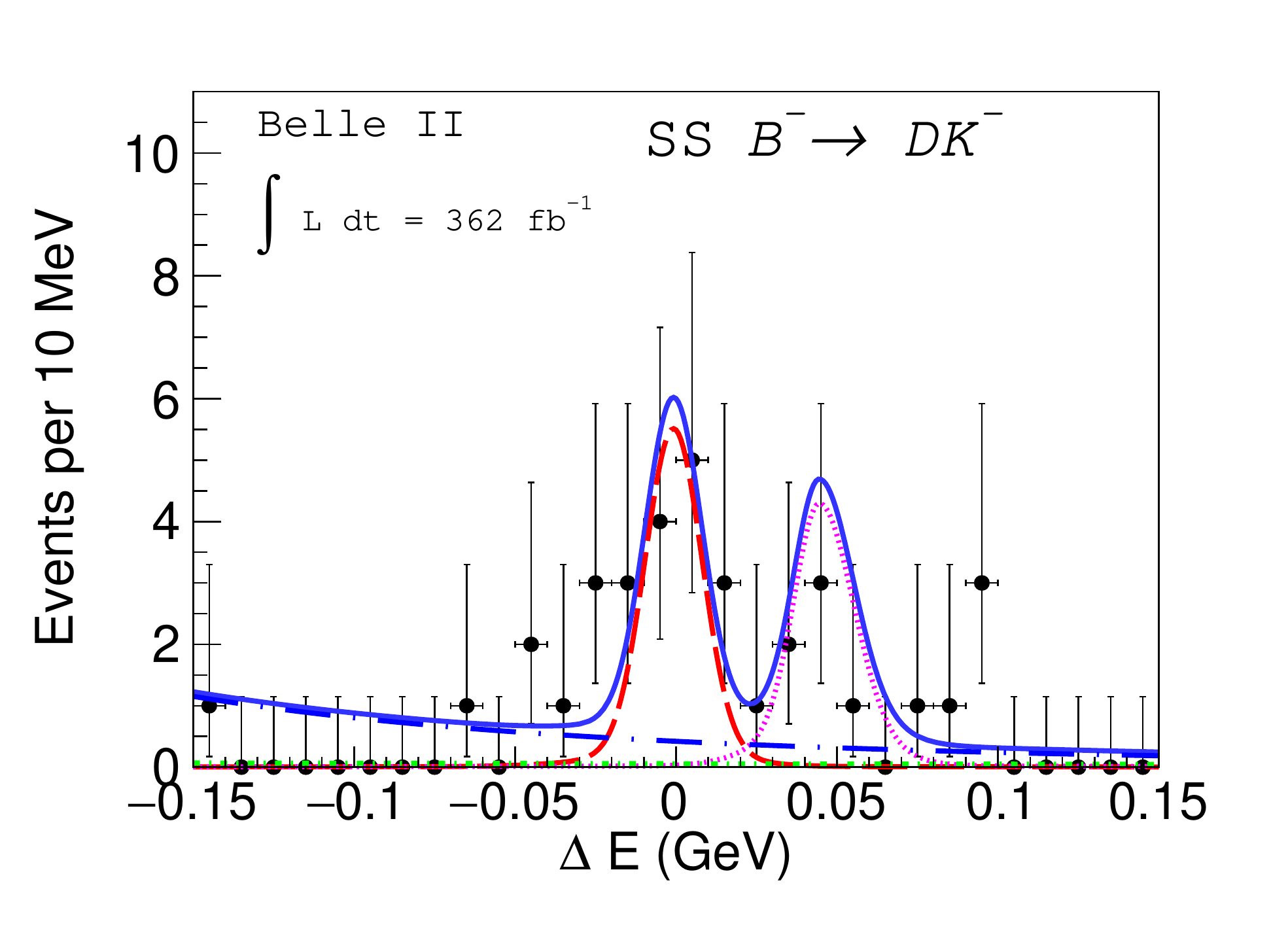}\put(40,55){}\end{overpic}
  \begin{overpic}[width=0.49\textwidth]{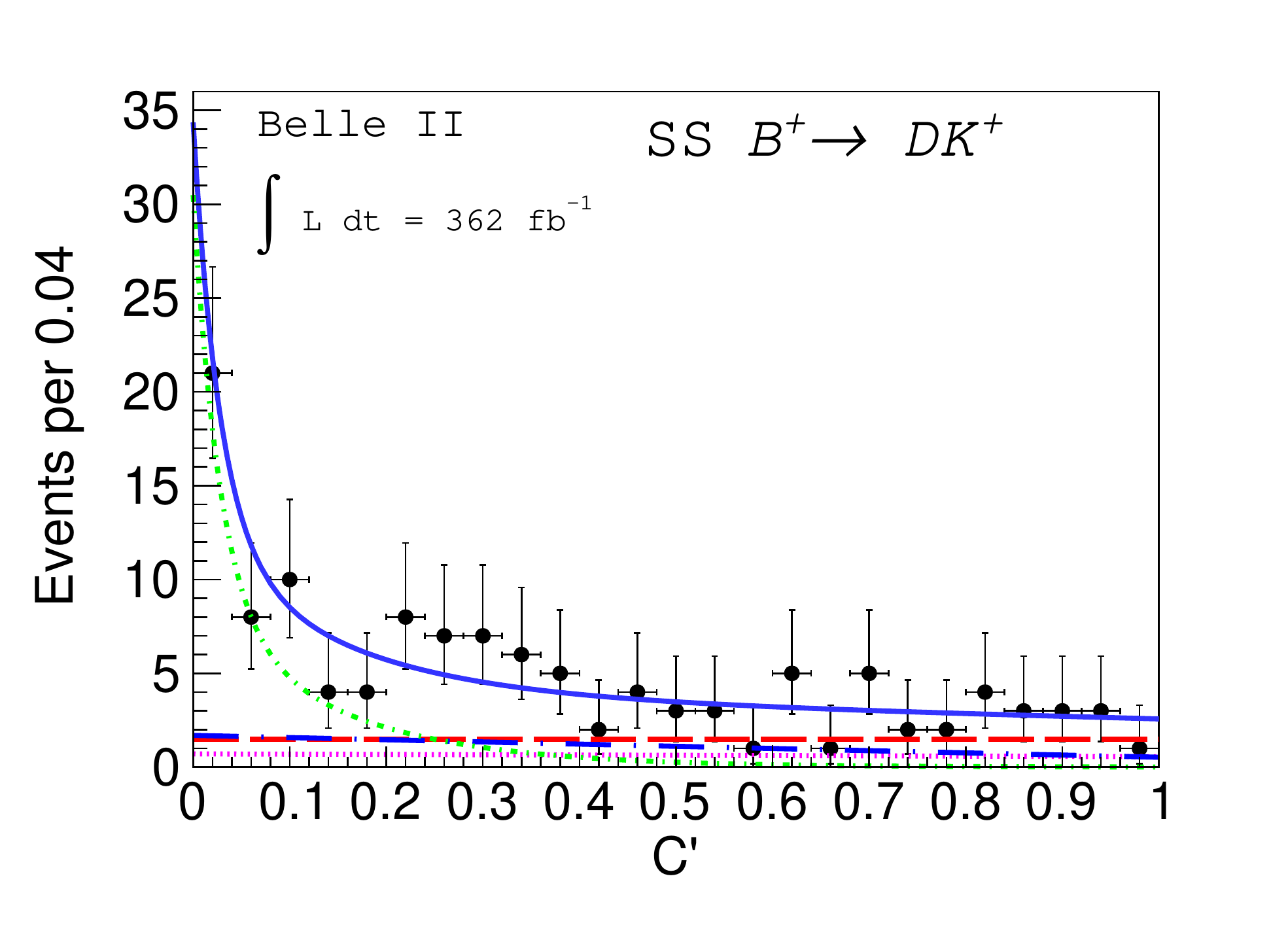}\put(40,55){}\end{overpic}
  \begin{overpic}[width=0.49\textwidth]{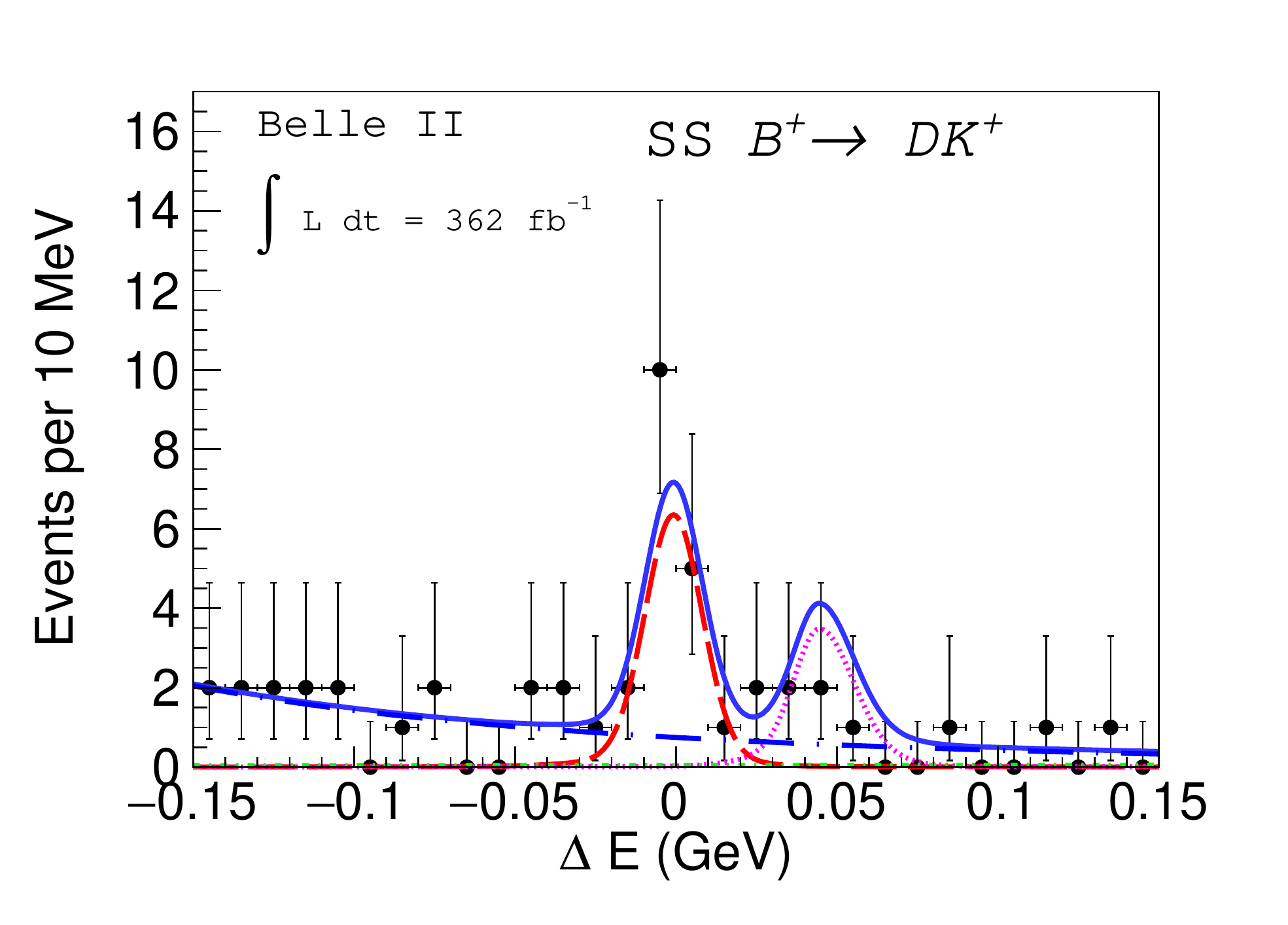}\put(40,55){}\end{overpic}
  \begin{overpic}[width=0.49\textwidth]{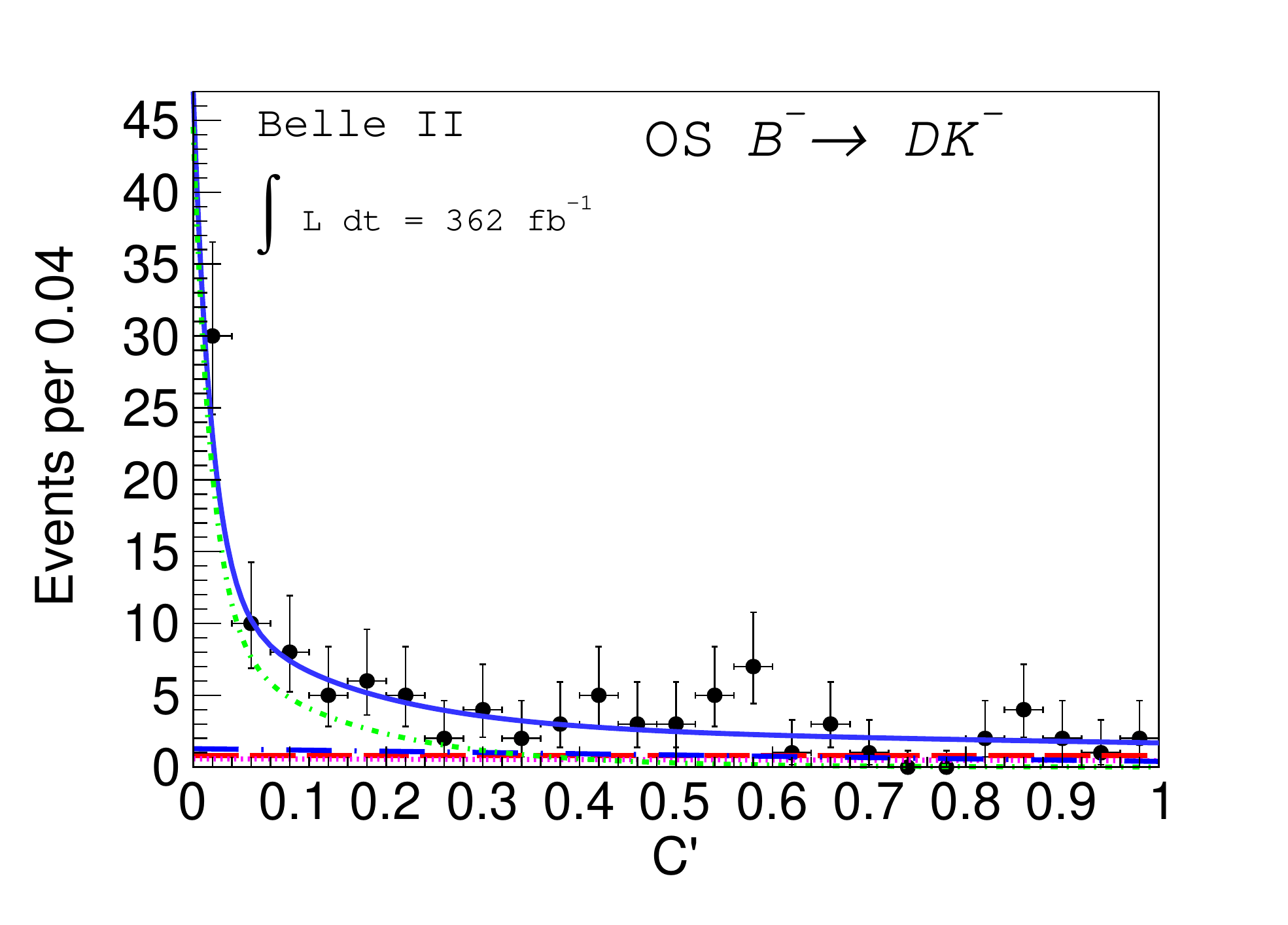}\put(40,55){}\end{overpic}
  \begin{overpic}[width=0.49\textwidth]{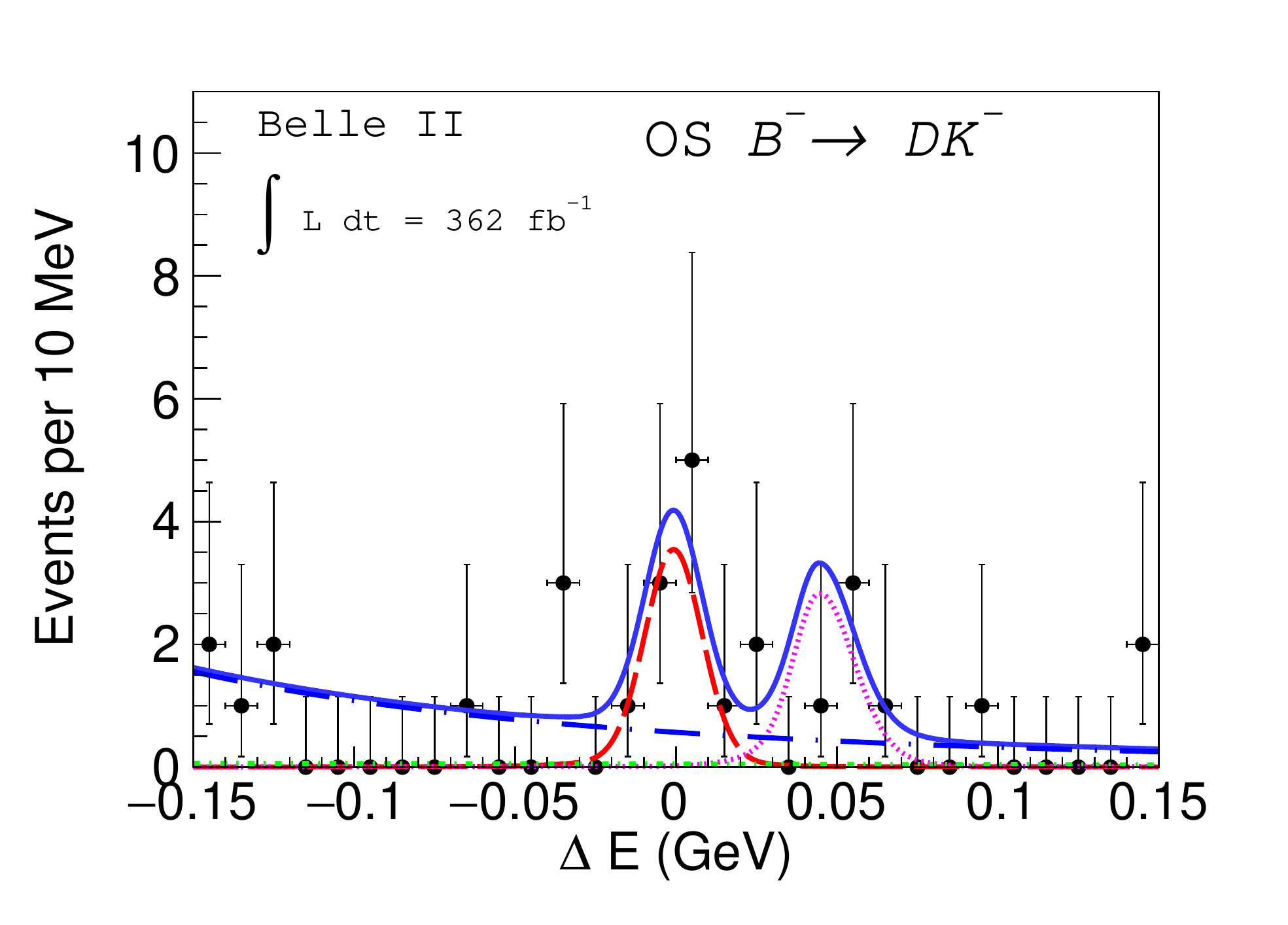}\put(40,55){}\end{overpic}
  \begin{overpic}[width=0.49\textwidth]{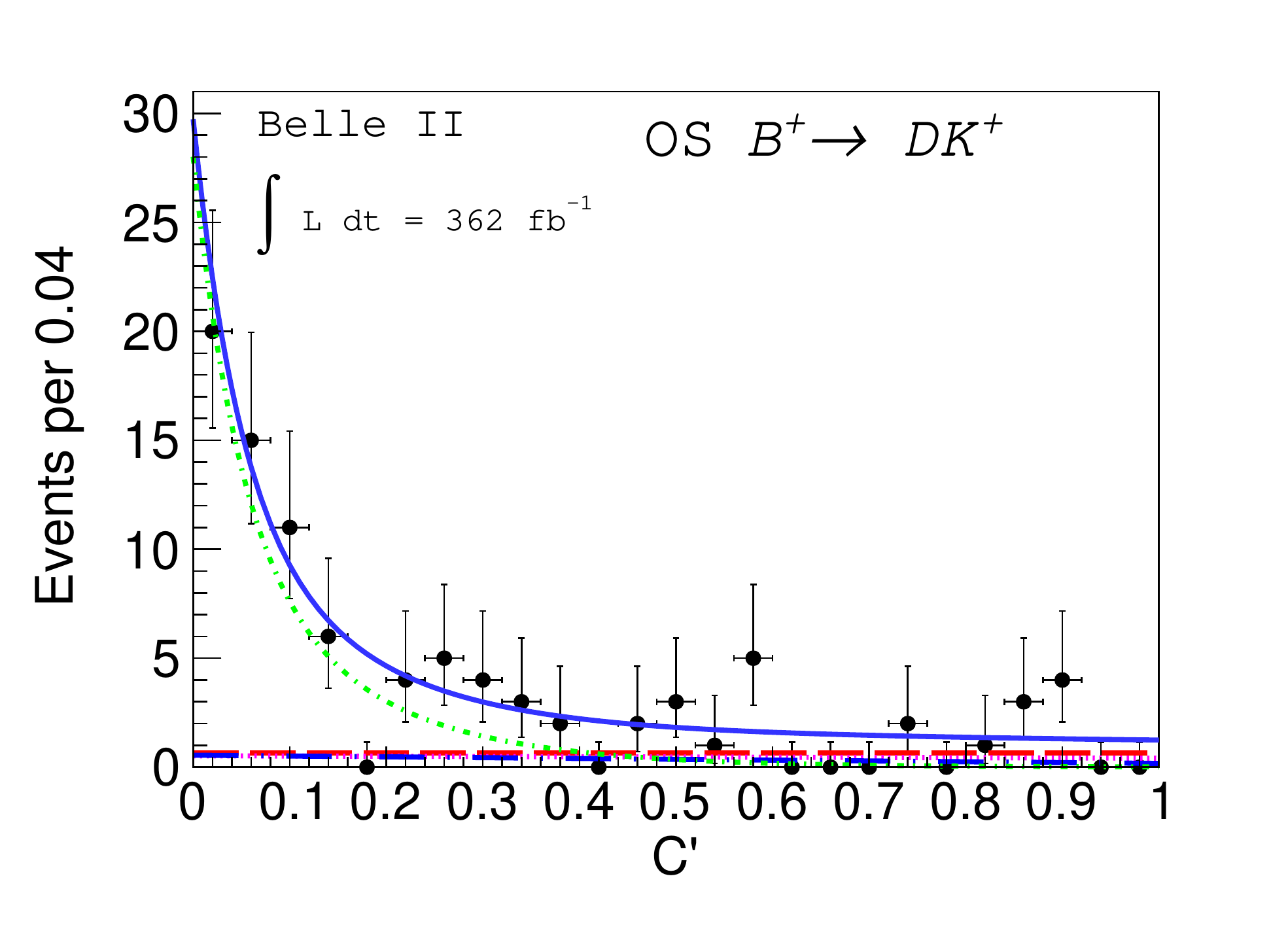}\put(40,55){}\end{overpic}
  \begin{overpic}[width=0.49\textwidth]{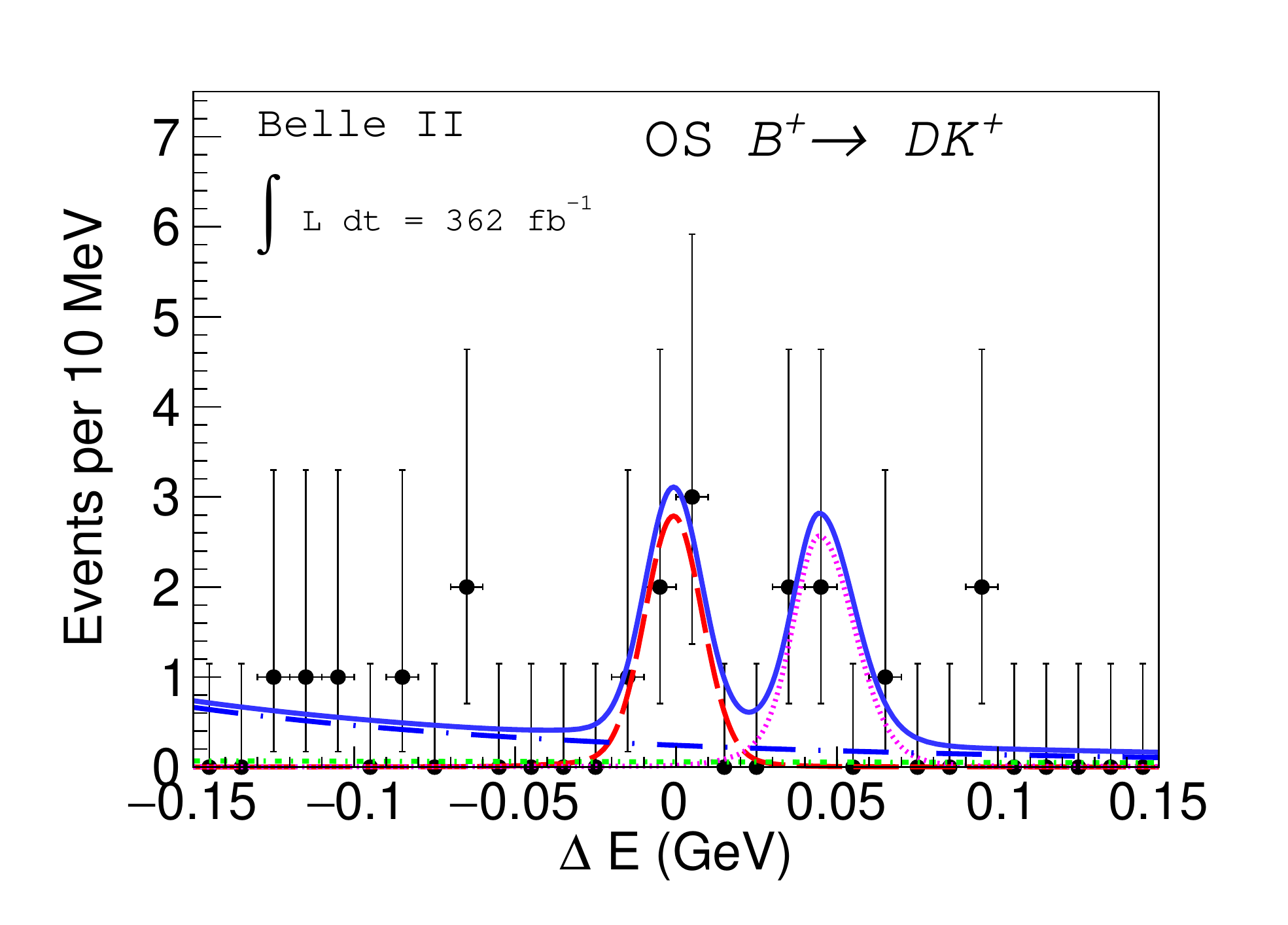}\put(40,55){}\end{overpic}

  \caption{\FullResultsFigureCaption{\Kpm}{Belle~II}}
  \label{fig:PHSPB2DK}
\end{figure}

\begin{figure}[H]
  \centering

  \begin{overpic}[width=0.49\textwidth]{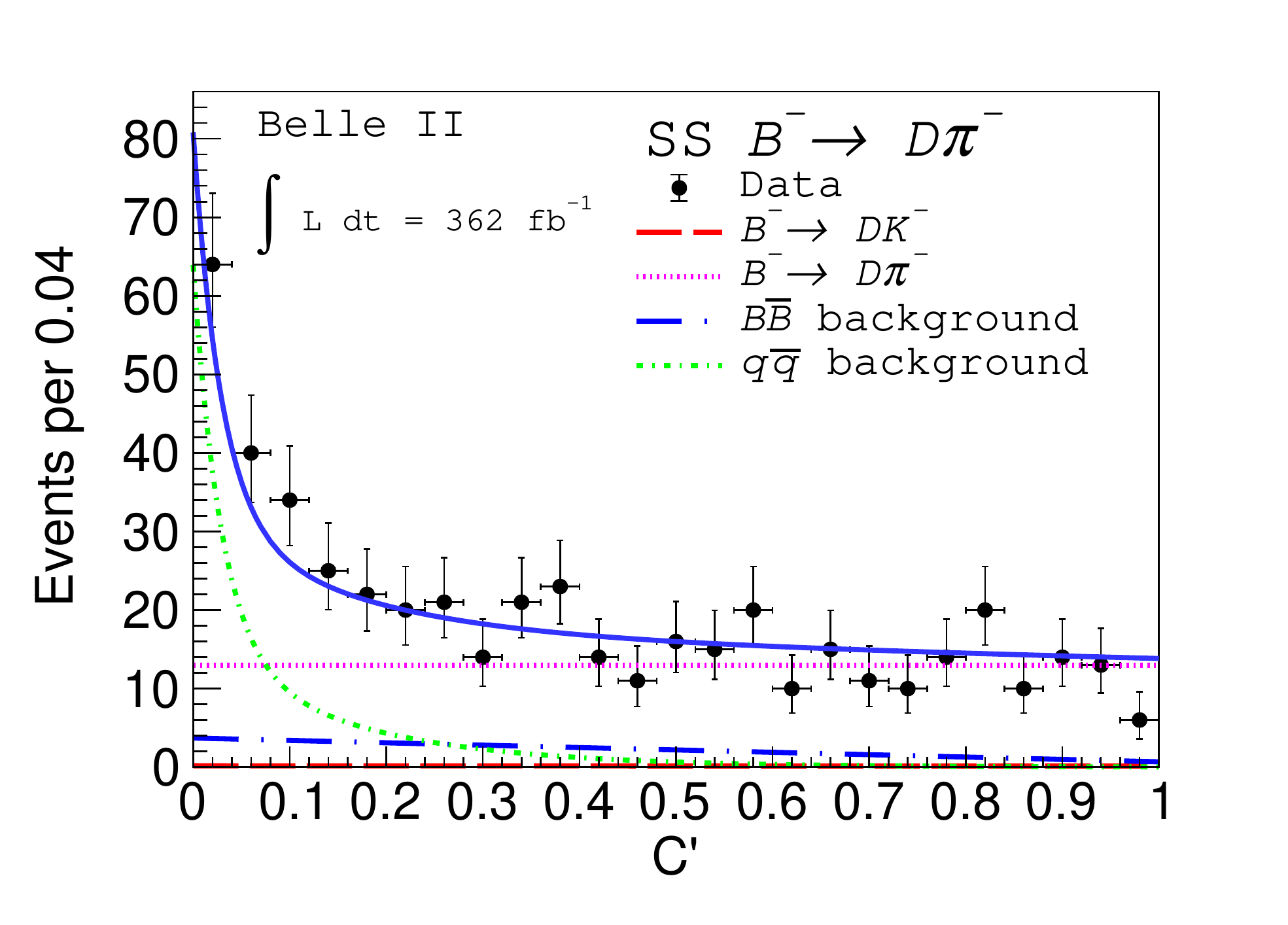}\put(40,55){}\end{overpic}
  \begin{overpic}[width=0.49\textwidth]{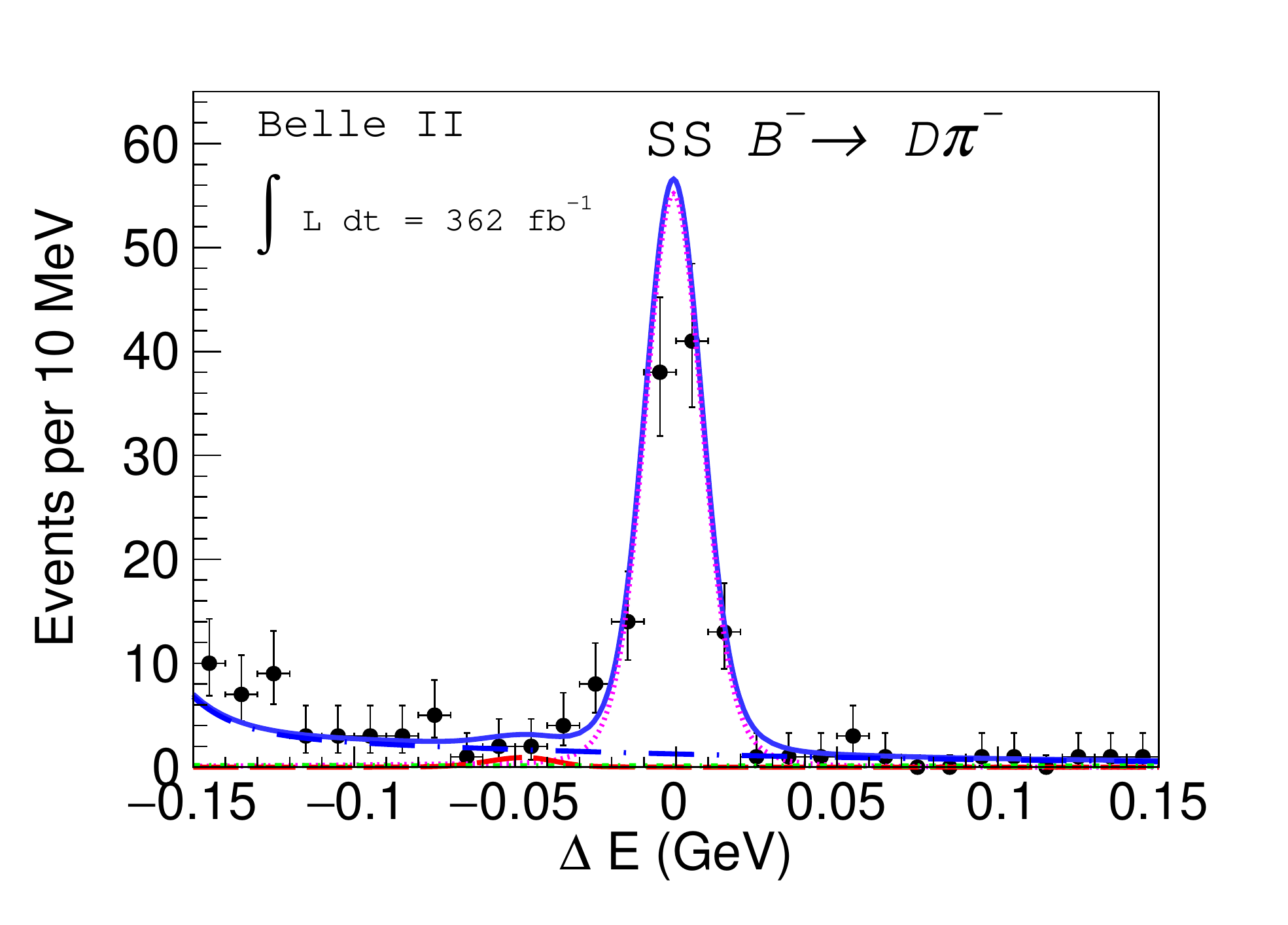}\put(40,55){}\end{overpic}
  \begin{overpic}[width=0.49\textwidth]{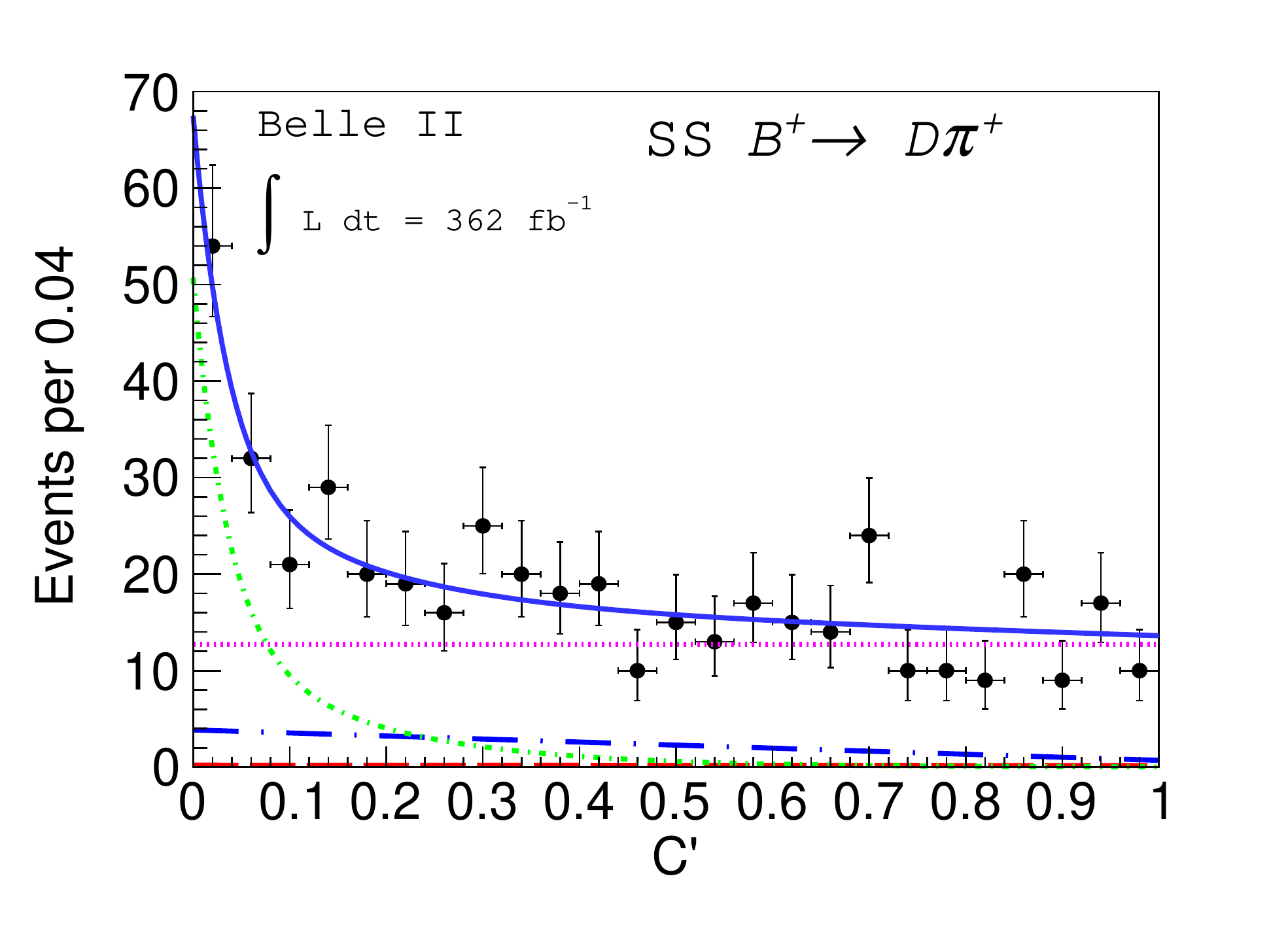}\put(40,55){}\end{overpic}
  \begin{overpic}[width=0.49\textwidth]{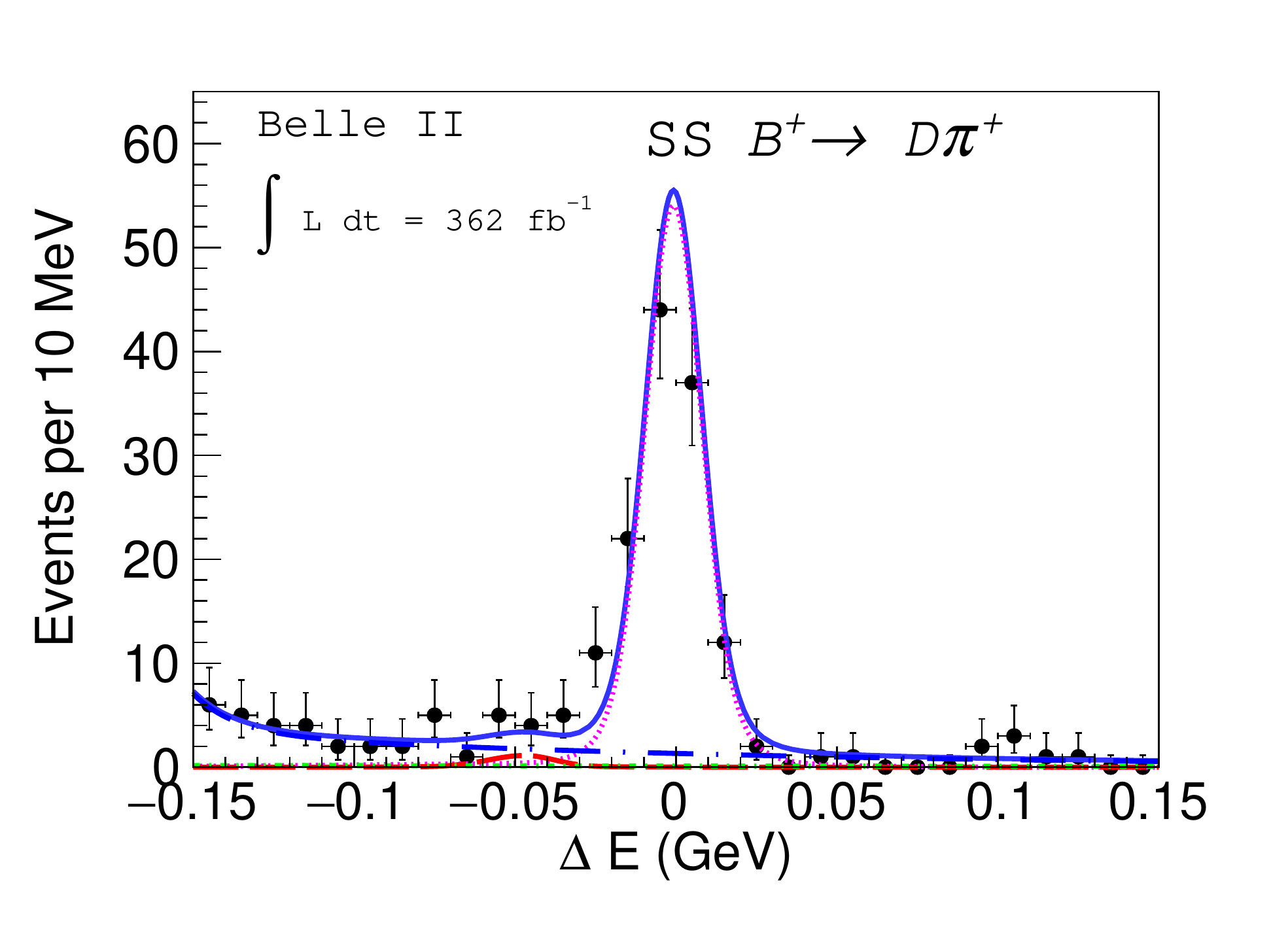}\put(40,55){}\end{overpic}
  \begin{overpic}[width=0.49\textwidth]{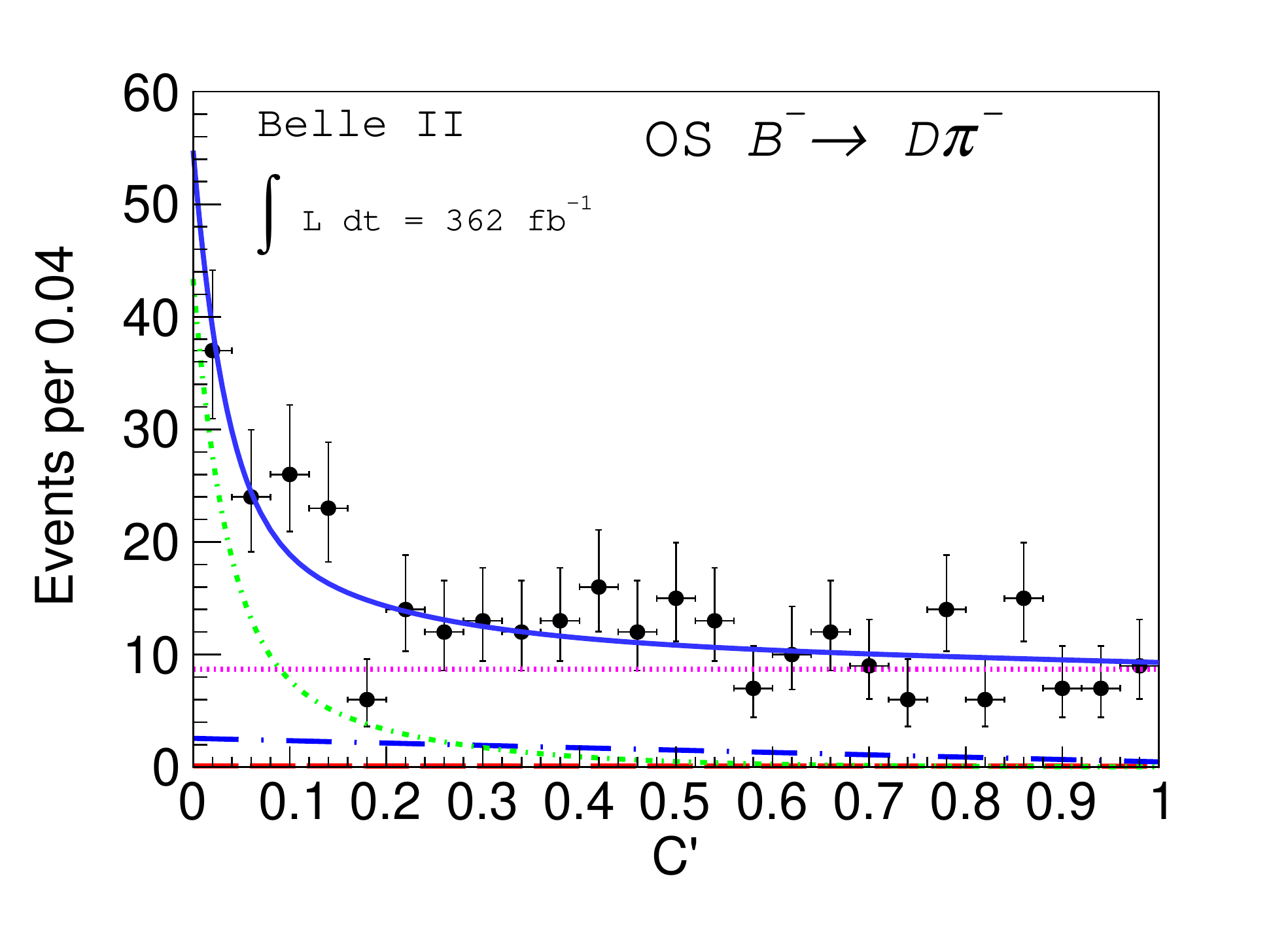}\put(40,55){}\end{overpic}
  \begin{overpic}[width=0.49\textwidth]{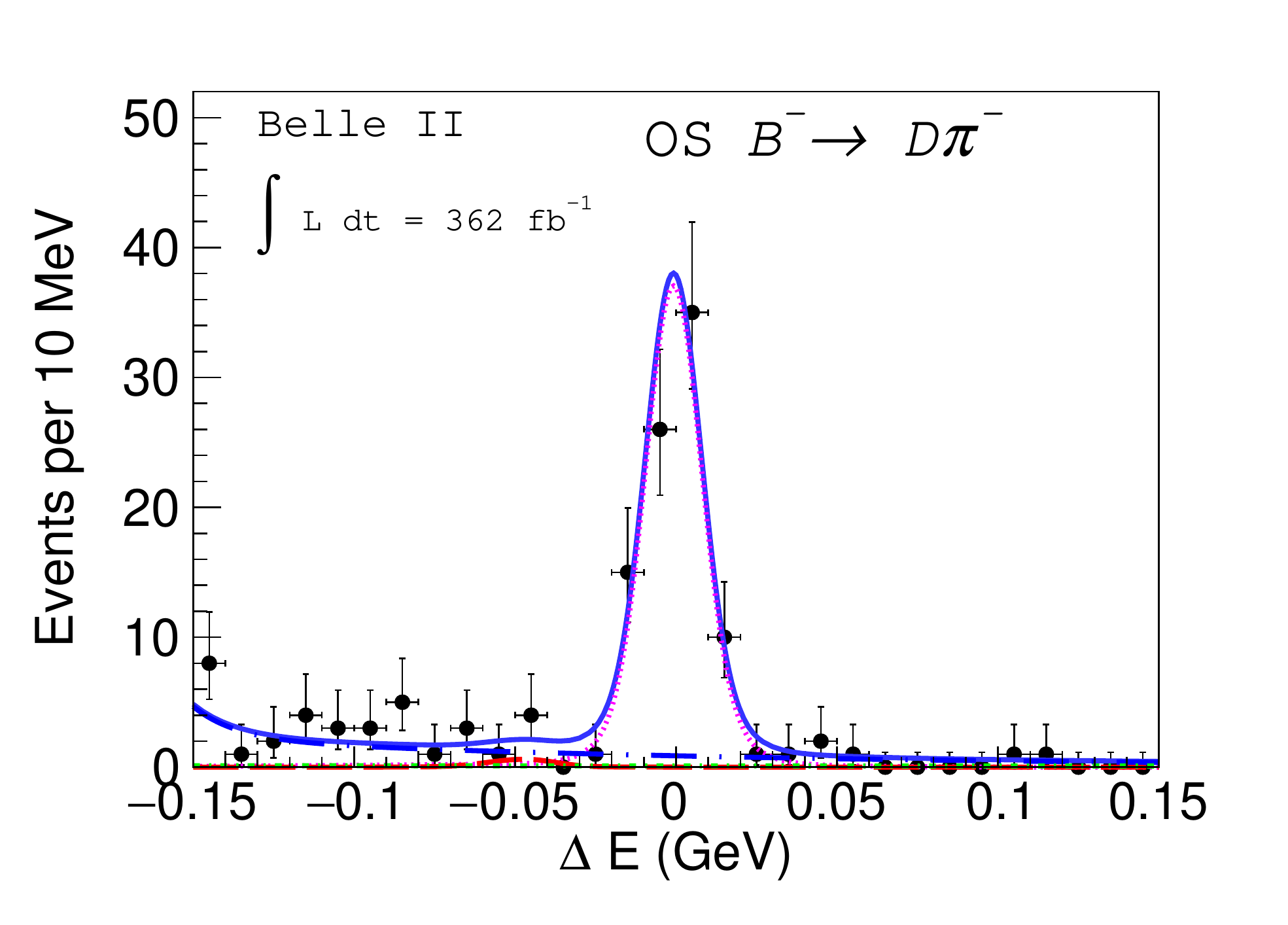}\put(40,55){}\end{overpic}
  \begin{overpic}[width=0.49\textwidth]{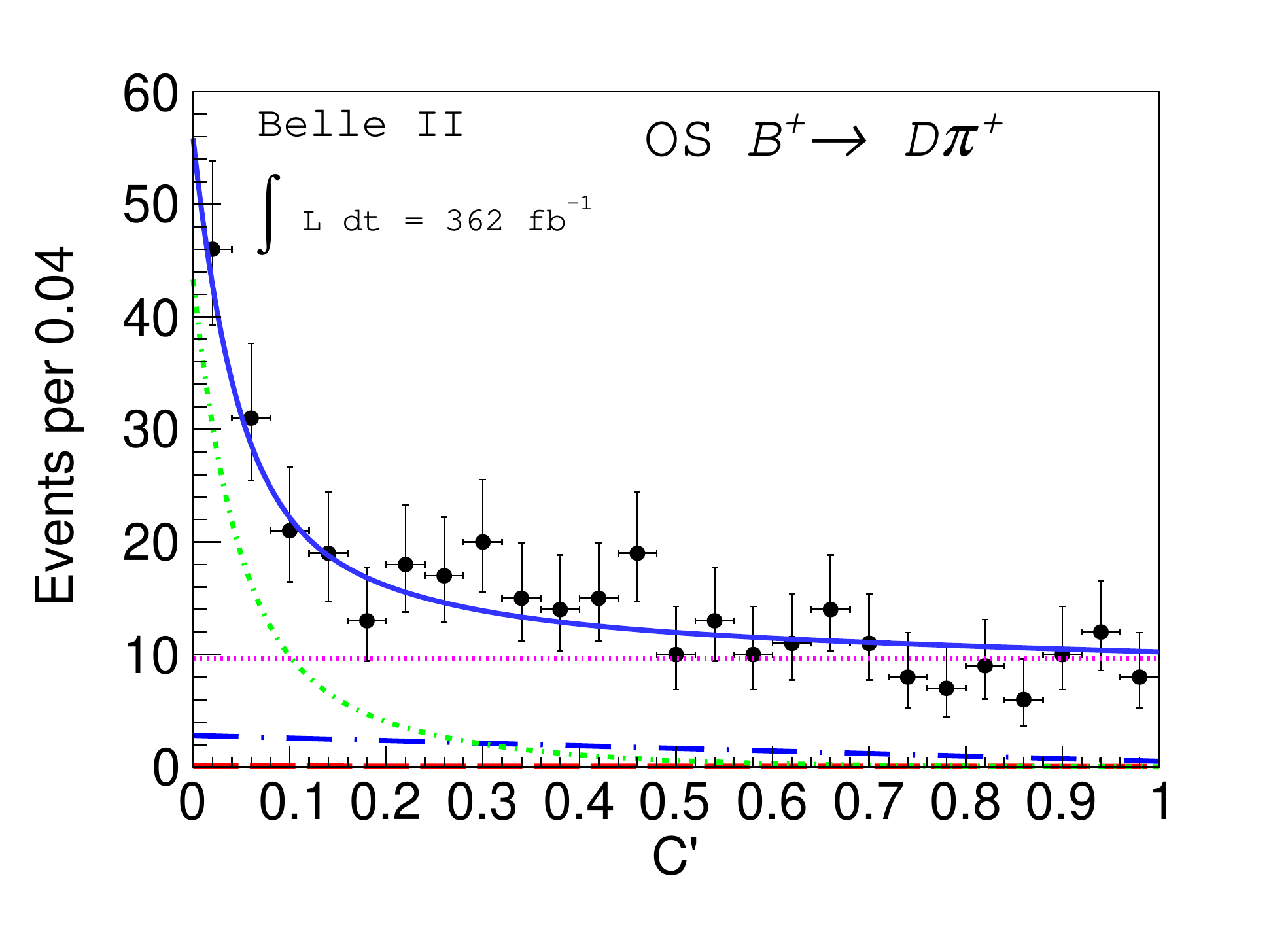}\put(40,55){}\end{overpic}
  \begin{overpic}[width=0.49\textwidth]{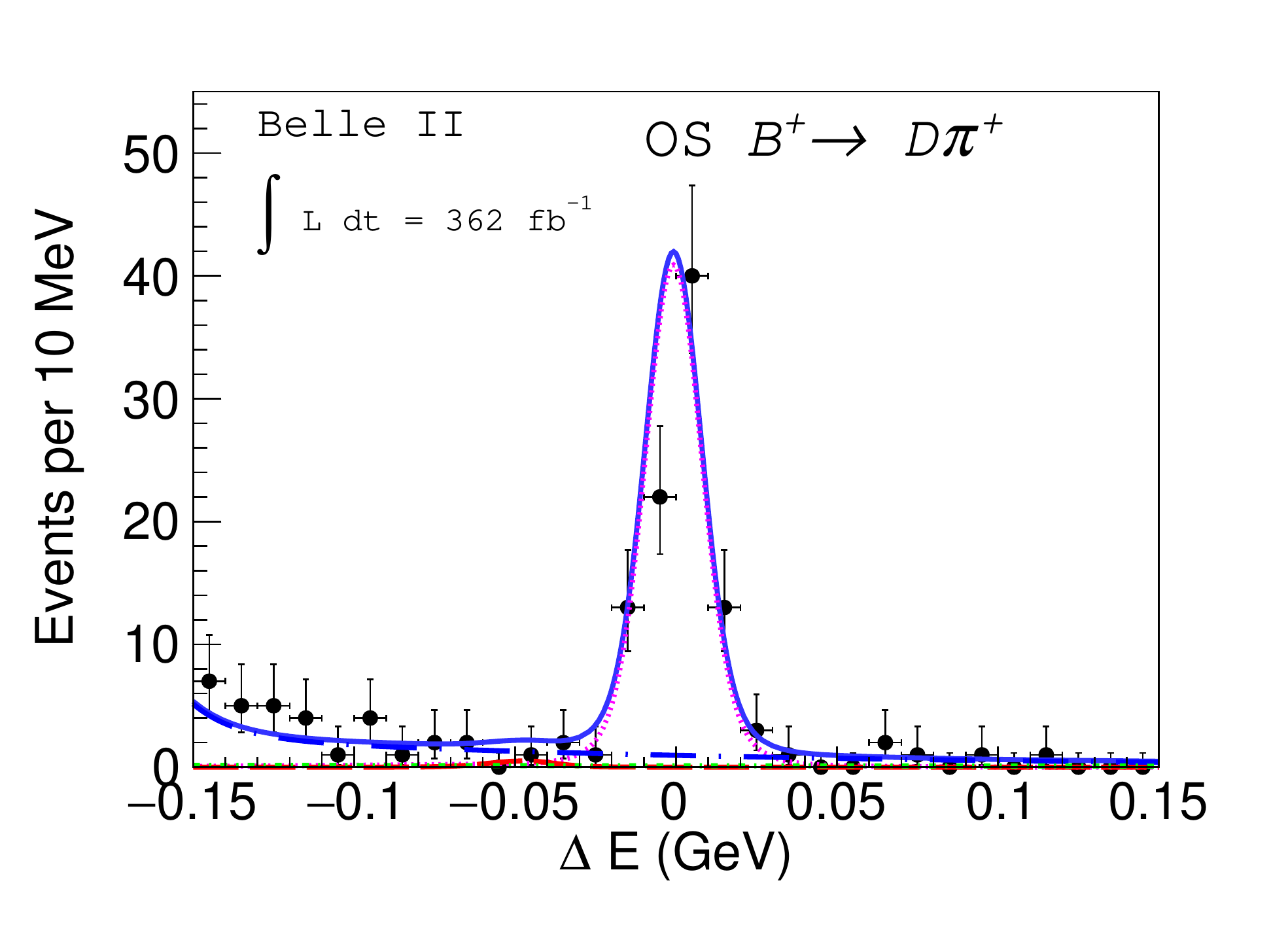}\put(40,55){}\end{overpic}

  \caption{\FullResultsFigureCaption{\pipm}{Belle~II}}
  \label{fig:PHSPB2DPi}
\end{figure}

\begin{figure}[H]
  \centering

  \begin{overpic}[width=0.49\textwidth]{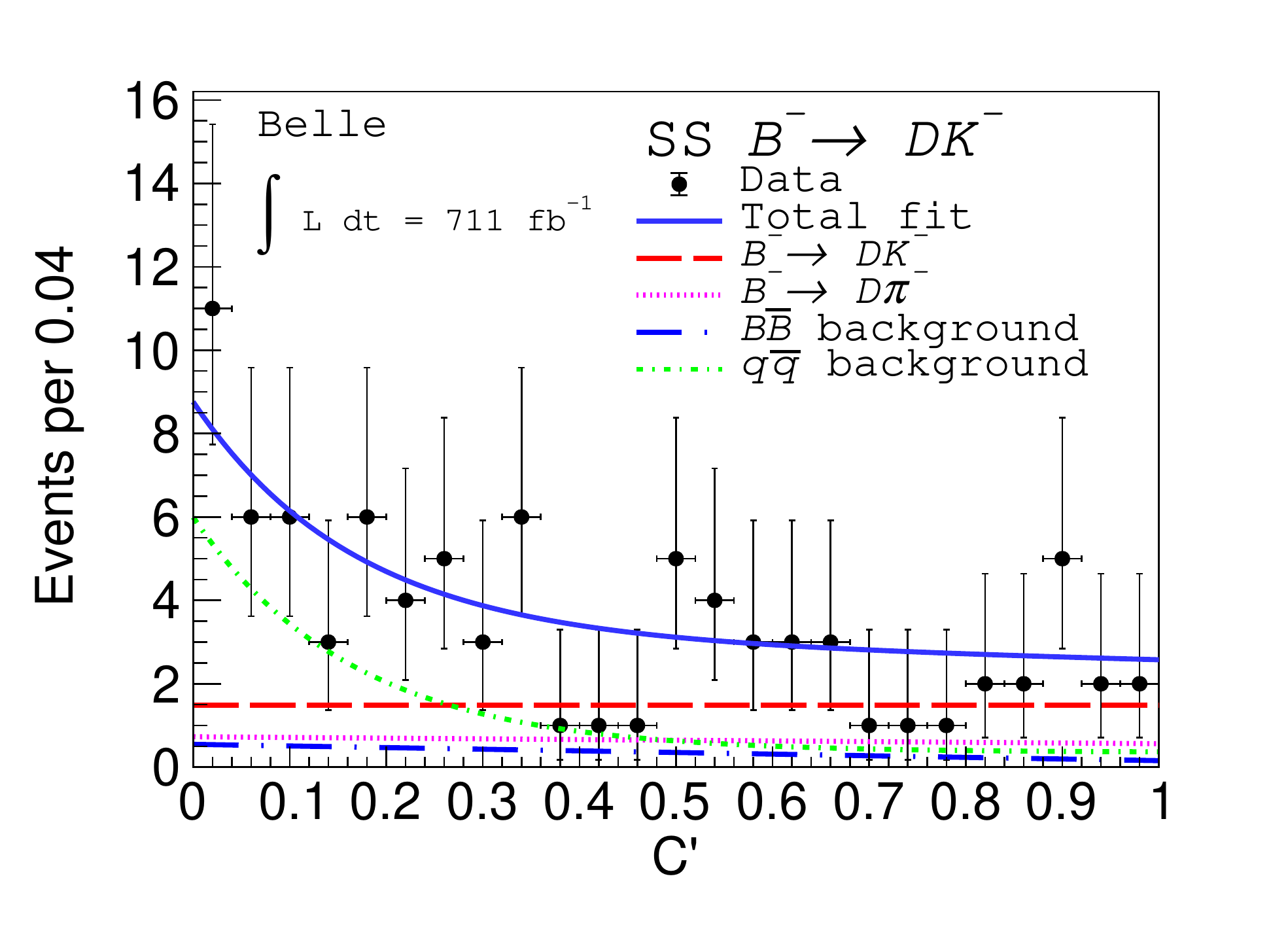}\put(40,55){}\end{overpic}
  \begin{overpic}[width=0.49\textwidth]{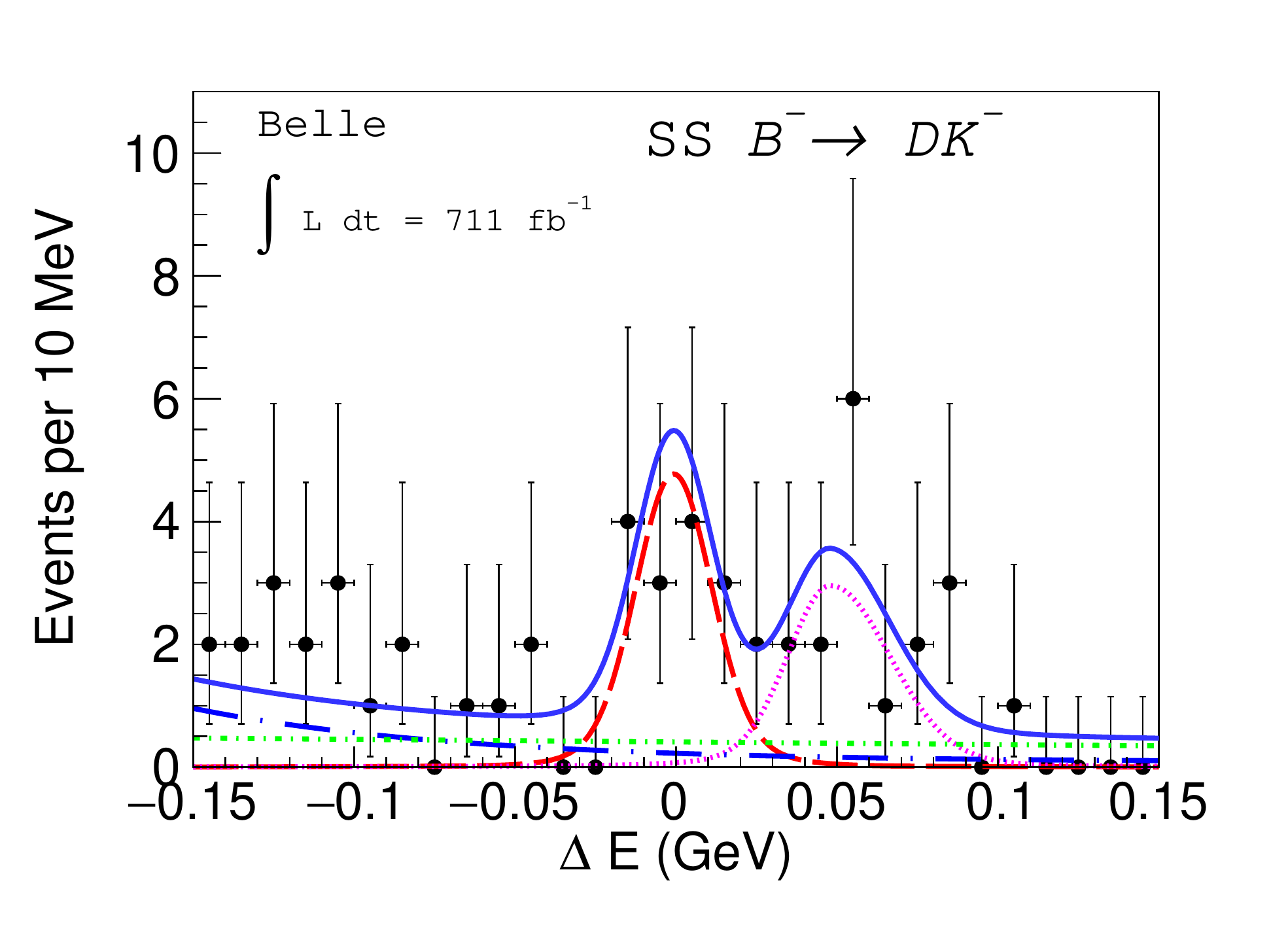}\put(40,55){}\end{overpic}
  \begin{overpic}[width=0.49\textwidth]{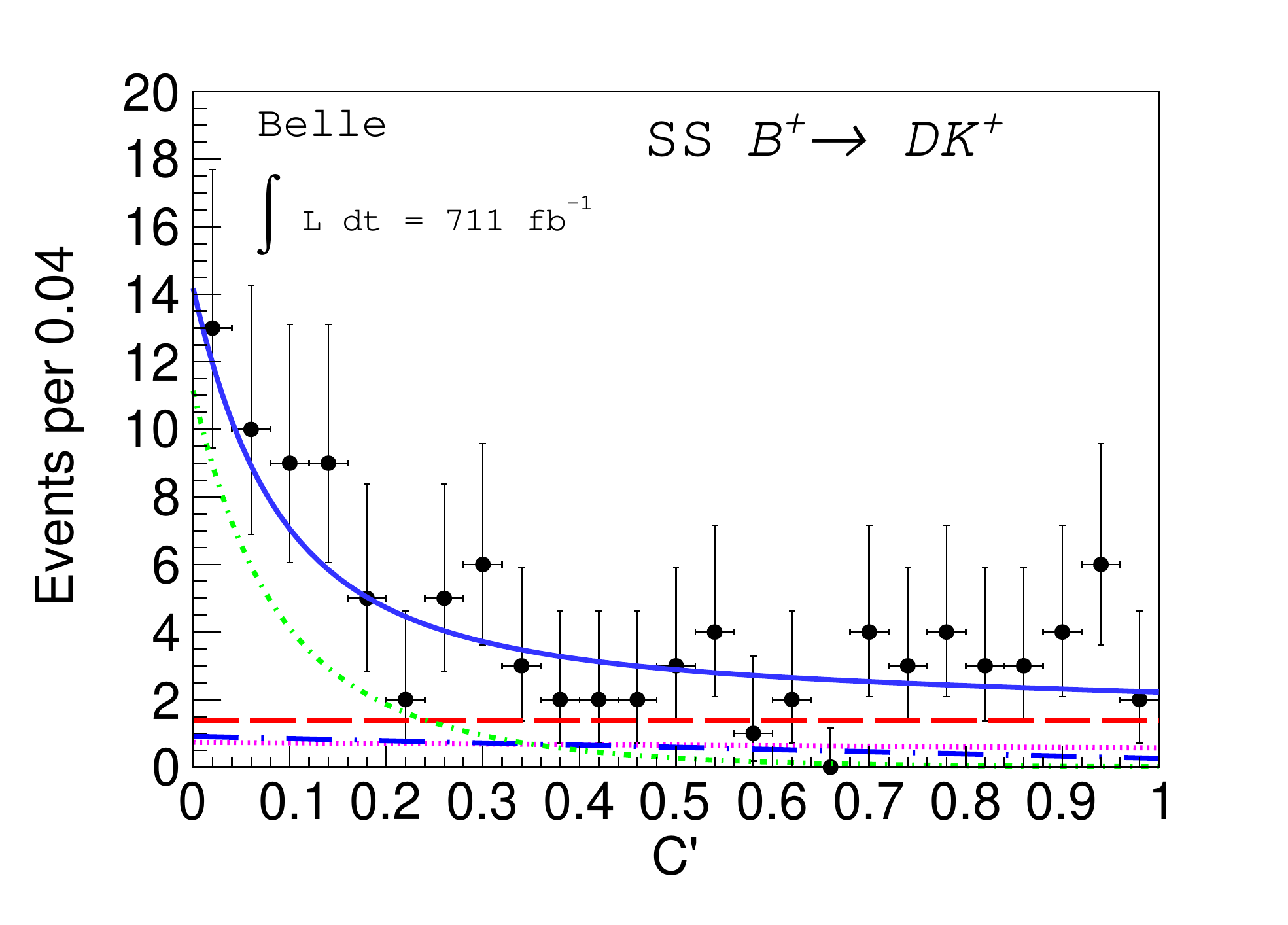}\put(40,55){}\end{overpic}
  \begin{overpic}[width=0.49\textwidth]{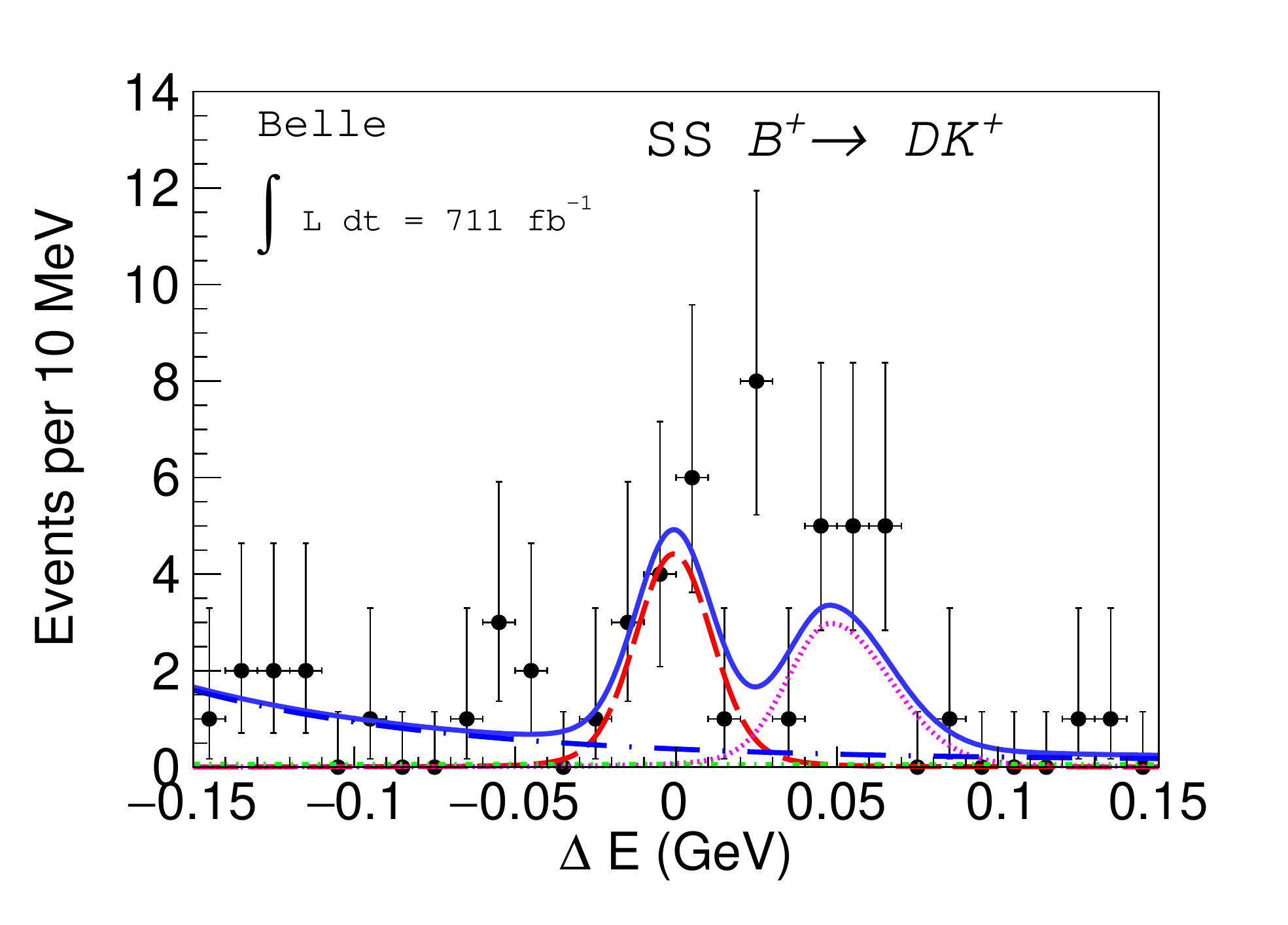}\put(40,55){}\end{overpic}
  \begin{overpic}[width=0.49\textwidth]{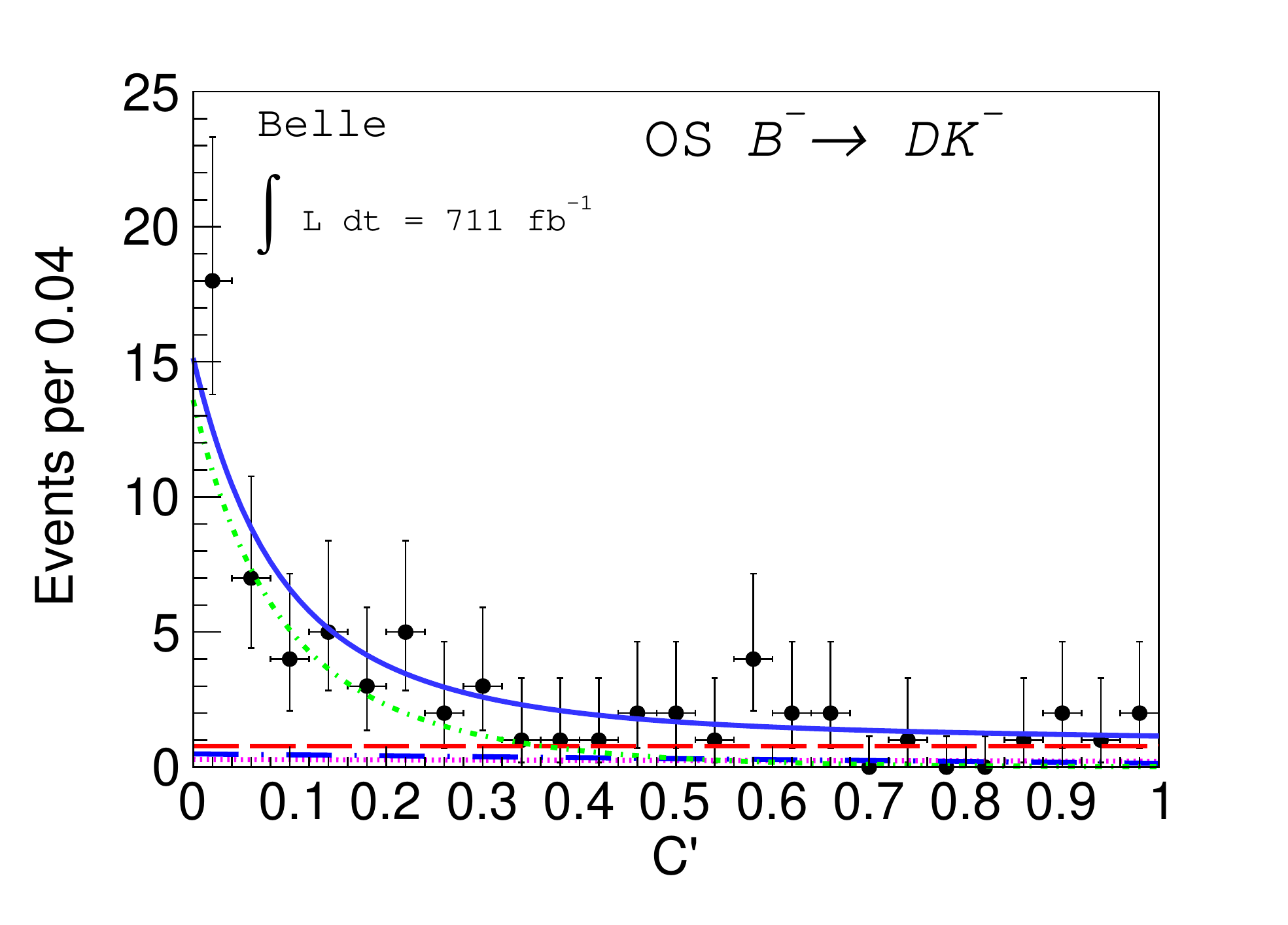}\put(40,55){}\end{overpic}
  \begin{overpic}[width=0.49\textwidth]{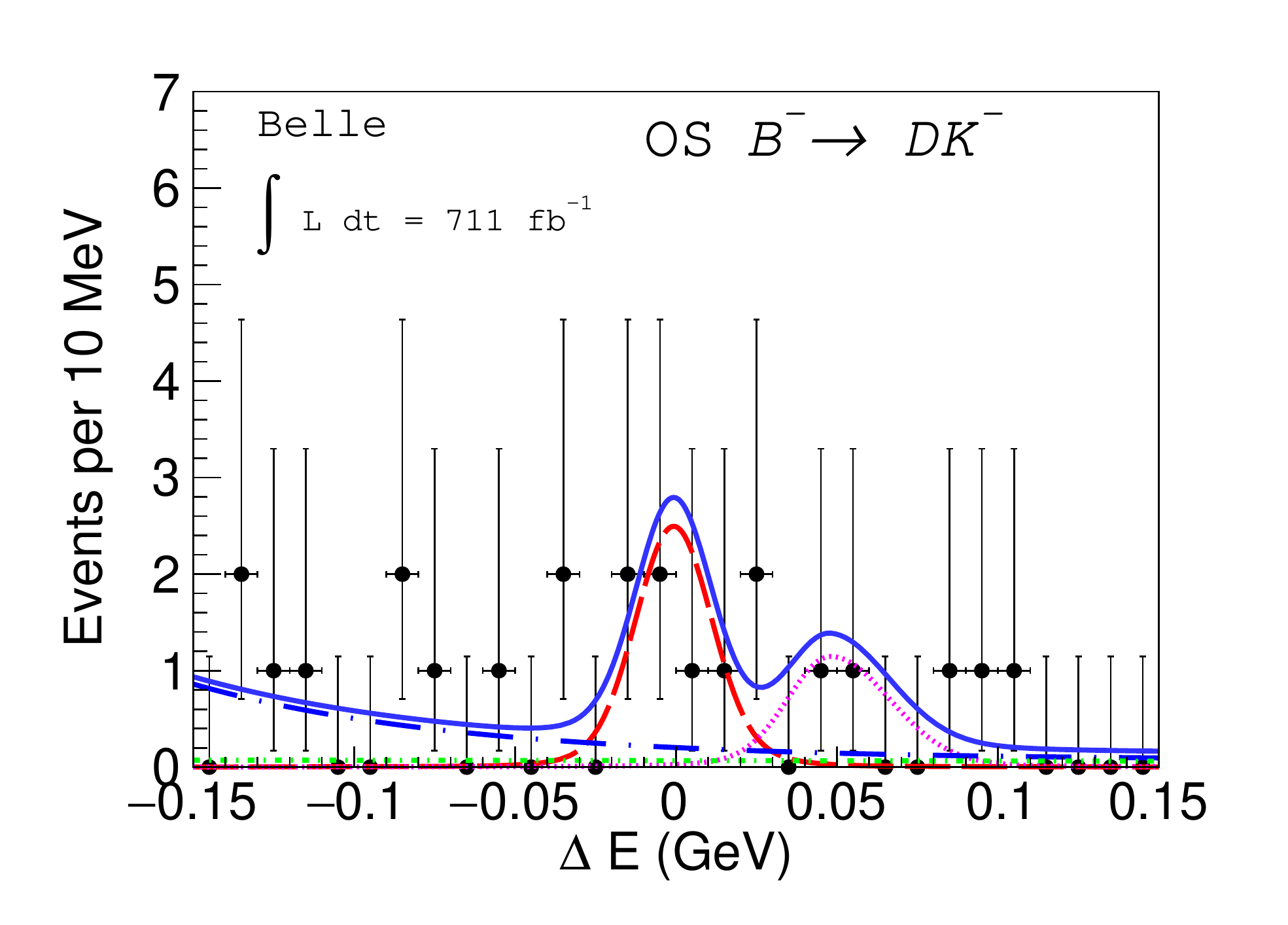}\put(40,55){}\end{overpic}
  \begin{overpic}[width=0.49\textwidth]{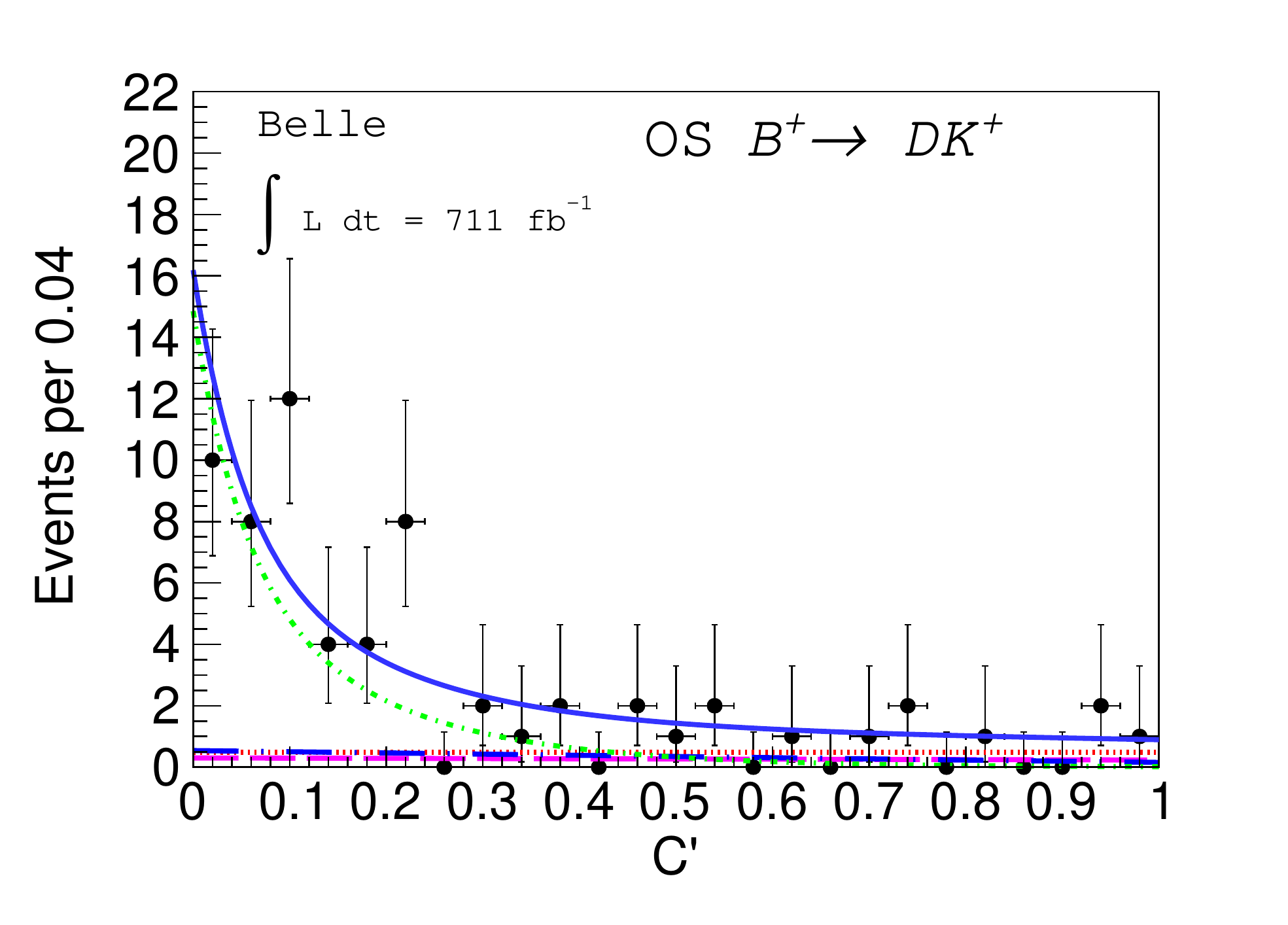}\put(40,55){}\end{overpic}
  \begin{overpic}[width=0.49\textwidth]{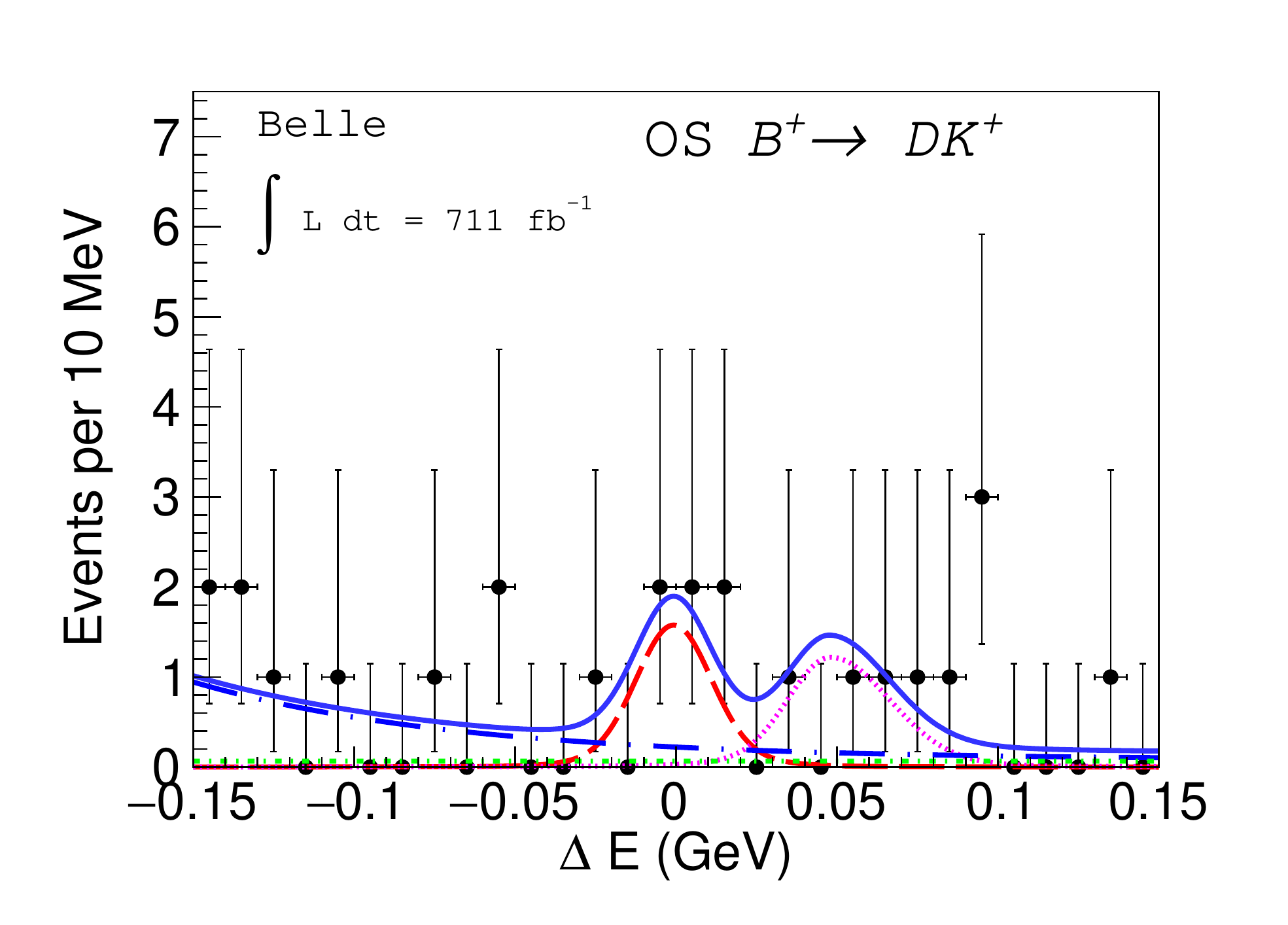}\put(40,55){}\end{overpic}

  \caption{\KstarResultsFigureCaption{\Kpm}{Belle}}
  \label{fig:KstarKB1DK}
\end{figure}

\begin{figure}[H]
  \centering

  \begin{overpic}[width=0.49\textwidth]{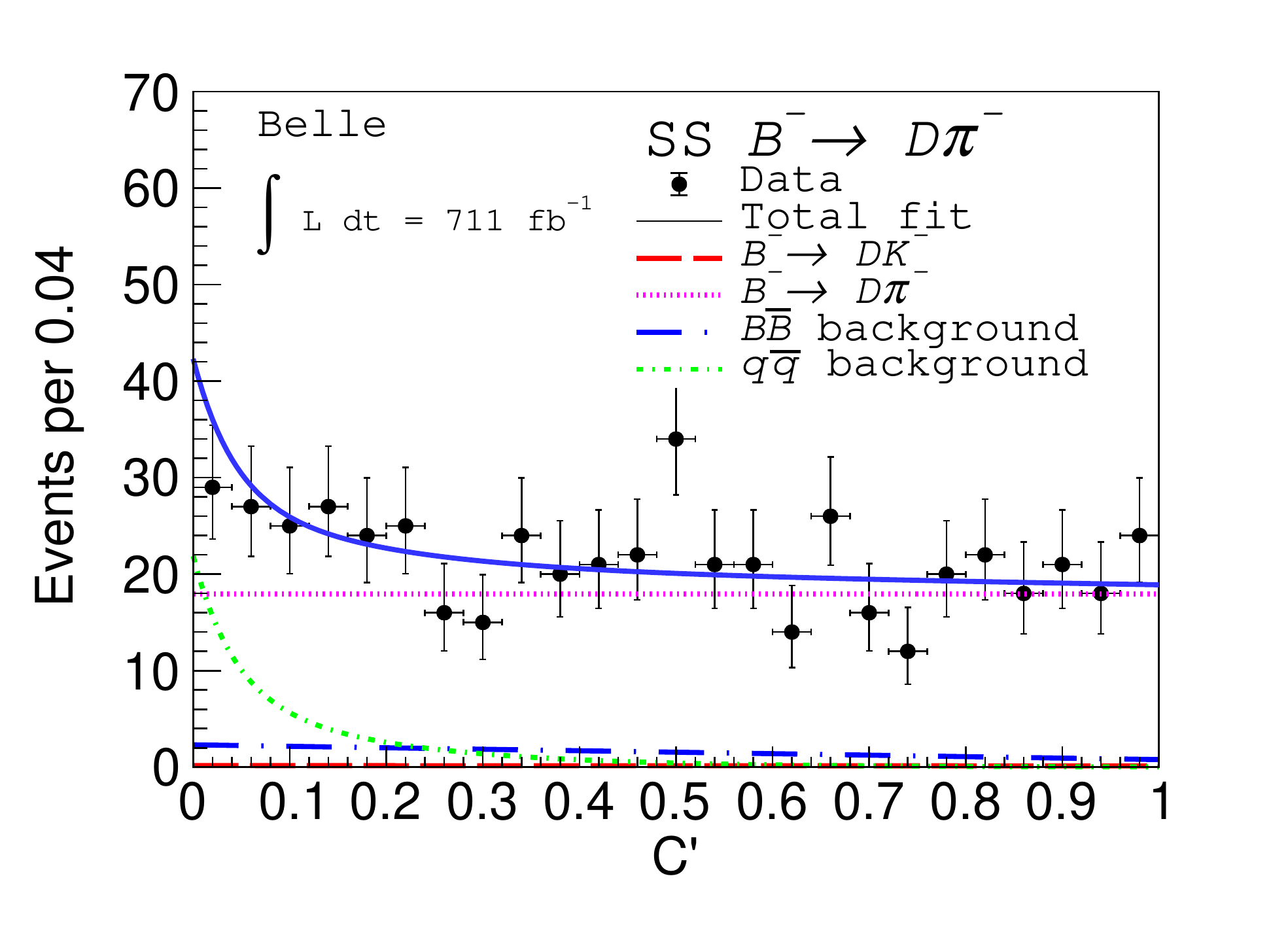}\put(40,55){}\end{overpic}
  \begin{overpic}[width=0.49\textwidth]{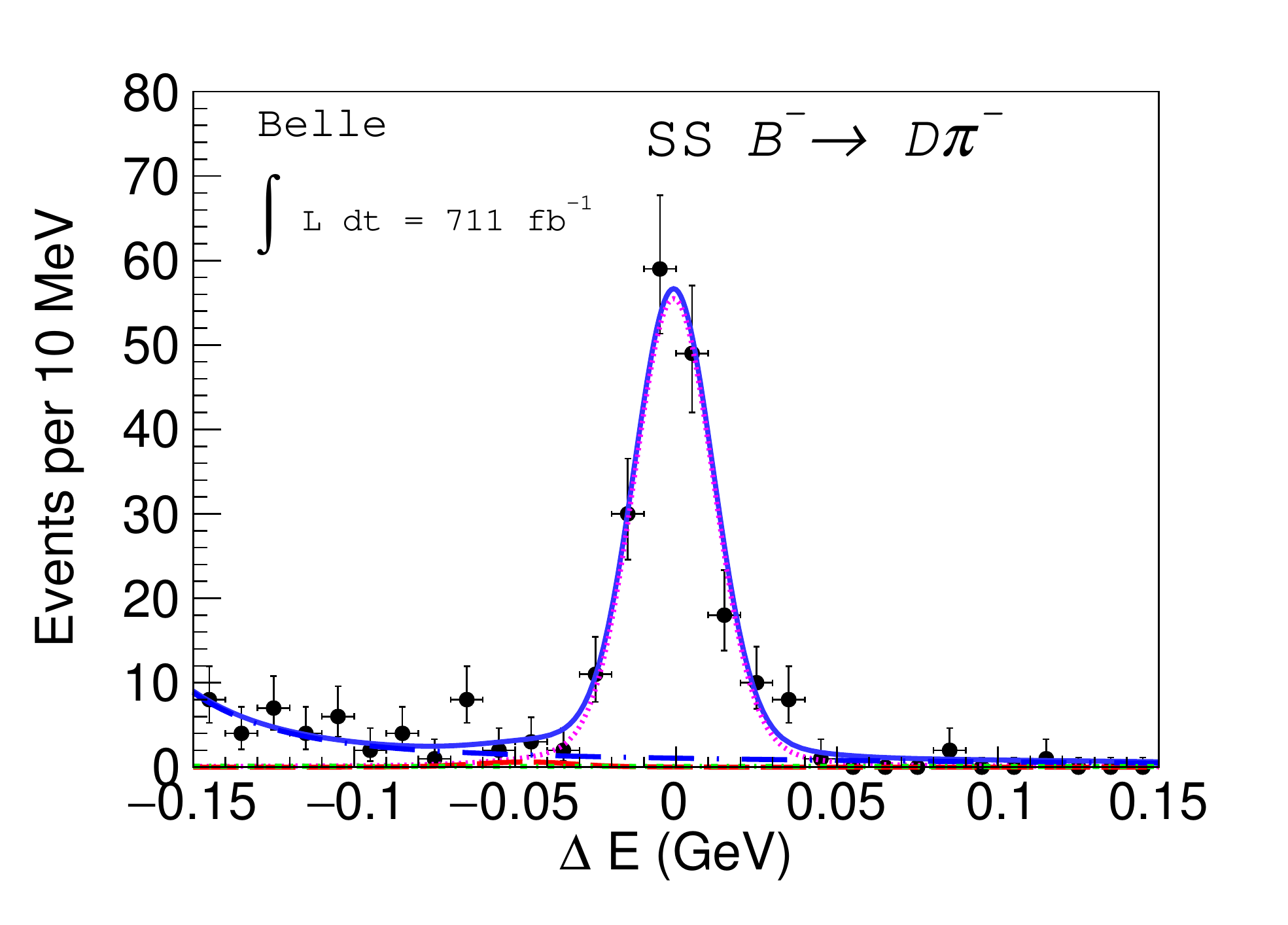}\put(40,55){}\end{overpic}
  \begin{overpic}[width=0.49\textwidth]{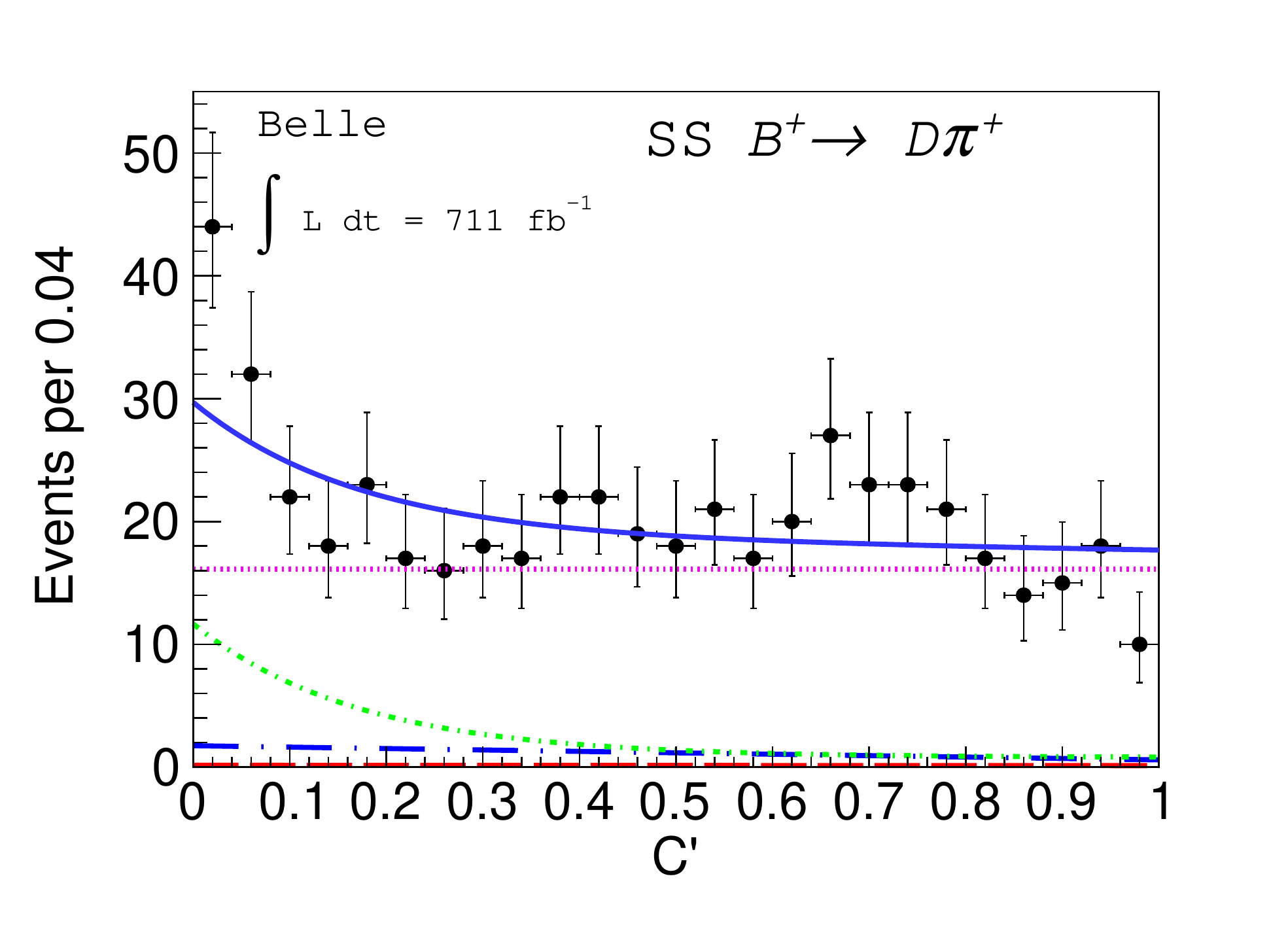}\put(40,55){}\end{overpic}
  \begin{overpic}[width=0.49\textwidth]{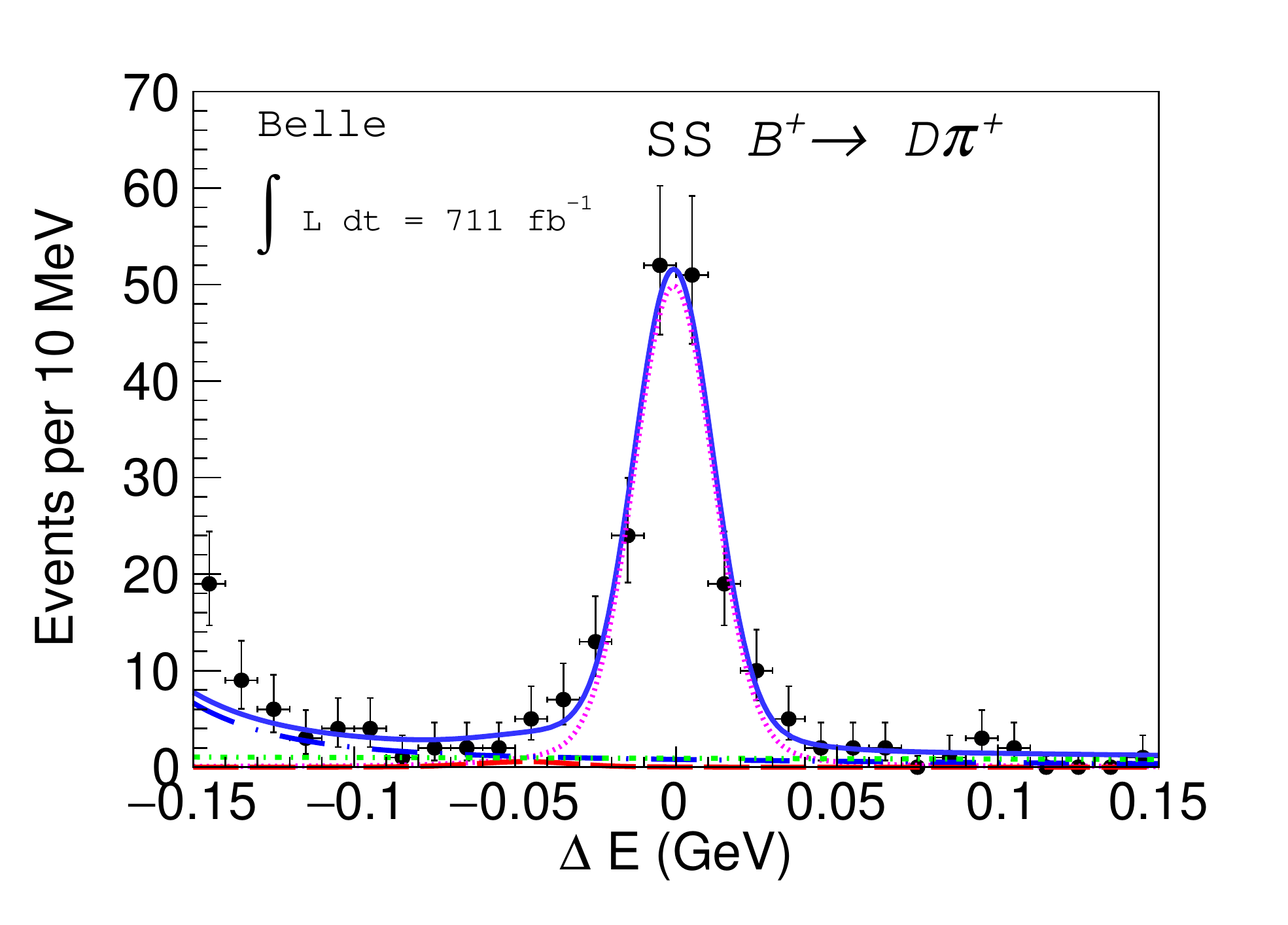}\put(40,55){}\end{overpic}
  \begin{overpic}[width=0.49\textwidth]{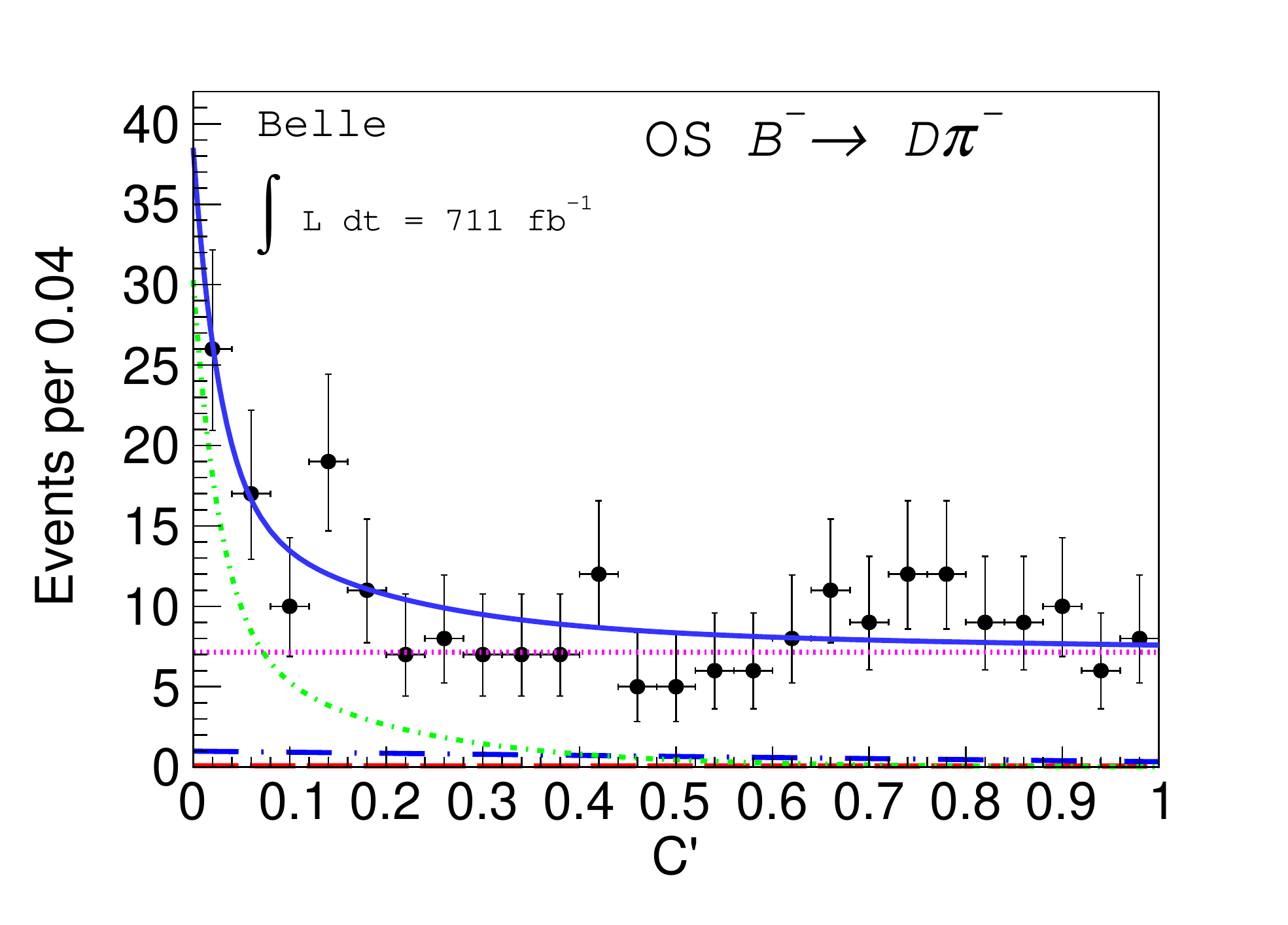}\put(40,55){}\end{overpic}
  \begin{overpic}[width=0.49\textwidth]{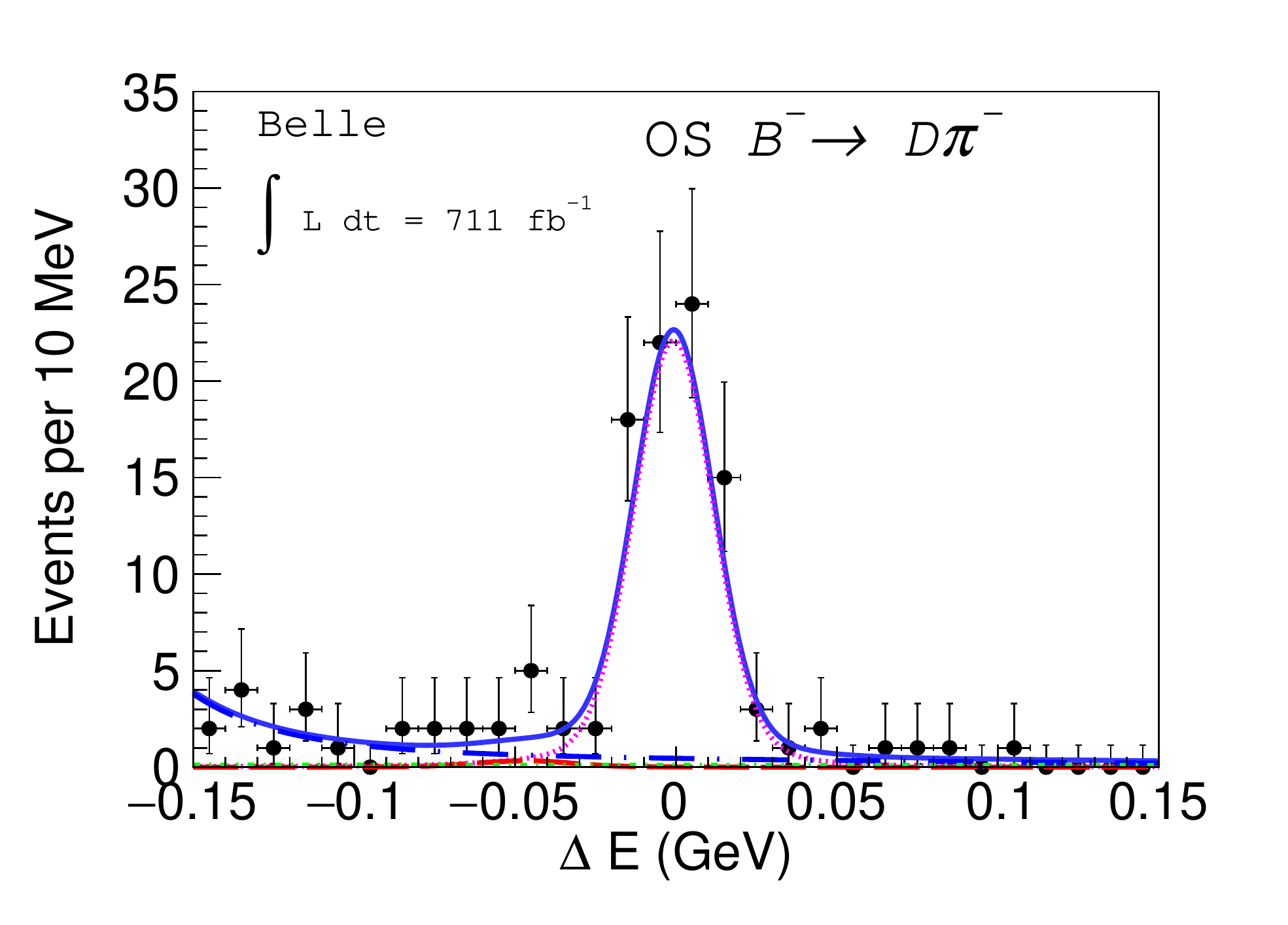}\put(40,55){}\end{overpic}
  \begin{overpic}[width=0.49\textwidth]{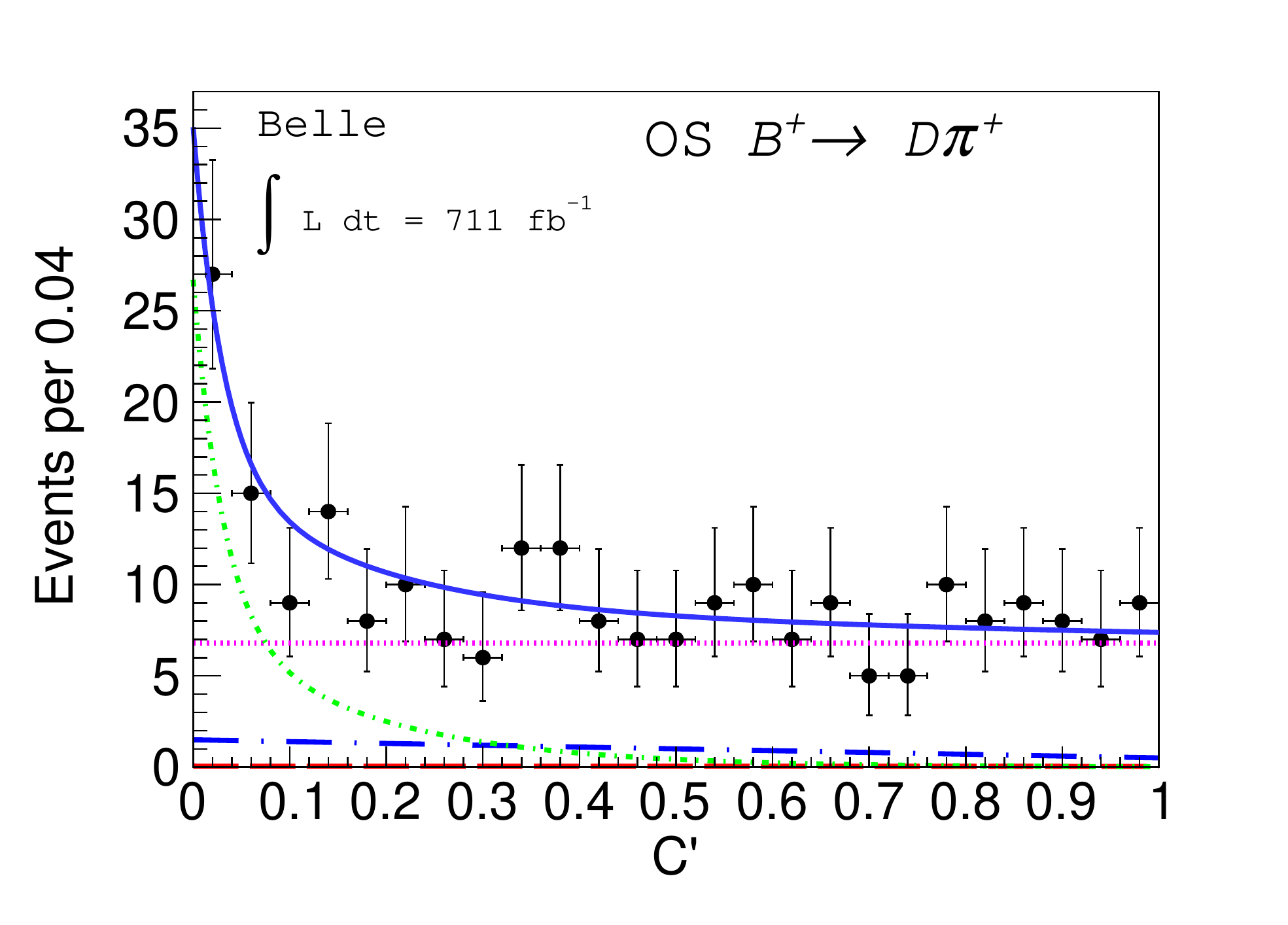}\put(40,55){}\end{overpic}
  \begin{overpic}[width=0.49\textwidth]{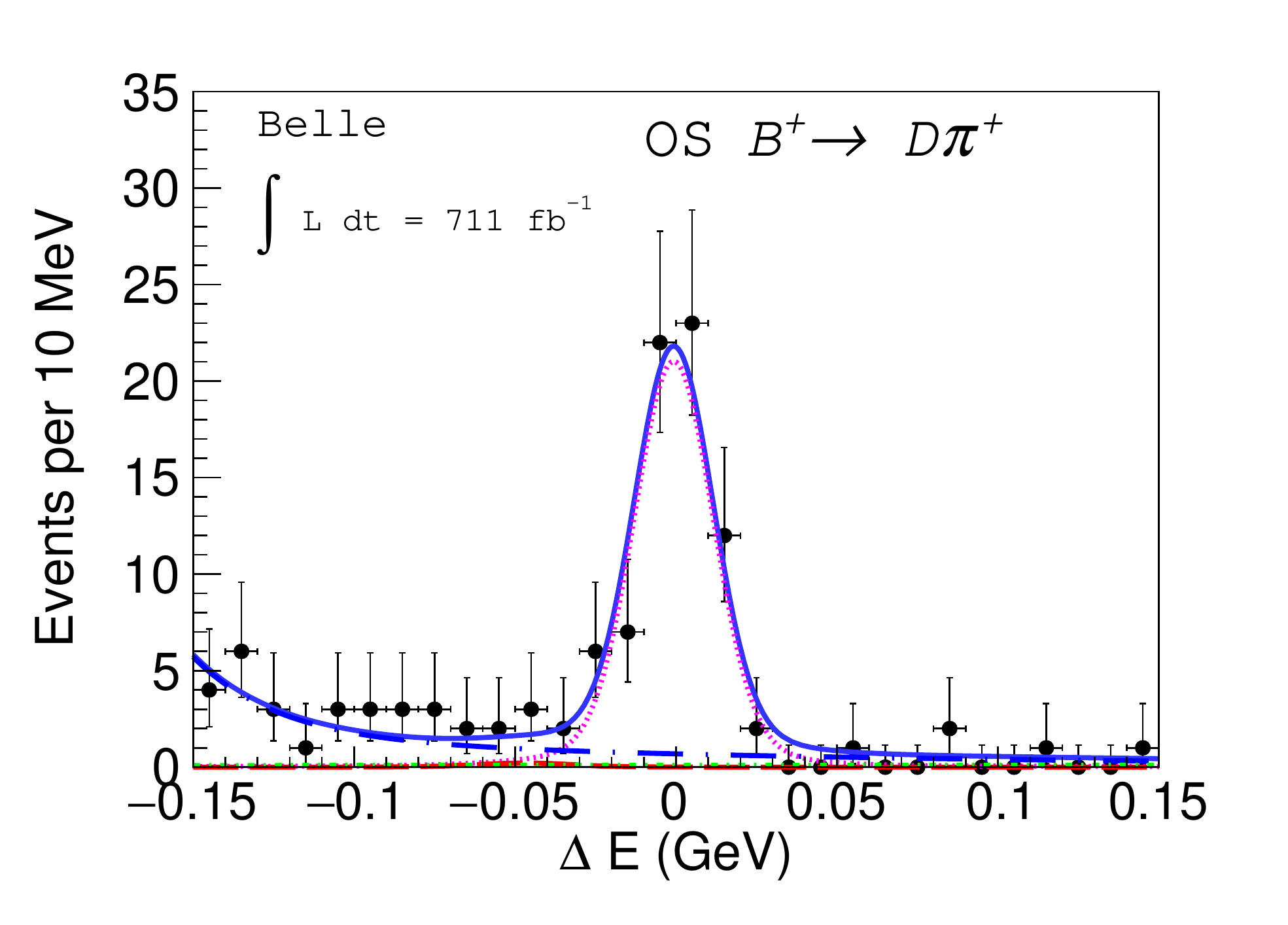}\put(40,55){}\end{overpic}

  \caption{\KstarResultsFigureCaption{\pipm}{Belle}}
  \label{fig:KstarKB1DPi}
\end{figure}

\begin{figure}[H]
  \centering

  \begin{overpic}[width=0.49\textwidth]{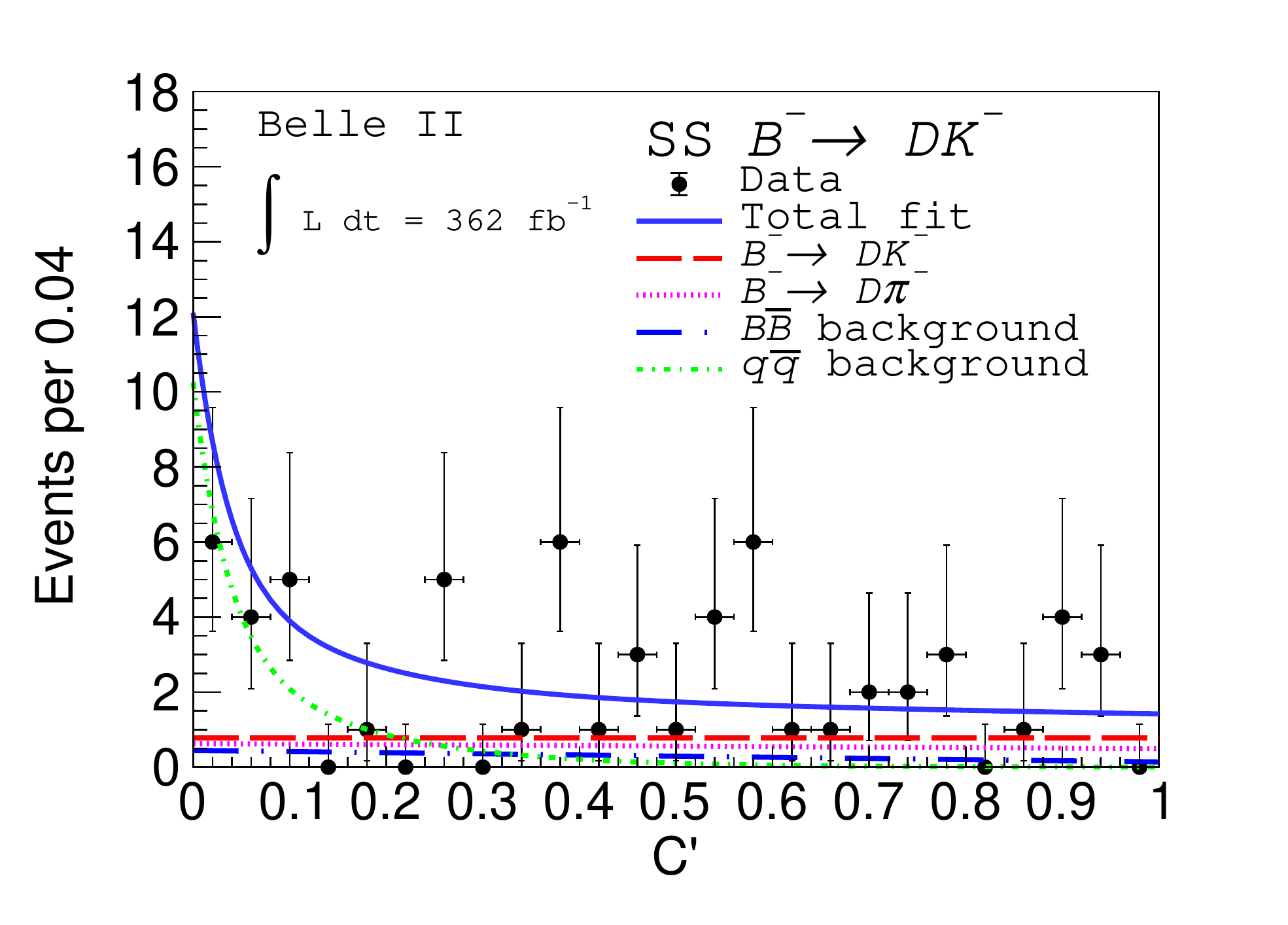}\put(40,55){}\end{overpic}
  \begin{overpic}[width=0.49\textwidth]{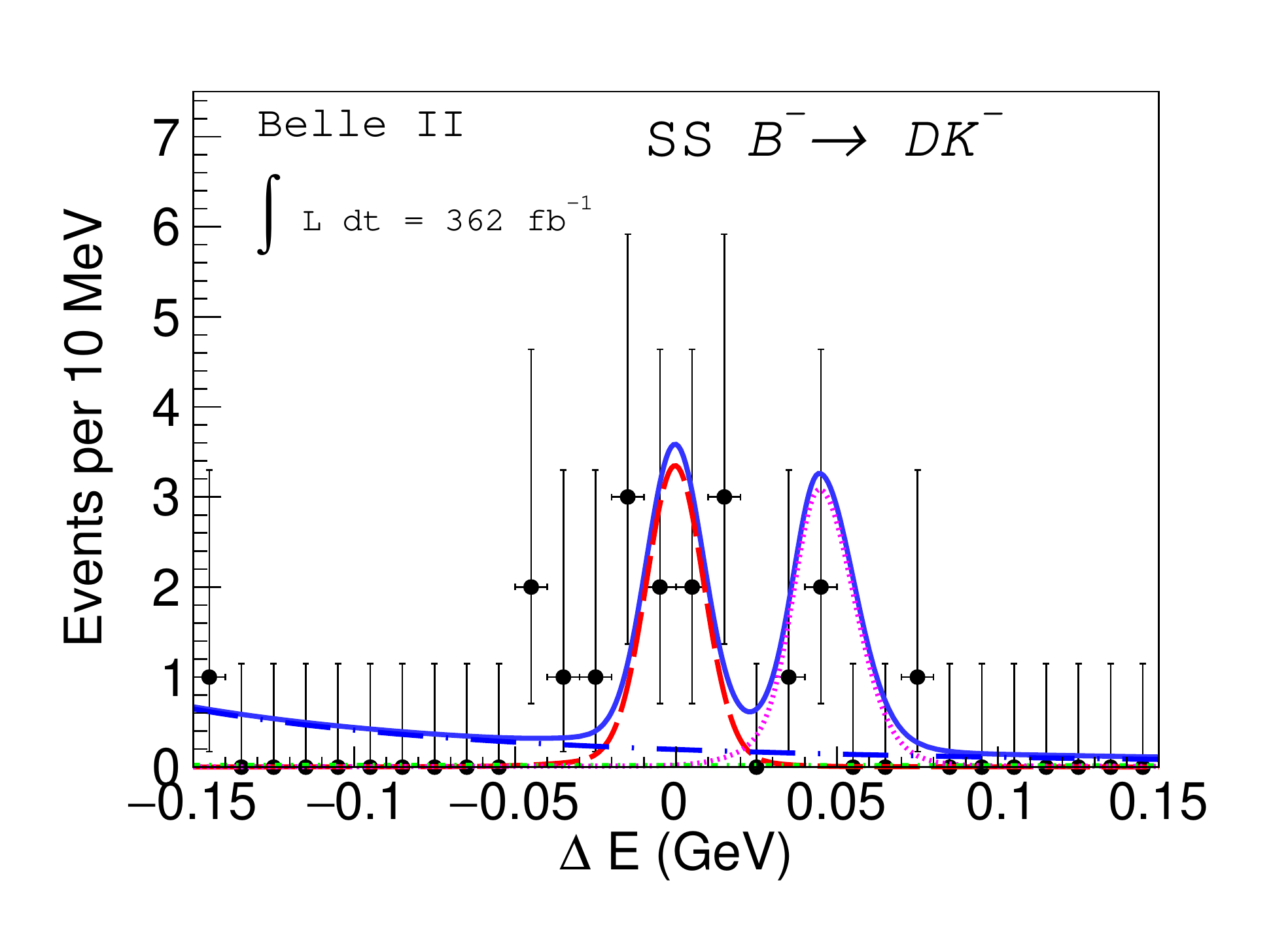}\put(40,55){}\end{overpic}
  \begin{overpic}[width=0.49\textwidth]{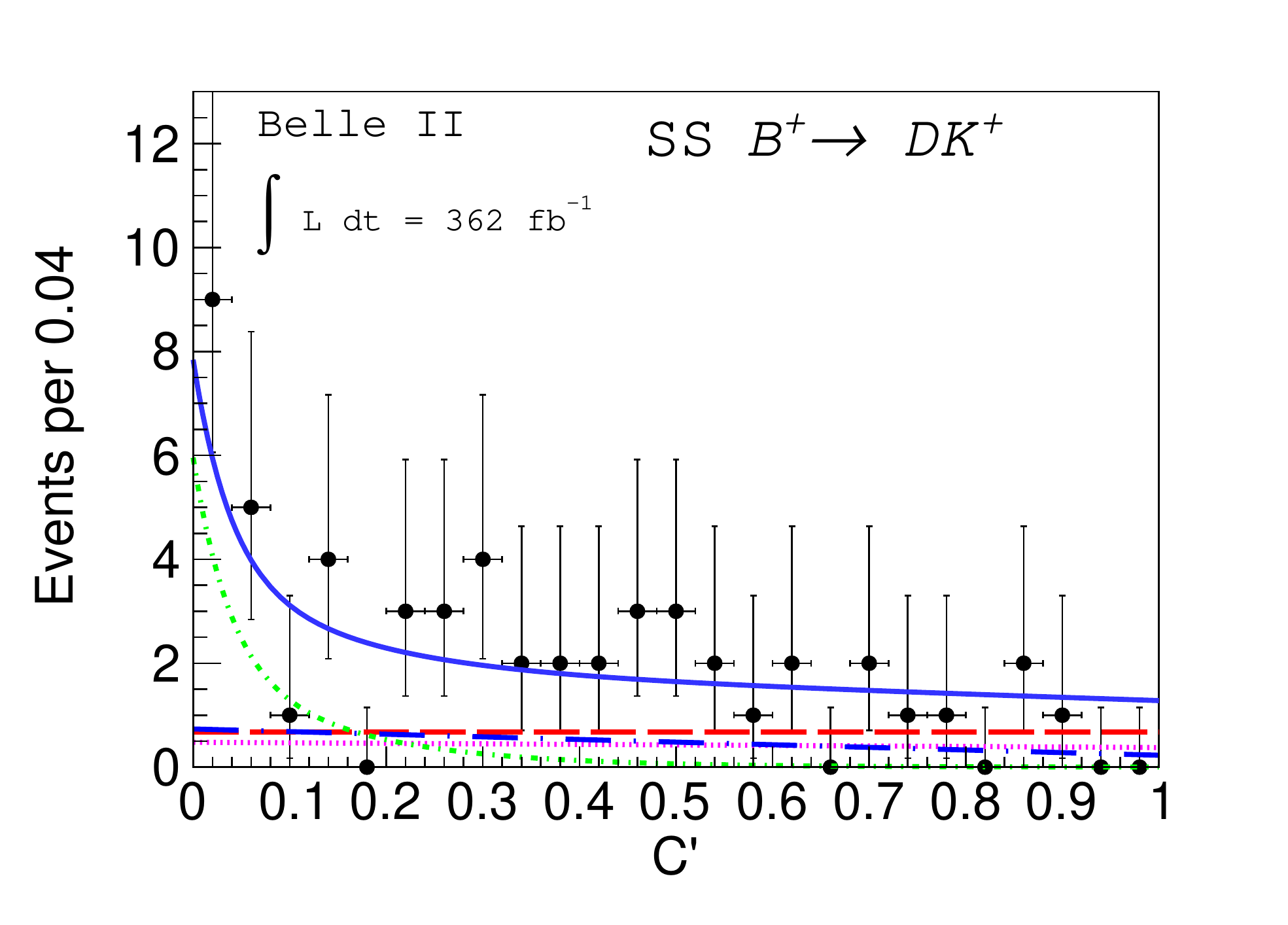}\put(40,55){}\end{overpic}
  \begin{overpic}[width=0.49\textwidth]{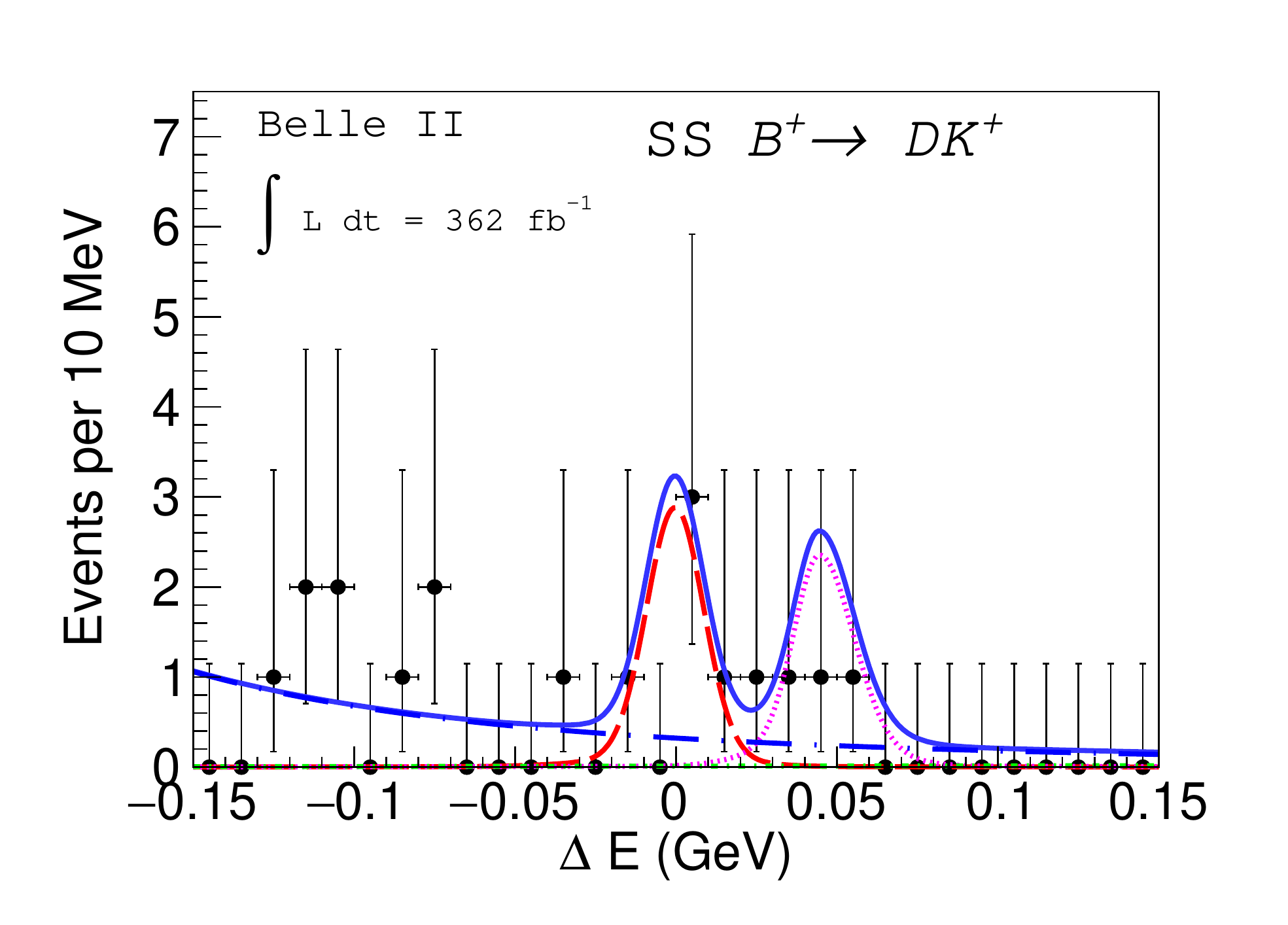}\put(40,55){}\end{overpic}
  \begin{overpic}[width=0.49\textwidth]{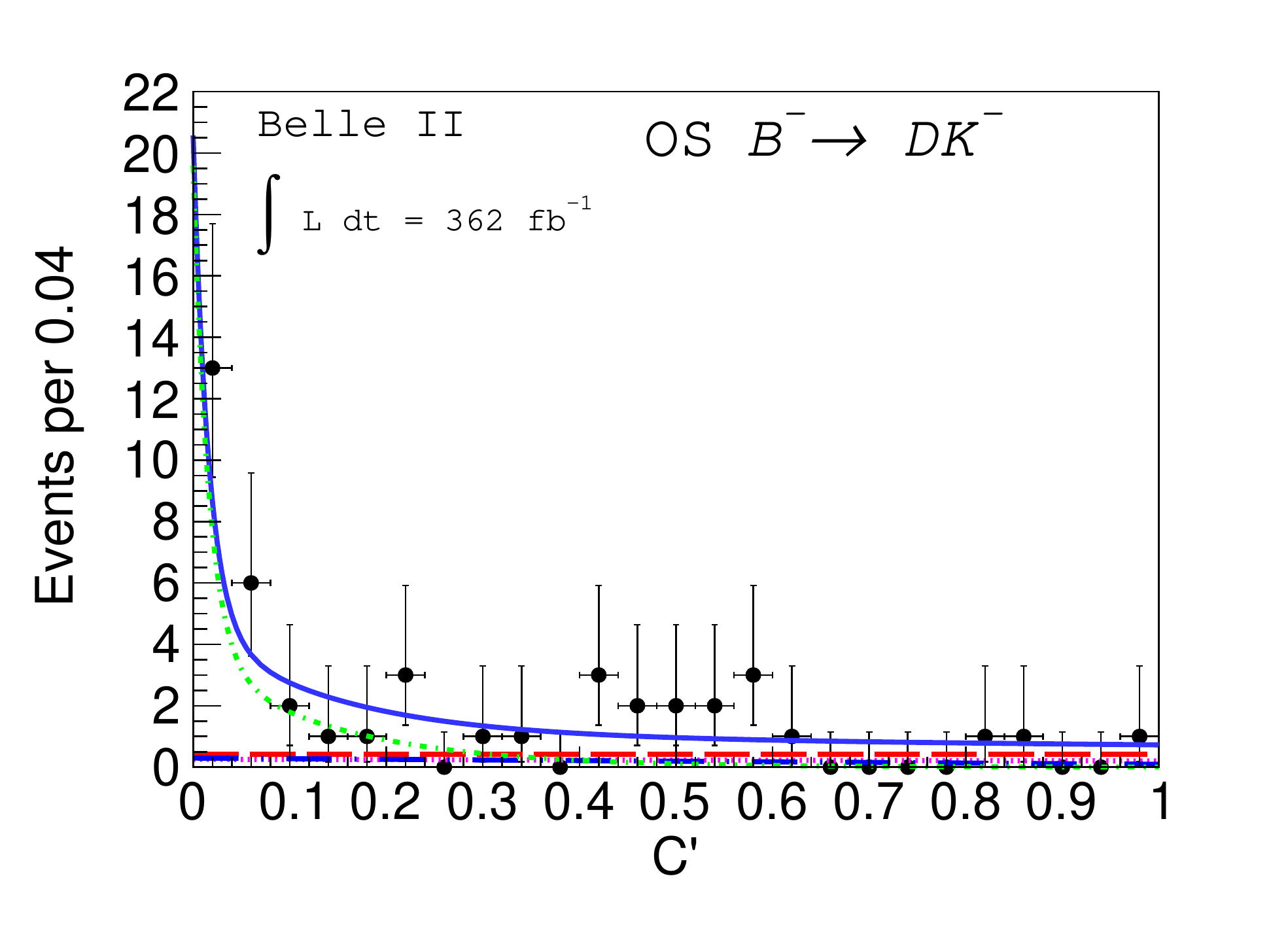}\put(40,55){}\end{overpic}
  \begin{overpic}[width=0.49\textwidth]{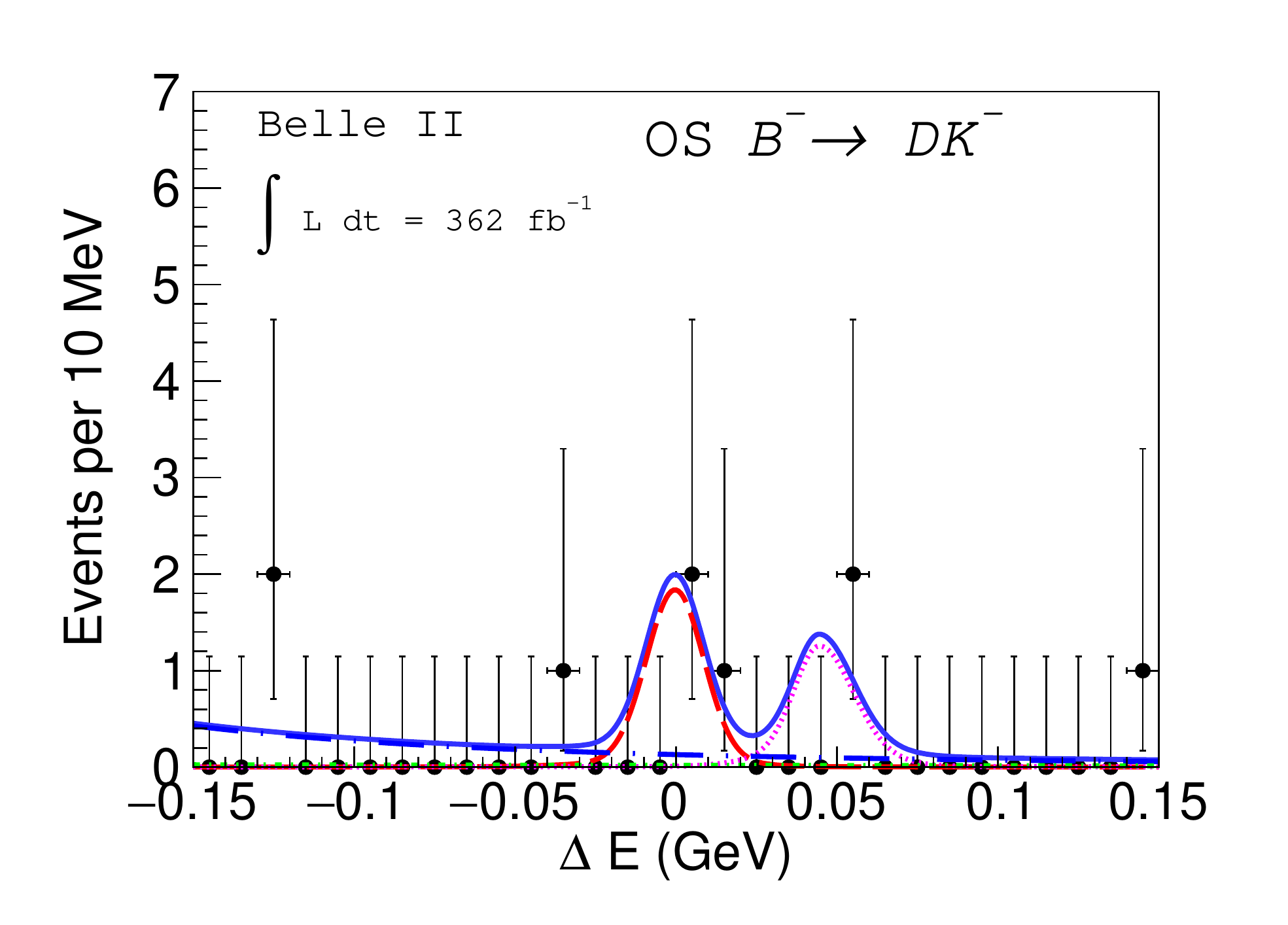}\put(40,55){}\end{overpic}
  \begin{overpic}[width=0.49\textwidth]{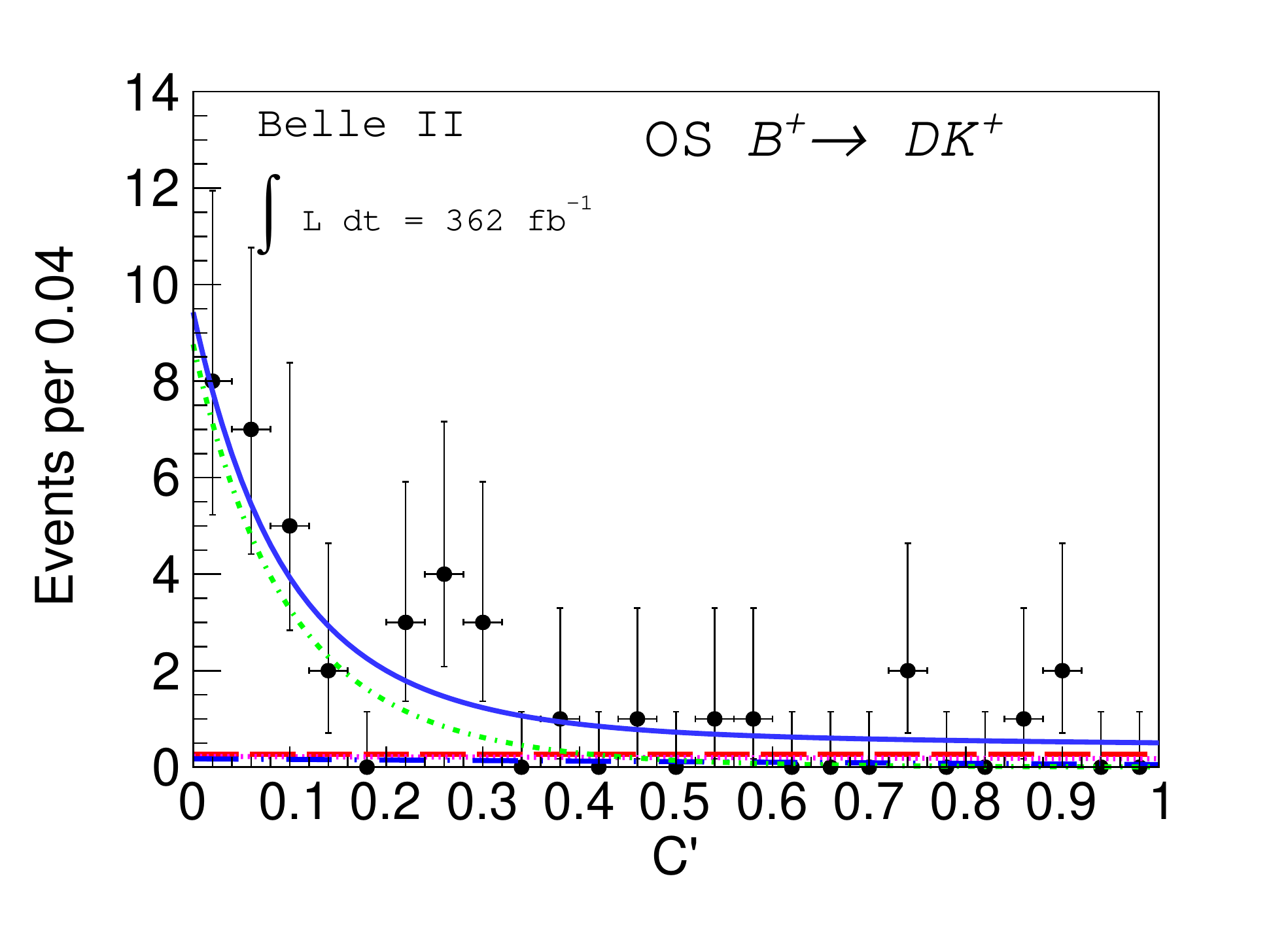}\put(40,55){}\end{overpic}
  \begin{overpic}[width=0.49\textwidth]{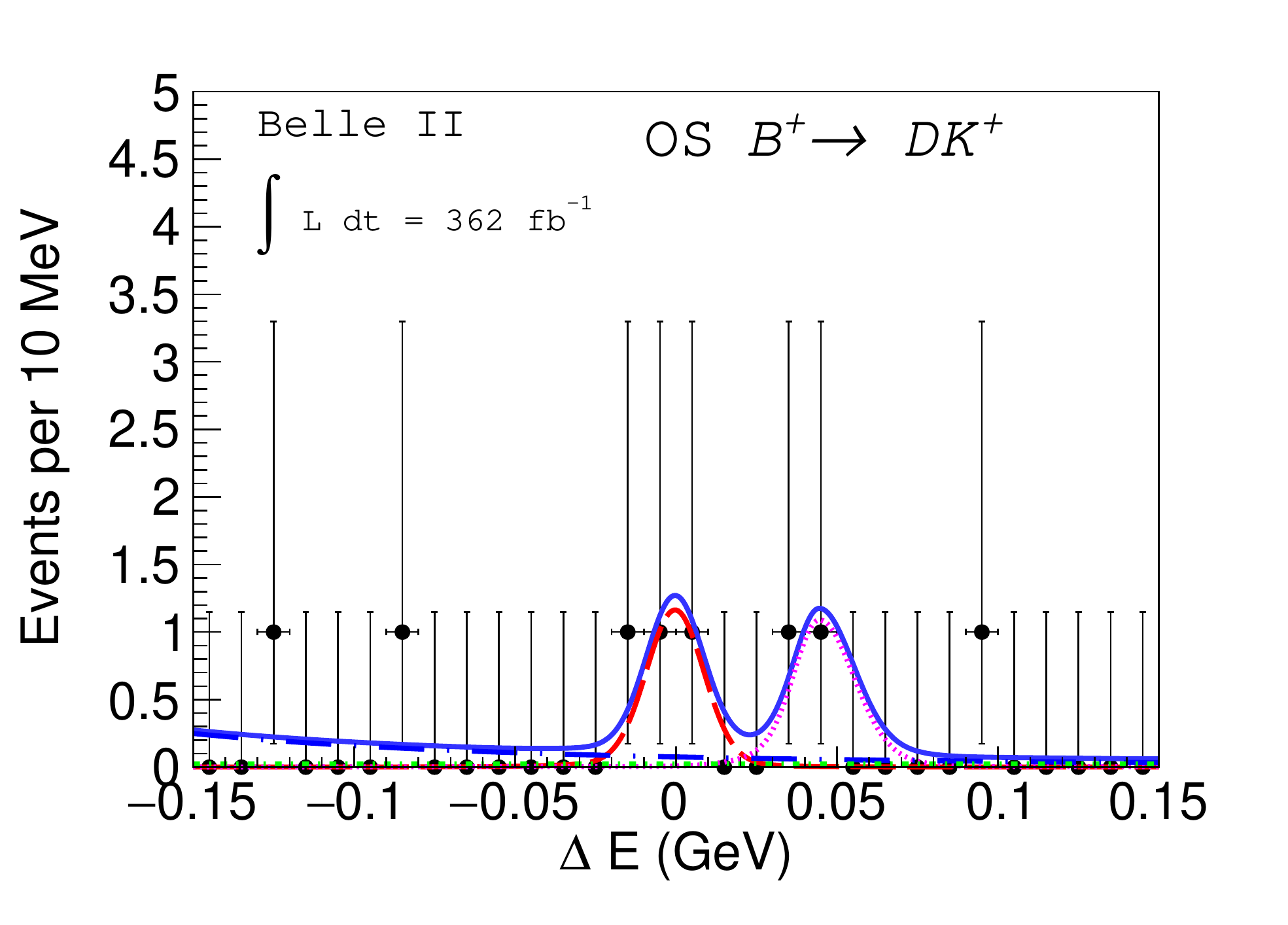}\put(40,55){}\end{overpic}

  \caption{\KstarResultsFigureCaption{\Kpm}{Belle~II}}
  \label{fig:KstarKB2DK}
\end{figure}

\begin{figure}[H]
  \centering

  \begin{overpic}[width=0.49\textwidth]{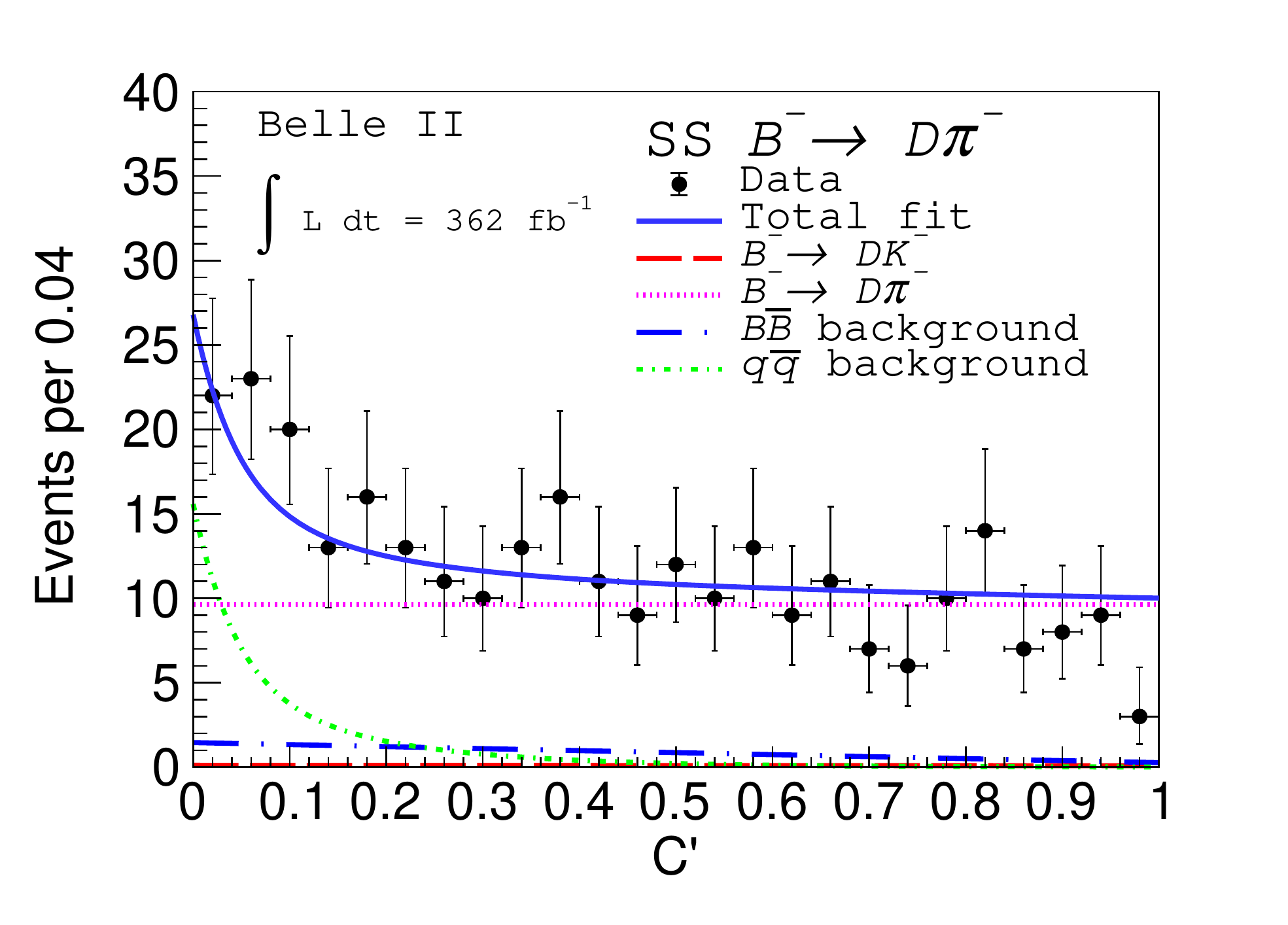}\put(40,55){}\end{overpic}
  \begin{overpic}[width=0.49\textwidth]{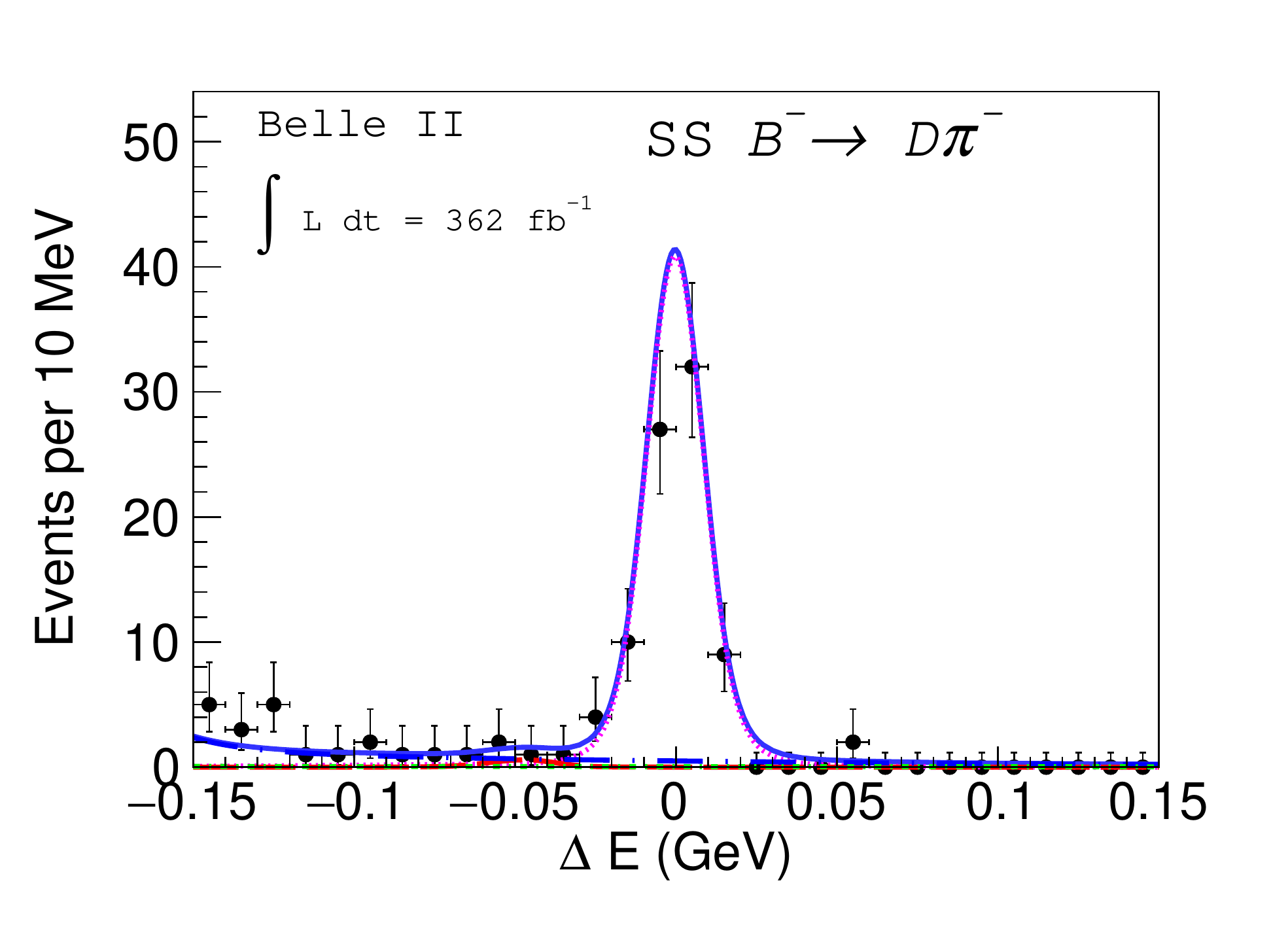}\put(40,55){}\end{overpic}
  \begin{overpic}[width=0.49\textwidth]{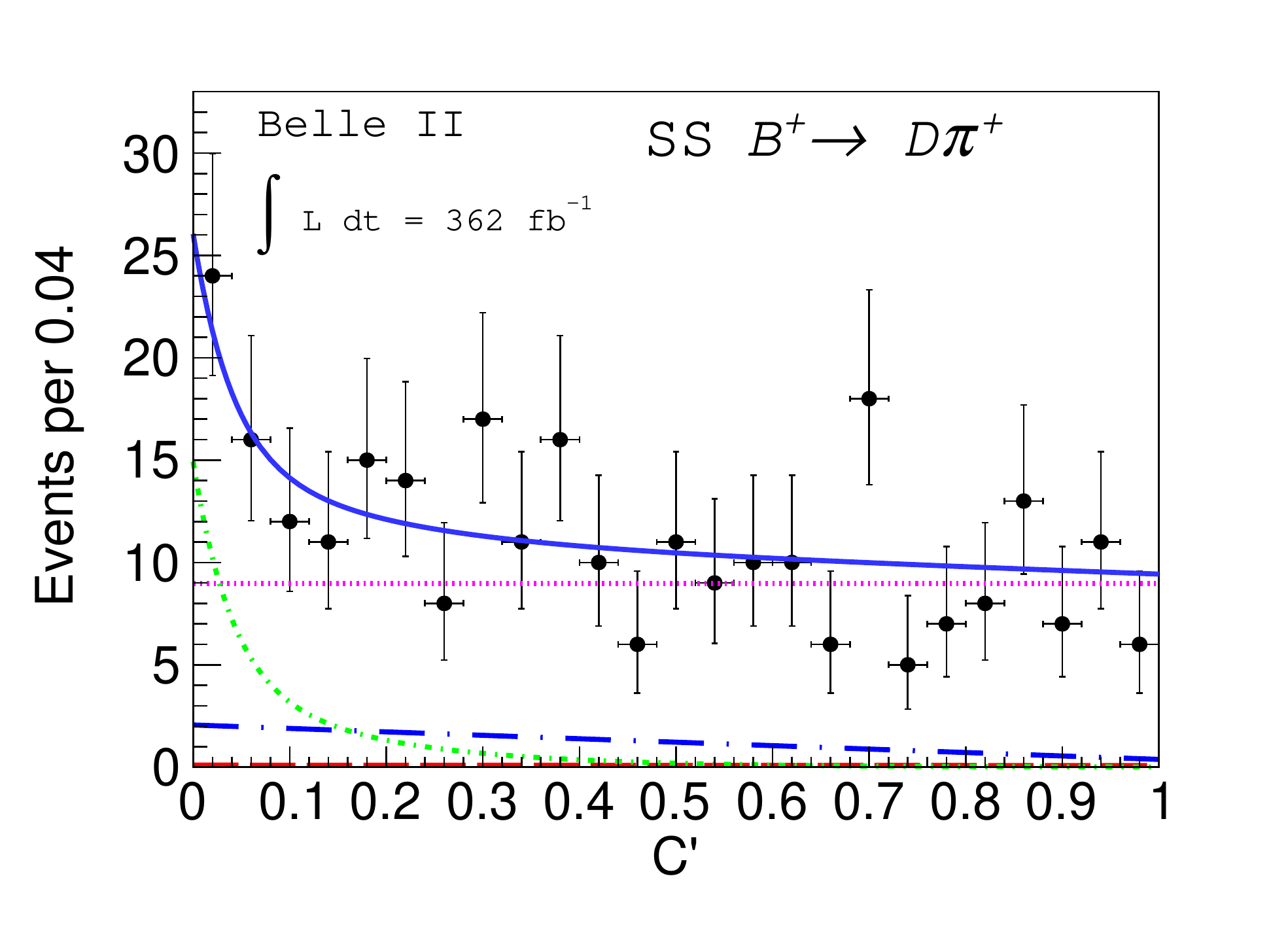}\put(40,55){}\end{overpic}
  \begin{overpic}[width=0.49\textwidth]{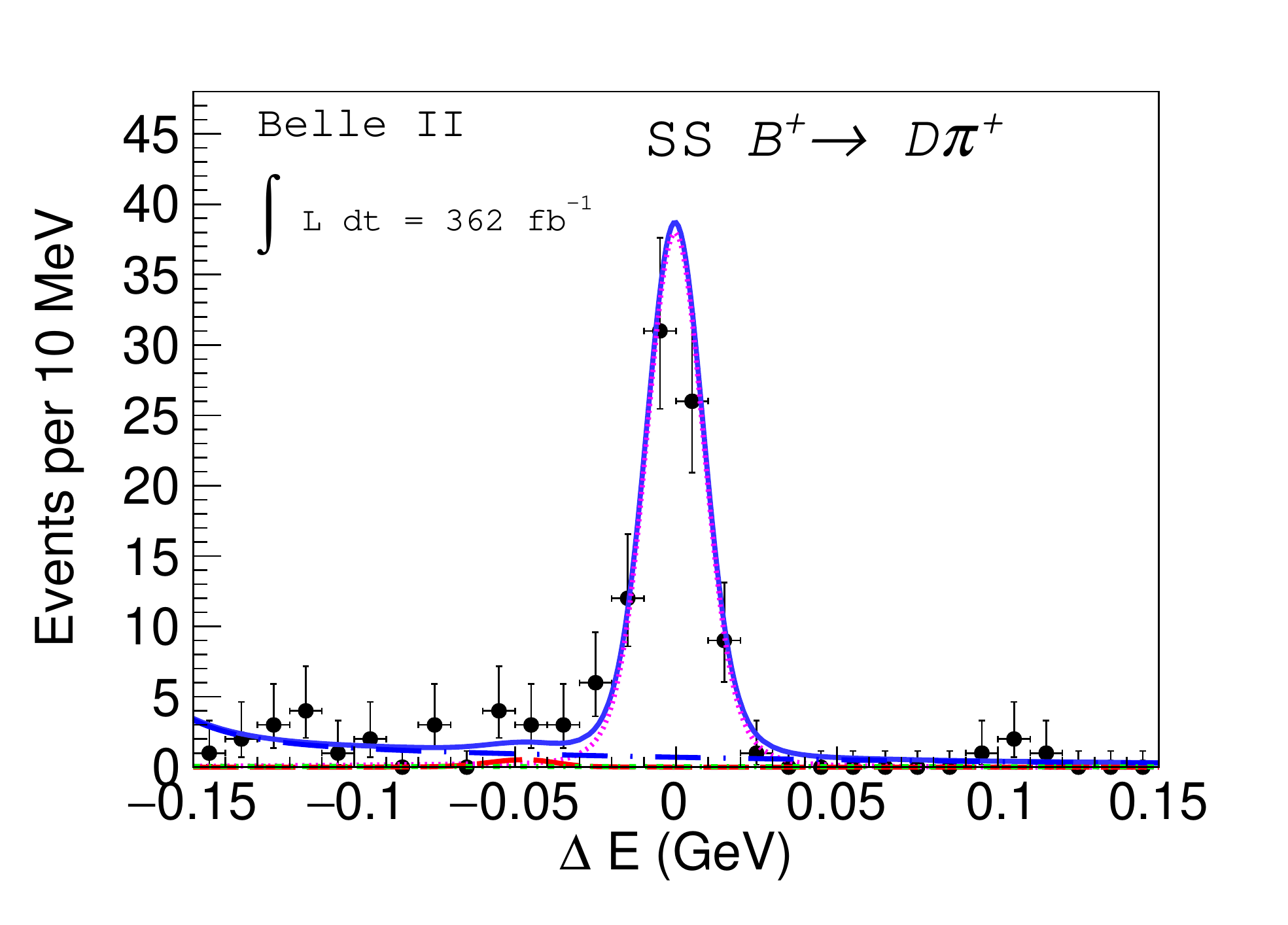}\put(40,55){}\end{overpic}
  \begin{overpic}[width=0.49\textwidth]{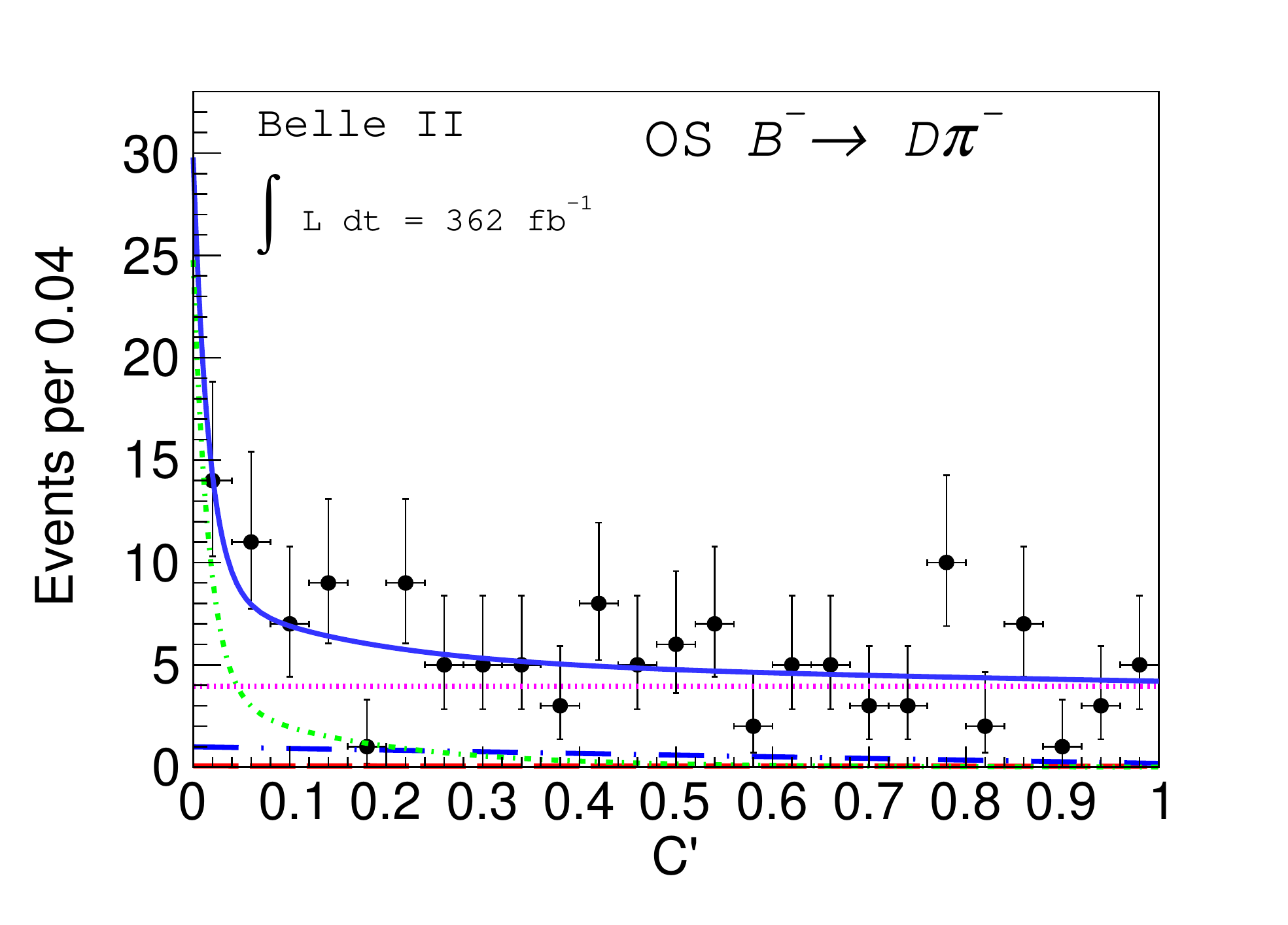}\put(40,55){}\end{overpic}
  \begin{overpic}[width=0.49\textwidth]{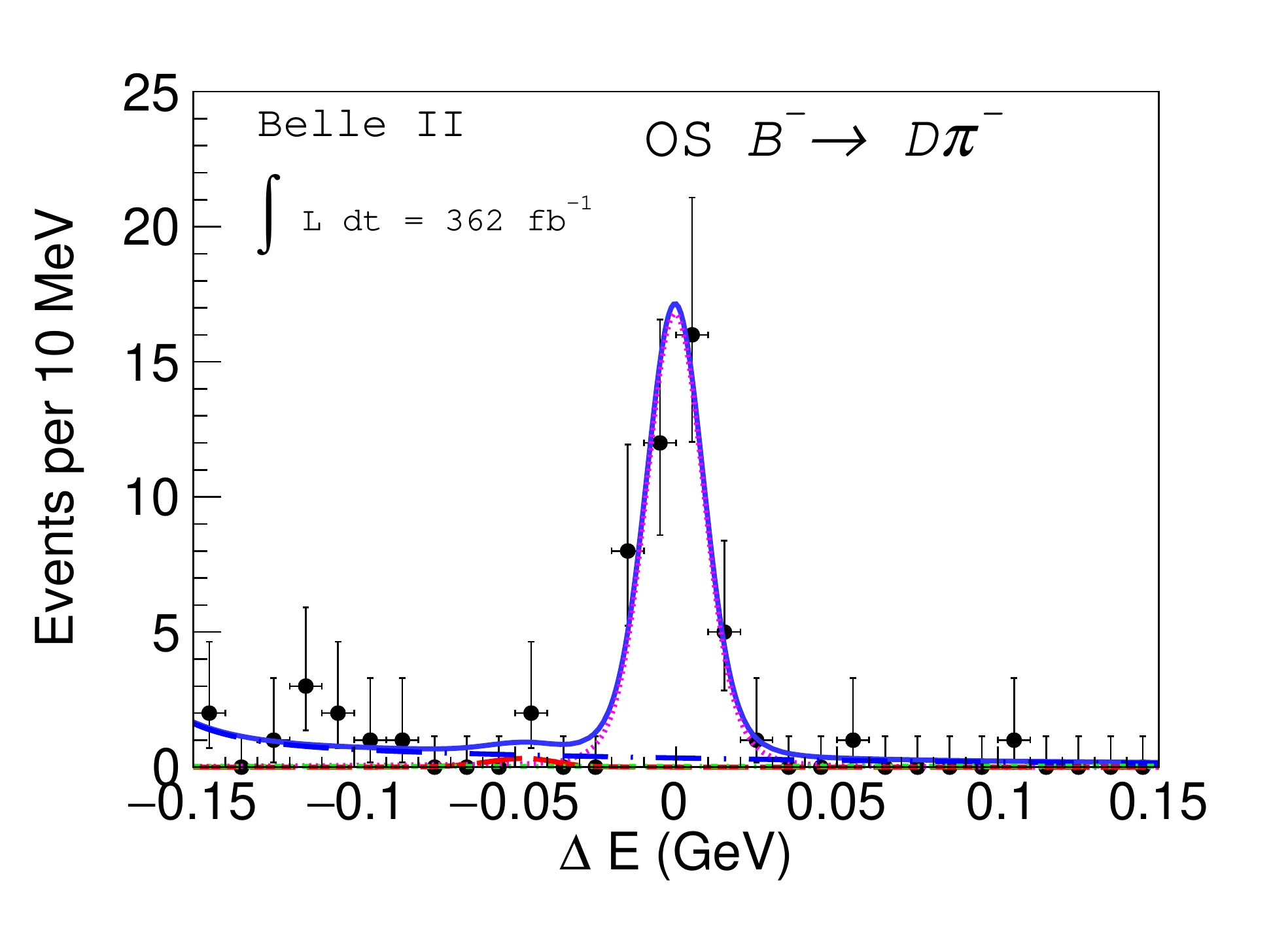}\put(40,55){}\end{overpic}
  \begin{overpic}[width=0.49\textwidth]{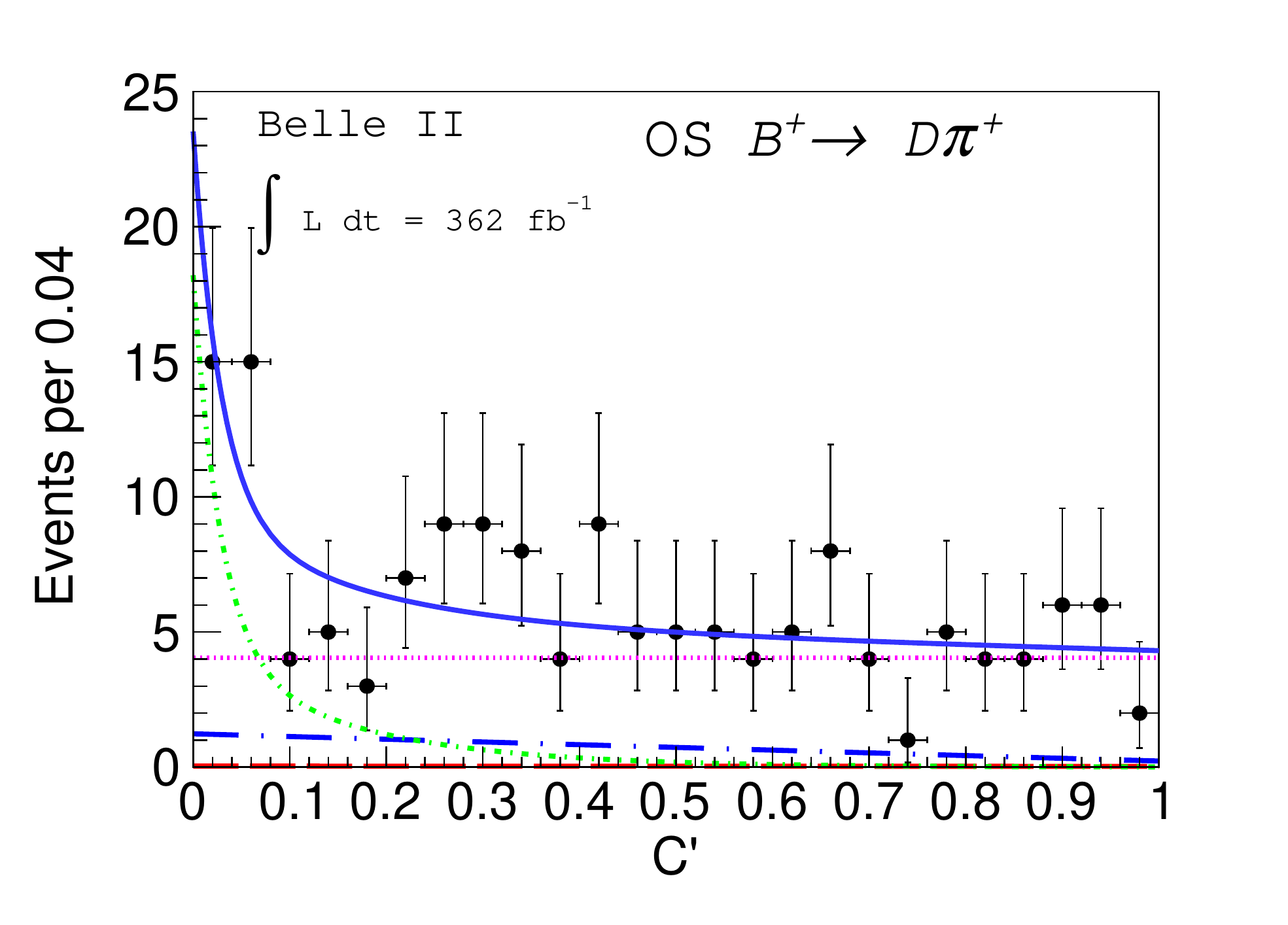}\put(40,55){}\end{overpic}
  \begin{overpic}[width=0.49\textwidth]{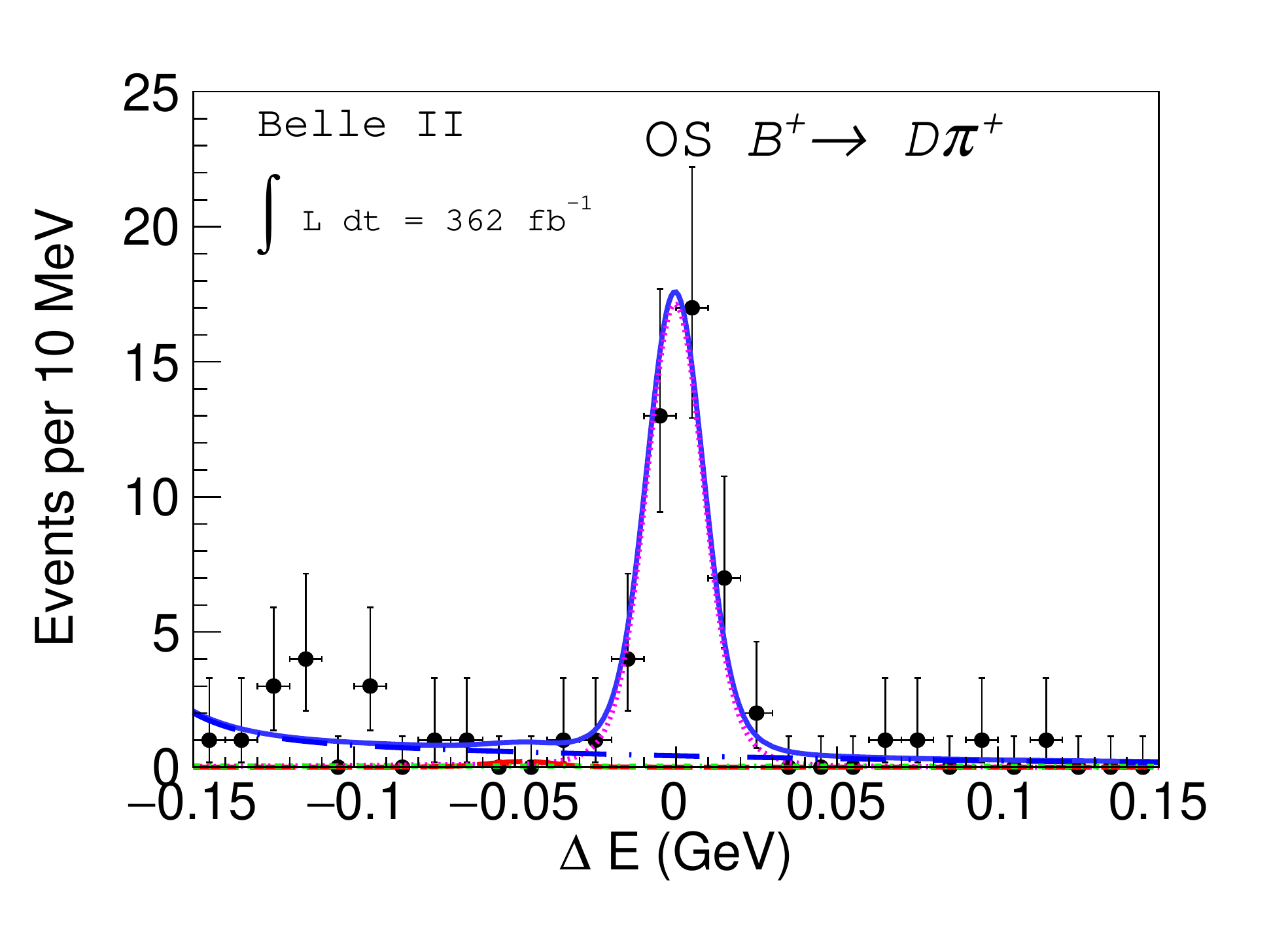}\put(40,55){}\end{overpic}

  \caption{\KstarResultsFigureCaption{\pipm}{Belle~II}}
  \label{fig:KstarKB2DPi}
\end{figure}

\section{Systematic uncertainties}
\label{sec:syst}

We consider the sources of systematic uncertainties listed in
table ~\ref{tab:B1B2DataSys}. For the first three sources in the table, 
we estimate the systematic effects associated with the fixed efficiencies in the fit and 
the uncertainties in the choice of model parameters.
We vary the values for the fixed parameters one thousand times, 
sampling them from a multivariate Gaussian distribution
with the known uncertainties and correlations.
We repeat the fit for each variation and inspect the distributions of results. 
If they are approximately Gaussian, we take the standard deviations as systematic uncertainties.
If they are non-Gaussian, we conservatively take the full ranges of the distributions
as the systematic uncertainties.
For the $\Delta E$ shape of the continuum component, which does not have any
fixed parameters, we repeat the fit using a second-order polynomial function as 
an alternative model and take the changes of the fit results as systematic
uncertainties.

In our fits, we assume equal efficiencies for detecting and reconstructing
$\PD\to\KS\Km\pip$ and $\PD\to\KS\Kp\pim$ decays.
In simulation, the ratio of the former to the latter is 0.98. We repeat
the analysis using this value and assign the differences in the
results as systematic uncertainties. Our simulated samples are generated
assuming the $\PD$ decay products are evenly distributed in the
available phase space, which is not the distribution expected in the
data. However, the efficiency ratios calculated with an alternative decay
model of $\PD\to\Kstarpm\Kmp$, which is the dominant process 
in $\PD\to\KS\Kmp\pipm$~\cite{ParticleDataGroup:2022pth}, 
are equivalent to the nominal ones.

\begin{table}[!t]
  \centering
  
  \caption{Systematic and statistical uncertainties in units of $10^{-2}$.
	The entries ``---'' indicate uncertainties smaller than $10^{-4}$.}
  
  \label{tab:B1B2DataSys}

  \vspace\baselineskip
  
  
  \begin{tabular}{l *{7}{c}}
    \hline
    \hline
    
    & {$A_{\text{SS}}^{{\it D K}}$}
    & {$A_{\text{OS}}^{{\it D K}}$}
    & {$A_{\text{SS}}^{{\it D \pi}}$}
    & {$A_{\text{OS}}^{{\it D \pi}}$}
    & {$R_{\text{SS}}^{{\it D K}/{\it D \pi}}$}
    & {$R_{\text{OS}}^{{\it D K}/{\it D \pi}}$}
    & {$R_{\text{SS/OS}}^{{\it D \pi}}$}
    \\

    \hline
    \hline
    
    \multicolumn{8}{c}{Full $\PD$ phase space}
    \\

    \hline
    
    $\epsilon_{\Kpm},\epsilon_{\pipm}$
    & 0.38
    & \phantom{1}0.56
    & 0.19
    & 0.14
    & 0.05
    & 0.06
    & 0.09
    \\
    
    $\delta$
    & ---
    & \phantom{1}0.03
    & ---
    & ---
    & 0.04
    & 0.03
    & 0.02
    \\
    
    Model
    & 0.62
    & \phantom{1}0.78
    & 0.02
    & 0.02
    & 0.30
    & 0.22
    & 0.07
    \\

    $\epsilon_{\KS\Km\pip} / \epsilon_{\KS\Kp\pim}$
    & 0.82
    & \phantom{1}0.83
    & 0.82
    & 0.83
    & 0.01
    & 0.01
    & 0.02
    \\

    \hline
    
    Total syst.\ unc.
    & 1.1\phantom{2}
    & \phantom{1}1.3\phantom{0}
    & 0.9\phantom{0}
    & 0.9\phantom{0}
    & 0.4\phantom{0}
    & 0.3\phantom{0}
    & 0.2\phantom{0}
    \\
    
    \hline
    
    Stat.\ unc.
    & 9.1\phantom{2}
    & 13.3\phantom{2}
    & 2.6\phantom{2}
    & 3.1\phantom{2}
    & 1.2\phantom{2}
    & 1.3\phantom{2}
    & 5.7\phantom{2}
    \\
    
    \hline
    \hline

    \multicolumn{8}{c}{$\Kstarpm$ region} \\

    \hline
    
    $\epsilon_{\Kpm},\epsilon_{\pipm}$
    & \phantom{1}0.37
    & \phantom{1}0.61
    & 0.17
    & 0.15
    & 0.03
    & 0.08
    & \phantom{1}0.13
    \\
    
    $\delta$
    & \phantom{1}0.02
    & \phantom{1}0.02
    & 0.01
    & 0.01
    & 0.03
    & 0.04
    & \phantom{1}0.04
    \\
    
    Model
    & \phantom{1}1.04
    & \phantom{1}0.97
    & 0.20
    & 0.03
    & 0.46
    & 0.49
    & \phantom{1}0.61
    \\

    $\epsilon_{\KS\Km\pip} / \epsilon_{\KS\Kp\pim}$
    & \phantom{1}1.6\phantom{2}
    & \phantom{1}0.8\phantom{2}
    & 1.6\phantom{2}
    & 0.8\phantom{2}
    & 0.1\phantom{2}
    & 0.1\phantom{2}
    & \phantom{1}1.7\phantom{2}
    \\
    
    \hline
    
    Total syst.\ unc.
    & \phantom{1}2.0\phantom{2}
    & \phantom{1}1.4\phantom{2}
    & 1.6\phantom{2}
    & 0.9\phantom{2}
    & 0.5\phantom{2}
    & 0.6\phantom{2}
    & \phantom{1}1.9\phantom{2}
    \\
    
    \hline
    Stat.\ unc.
    & 11.9\phantom{2}
    & 18.4\phantom{2}
    & 2.9\phantom{2}
    & 4.6\phantom{2}
    & 1.2\phantom{2}
    & 2.0\phantom{2}
    & 13.2\phantom{2}
    \\
    
    \hline
    \hline
  \end{tabular}
\end{table}

\section{Results}
\label{sec:results}

The results from the Belle and Belle~II data for the full $\PD$ phase space are
\begin{alignat}{1}
  A_{\text{SS}}^{{\it D K}}           &= -0.089 \pm 0.091 \pm 0.011,\\
  A_{\text{OS}}^{{\it D K}}           &=  0.109 \pm 0.133 \pm 0.013,\\
  A_{\text{SS}}^{{\it D \pi}}          &=  0.018 \pm 0.026 \pm 0.009,\\
  A_{\text{OS}}^{{\it D \pi}}          &= -0.028 \pm 0.031 \pm 0.009,\\
  R_{\text{SS}}^{{\it D K}/{\it D \pi}}    &=  0.122 \pm 0.012 \pm 0.004,\\
  R_{\text{OS}}^{{\it D K}/{\it D \pi}}    &=  0.093 \pm 0.013 \pm 0.003,\\
  R_{\text{SS}/\text{OS}}^{{\it D \pi}} &=  1.428 \pm 0.057 \pm 0.002,
\end{alignat}
and for the $\Kstarpm$ region are
\begin{alignat}{1}
  A_{\text{SS}}^{{\it D K}}           &= 0.055 \pm 0.119 \pm 0.020,\\
  A_{\text{OS}}^{{\it D K}}           &= 0.231 \pm 0.184 \pm 0.014,\\
  A_{\text{SS}}^{{\it D \pi}}           &= 0.046 \pm 0.029 \pm 0.016,\\
  A_{\text{OS}}^{{\it D \pi}}           &= 0.009 \pm 0.046 \pm 0.009,\\
  R_{\text{SS}}^{{\it D K}/{\it D \pi}}    &= 0.093 \pm 0.012 \pm 0.005,\\
  R_{\text{OS}}^{{\it D K}/{\it D \pi}}    &= 0.103 \pm 0.020 \pm 0.006,\\
  R_{\text{SS}/\text{OS}}^{{\it D \pi}} &= 2.412 \pm 0.132 \pm 0.019,
\end{alignat}
where the first uncertainty is statistical and the second is systematic.
Tables~\ref{tab:PHSPStatCor}--\ref{tab:KstarKDataSystCor} list
the statistical and systematic correlations of all results.
Our results are consistent with LHCb's results~\cite{LHCb:2020vut},
with worse precision due to a smaller sample size.
This study is also performed with the Belle data set alone and the results are reported in Appendix~\ref{App:BelleResult}.

Our results alone do not allow for an unambiguous determination of $\phi_3$,
but combined with other results in global fits they constrain it.
The enhancement of the $D\to\KS\Kpm\pimp$'s coherence factor $\kappa_D$ 
in the $\Kstarpm$ region indicated by the current CLEO measurement~\cite{CLEO:2012obf} 
suggests the possibility of an enhancement of the $\CP$-violating asymmetry in that region. However, the current precisions of both GLS results from this paper and strong-phase difference results prevent a conclusive statement if such enhancement is sufficient to compensate for the extra efficiency loss due to the phase space restriction.

\section{Summary}
\label{sec:summary}

We measure the $\CP$ asymmetries and branching-fraction ratios for
$\Bpm\to\PD\Kpm$ and $\Bpm\to\PD\pipm$ decays with 
$\PD\to\KS\Kpm\pimp$ using the full Belle data set containing
\num{772e6}~$\BBbar$ pairs and a Belle~II data set containing
\num{387e6}~$\BBbar$ pairs.
We extract these observables simultaneously through a simultaneous fit across data sets and channels
for the full $\PD$ phase space and in the region of the $\Kstarpm$. 
These results, combined with other $\phi_3$-related results, constrain the unitarity triangle angle $\phi_3$.

\newcommand{\CorrelationTableCaption}[3]{#1 correlations for results in the #2 using #3.}
\newcommand{\FullCorrelationTableCaption}[2]{\CorrelationTableCaption{#1}{full $\PD$ phase space}{#2}}
\newcommand{\KstarCorrelationTableCaption}[2]{\CorrelationTableCaption{#1}{region of the $\Kstarpm$}{#2}}

\begin{table}[!t]
  \centering

  \caption{\FullCorrelationTableCaption{Statistical}{Belle and Belle~II data}}

  \label{tab:PHSPStatCor}

  \vspace\baselineskip
  
  \sisetup{round-mode=places,round-precision=3}
  
  \begin{tabular}{l|*{6}{S}}
    \hline
    \hline
    & {$A_{\text{OS}}^{{\it D K}}$}
    & {$A_{\text{SS}}^{{\it D \pi}}$}
    & {$A_{\text{OS}}^{{\it D \pi}}$}
    & {$R_{\text{SS}}^{{\it D K}/{\it D \pi}}$}
    & {$R_{\text{OS}}^{{\it D K}/{\it D \pi}}$}
    & {$R_{\text{SS/OS}}^{{\it D \pi}}$}
    \\

    \hline

    $A_{\text{SS}}^{{\it D K}}$         & 0.0007 & -0.0095 & 0.0003 & 0.0006 & -0.0001 & 0.0027 \\  
    $A_{\text{OS}}^{{\it D K}}$         &  & 0.0004 & -0.0063 & 0.0025 & 0.0078 & -0.0010 \\  
    $A_{\text{SS}}^{{\it D \pi}}$        &  &  & 0.0002 & 0.0056 & -0.0014 & -0.0044 \\  
    $A_{\text{OS}}^{{\it D \pi}}$        &  &  &  & 0.0007 & 0.0006 & 0.0005 \\  
    $R_{\text{SS}}^{{\it D K}/{\it D \pi}}$  &  &  &  &  & 0.0383 & -0.1761 \\  
    $R_{\text{OS}}^{{\it D K}/{\it D \pi}}$   &  &  &  &  &  & 0.1777 \\  

    \hline
    \hline
  \end{tabular}
\end{table}

\begin{table}[!t]
  \centering
  \caption{\FullCorrelationTableCaption{Systematic}{Belle and Belle~II data}}
  \label{tab:PHSPSysCor}

  \vspace\baselineskip
  
  \sisetup{round-mode=places,round-precision=3}

  \begin{tabular}{l|*{6}{S}}
    \hline
    \hline
    & {$A_{\text{OS}}^{{\it D K}}$}
    & {$A_{\text{SS}}^{{\it D \pi}}$}
    & {$A_{\text{OS}}^{{\it D \pi}}$}
    & {$R_{\text{SS}}^{{\it D K}/{\it D \pi}}$}
    & {$R_{\text{OS}}^{{\it D K}/{\it D \pi}}$}
    & {$R_{\text{SS/OS}}^{{\it D \pi}}$}
    \\

    \hline
		$A_{SS,DK}$ &  0.3895 & -0.0020 & 0.0645 & 0.4087 & 0.4469 & 0.1291 \\  
		$A_{OS,DK}$ &  &  0.0254 & 0.0013 & 0.2054 & 0.2340 & 0.0316 \\  
		$A_{SS,D\pi}$ &  &  &  0.0363 & 0.0517 & 0.0432 & -0.0482 \\  
		$A_{OS,D\pi}$ &  &  &  &  0.0943 & 0.0861 & 0.0627 \\  
		$R_{SS,DK/D\pi}$ &  &  &  &  &  0.9526 & -0.0717 \\  
		$R_{OS,DK/D\pi}$ &  &  &  &  &   & 0.0888 \\  

    \hline
    \hline
  \end{tabular}
\end{table}

\begin{table}[!t]
  \centering
  \caption{\KstarCorrelationTableCaption{Statistical}{Belle and Belle~II data}}
    \label{tab:KstarKDataStatCor}

    \vspace\baselineskip

    \sisetup{round-mode=places,round-precision=3}
    
    \begin{tabular}{l|@{}*{6}{S}}
      \hline
      \hline
      & {$A_{\text{OS}}^{{\it D K}}$}
      & {$A_{\text{SS}}^{{\it D \pi}}$}
      & {$A_{\text{OS}}^{{\it D \pi}}$}
      & {$R_{\text{SS}}^{{\it D K}/{\it D \pi}}$}
      & {$R_{\text{OS}}^{{\it D K}/{\it D \pi}}$}
      & {$R_{\text{SS/OS}}^{{\it D \pi}}$}
      \\
      
      \hline
      $A_{\text{SS}}^{{\it D K}}$       &   0.0026 & -0.0115 & 0.0006 & -0.0515 & -0.0130 & 0.0018 \\
      $A_{\text{OS}}^{{\it D K}}$       &   &  0.0008 & -0.0107 & -0.0037 & -0.0339 & 0.0016 \\
      $A_{\text{SS}}^{{\it D \pi}}$        &  &  &  0.0003 & 0.0020 & -0.0038 & -0.0109 \\
      $A_{\text{OS}}^{{\it D \pi}}$        &  &  &  &  -0.0019 & -0.0019 & 0.0136 \\
      $R_{\text{SS}}^{{\it D K}/{\it D \pi}}$ &  &  &  &  &  0.0338 & -0.1323 \\
      $R_{\text{OS}}^{{\it D K}/{\it D \pi}}$ &  &  &  &  &  &  0.2079 \\
      \hline
      \hline
    \end{tabular}
\end{table}

\begin{table}[!t]
  \centering
  \caption{\KstarCorrelationTableCaption{Systematic}{Belle and Belle~II data}}
  \label{tab:KstarKDataSystCor}

  \vspace\baselineskip
  
  \sisetup{round-mode=places,round-precision=3}
  
  \begin{tabular}{l|*{6}{S}}
    \hline
    \hline
    & {$A_{\text{OS}}^{{\it D K}}$}
    & {$A_{\text{SS}}^{{\it D \pi}}$}
    & {$A_{\text{OS}}^{{\it D \pi}}$}
    & {$R_{\text{SS}}^{{\it D K}/{\it D \pi}}$}
    & {$R_{\text{OS}}^{{\it D K}/{\it D \pi}}$}
    & {$R_{\text{SS/OS}}^{{\it D \pi}}$}
    \\

    \hline
		$A_{SS,DK}$ & 0.1949 & 0.0461 & 0.0126 & 0.1199 & -0.0529 & 0.1913 \\  
		$A_{OS,DK}$ &  & 0.0379 & 0.0038 & 0.3436 & 0.2098 & 0.0061 \\  
		$A_{SS,D\pi}$ &  &  & 0.0233 & -0.0035 & -0.0369 & 0.0173 \\  
		$A_{OS,D\pi}$ &  &  &  & -0.0164 & -0.0232 & 0.0055 \\  
		$R_{SS,DK/D\pi}$ &  &  &  &  & 0.9143 & 0.0149 \\  
		$R_{OS,DK/D\pi}$ &  &  &  &  &  & -0.0969 \\  
    \hline
    \hline
  \end{tabular}
\end{table}

\appendix
\section{Belle data results}
\label{App:BelleResult}
The results using only Belle data for the full $\PD$ phase space are
\begin{alignat}{1}
  A_{\text{SS}}^{{\it D K}} &= -0.121 \pm 0.120 \pm 0.013,\\
  A_{\text{OS}}^{{\it D K}} &= -0.016 \pm 0.182 \pm 0.014,\\
  A_{\text{SS}}^{{\it D \pi}} &= 0.014 \pm 0.032 \pm 0.011,\\
  A_{\text{OS}}^{{\it D \pi}} &= 0.001 \pm 0.039 \pm 0.011,\\
  R_{\text{SS}}^{{\it D K}/{\it D \pi}} &= 0.112 \pm 0.014 \pm 0.002,\\
  R_{\text{OS}}^{{\it D K}/{\it D \pi}} &= 0.085 \pm 0.016 \pm 0.002,\\
  R_{\text{SS}/\text{OS}}^{{\it D \pi}} &= 1.472 \pm 0.074 \pm 0.002,
\end{alignat}
and for the $\Kstarpm$ region are
\begin{alignat}{1}
  A_{\text{SS}}^{{\it D K}} &= 0.028 \pm 0.163 \pm 0.022,\\
  A_{\text{OS}}^{{\it D K}} &= 0.220 \pm 0.245 \pm 0.014,\\
  A_{\text{SS}}^{{\it D \pi}} &= 0.041 \pm 0.036 \pm 0.011,\\
  A_{\text{OS}}^{{\it D \pi}} &= 0.041 \pm 0.059 \pm 0.011,\\
  R_{\text{SS}}^{{\it D K}/{\it D \pi}} &= 0.082 \pm 0.014 \pm 0.002,\\
  R_{\text{OS}}^{{\it D K}/{\it D \pi}} &= 0.097 \pm 0.027 \pm 0.002,\\
  R_{\text{SS}/\text{OS}}^{{\it D \pi}} &= 2.592 \pm 0.180 \pm 0.005.
\end{alignat}

Tables~\ref{tab:B1PHSPDataStatCor}--\ref{tab:B1KstarKDataSystCor} list
the statistical and systematic correlations. 
These results on Belle data only are provided to facilitate combinations with future updates of Belle II results.
\begin{table}[!t]
  \centering
  \caption{\FullCorrelationTableCaption{Statistical}{Belle data alone}}
  \label{tab:B1PHSPDataStatCor}

  \vspace\baselineskip
  
  \sisetup{round-mode=places,round-precision=3}
  
  \begin{tabular}{l|*{6}{S}}
    \hline
    \hline
    & {$A_{\text{OS}}^{{\it D K}}$}
    & {$A_{\text{SS}}^{{\it D \pi}}$}
    & {$A_{\text{OS}}^{{\it D \pi}}$}
    & {$R_{\text{SS}}^{{\it D K}/{\it D \pi}}$}
    & {$R_{\text{OS}}^{{\it D K}/{\it D \pi}}$}
    & {$R_{\text{SS/OS}}^{{\it D \pi}}$}
    \\
    
    \hline
    $A_{\text{SS}}^{{\it D K}}$       &  -0.0000 & -0.0154 & 0.0003 & 0.0271 & 0.0019 & 0.0041 \\
    $A_{\text{OS}}^{{\it D K}}$      &   &  -0.0003 & -0.0130 & 0.0037 & 0.0505 & 0.0003 \\
    $A_{\text{SS}}^{{\it D \pi}}$       &  &  &  0.0003 & 0.0085 & -0.0027 & -0.0075 \\
    $A_{\text{OS}}^{{\it D \pi}}$       &  &  &  &  -0.0017 & -0.0001 & 0.0074 \\
    $R_{\text{SS}}^{{\it D K}/{\it D \pi}}$ &  &  &  &  &  0.0482 & -0.1659 \\
    $R_{\text{OS}}^{{\it D K}/{\it D \pi}}$ &  &  &  &  &  &  0.1675 \\
    \hline
    \hline
  \end{tabular}
\end{table}

\begin{table}[!t]
  \centering
  \caption{\FullCorrelationTableCaption{Systematic}{Belle data alone}}
  \label{tab:B1PHSPDataSystCor}

  \vspace\baselineskip
  
  \sisetup{round-mode=places,round-precision=3}
  
  \begin{tabular}{l|*{6}{S}}
    \hline
    \hline
    & {$A_{\text{OS}}^{{\it D K}}$}
    & {$A_{\text{SS}}^{{\it D \pi}}$}
    & {$A_{\text{OS}}^{{\it D \pi}}$}
    & {$R_{\text{SS}}^{{\it D K}/{\it D \pi}}$}
    & {$R_{\text{OS}}^{{\it D K}/{\it D \pi}}$}
    & {$R_{\text{SS/OS}}^{{\it D \pi}}$}
    \\

    \hline
		$A_{SS,DK}$ & 0.3257 & -0.0152 & -0.0733 & -0.1787 & 0.0256 & 0.2979 \\  
		$A_{OS,DK}$ &  & 0.0252 & -0.0529 & -0.1246 & 0.0647 & 0.1630 \\  
		$A_{SS,D\pi}$ &  &  & 0.0582 & 0.0826 & 0.0586 & -0.0215 \\  
		$A_{OS,D\pi}$ &  &  &  & 0.0357 & -0.0258 & -0.0604 \\  
		$R_{SS,DK/D\pi}$ &  &  &  &  & 0.8038 & -0.3133 \\  
		$R_{OS,DK/D\pi}$ &  &  &  &  &  & 0.1832 \\  
    \hline
    \hline
  \end{tabular}
\end{table}

\begin{table}[!t]
  \centering
  \caption{\KstarCorrelationTableCaption{Statistical}{Belle data alone}}
  \label{tab:B1KstarKDataStatCor}

  \vspace\baselineskip
  
  \sisetup{round-mode=places,round-precision=3}
  
  \begin{tabular}{l|*{6}{S}}
    \hline
    \hline
    & {$A_{\text{OS}}^{{\it D K}}$}
    & {$A_{\text{SS}}^{{\it D \pi}}$}
    & {$A_{\text{OS}}^{{\it D \pi}}$}
    & {$R_{\text{SS}}^{{\it D K}/{\it D \pi}}$}
    & {$R_{\text{OS}}^{{\it D K}/{\it D \pi}}$}
    & {$R_{\text{SS/OS}}^{{\it D \pi}}$}
    \\

    \hline
		$A_{SS,DK}$ &  -0.0026 & -0.0134 & -0.0006 & -0.0716 & 0.0139 & -0.0012 \\  
		$A_{OS,DK}$ &  &  -0.0004 & -0.0214 & -0.0061 & -0.0188 & 0.0034 \\  
		$A_{SS,D\pi}$ &  &  &  0.0001 & 0.0050 & 0.0013 & -0.0086 \\  
		$A_{OS,D\pi}$ &  &  &  &  -0.0022 & -0.0055 & 0.0200 \\  
		$R_{SS,DK/D\pi}$ &  &  &  &  &  0.0359 & -0.1207 \\  
		$R_{OS,DK/D\pi}$ &  &  &  &  &  &  0.1963 \\  
    \hline
    \hline
  \end{tabular}
\end{table}

\begin{table}[!t]
  \centering
  \caption{\KstarCorrelationTableCaption{Systematic}{Belle data alone}}
  \label{tab:B1KstarKDataSystCor}

  \vspace\baselineskip
  
  \sisetup{round-mode=places,round-precision=3}
  
  \begin{tabular}{l|*{6}{S}}
    \hline
    \hline
    & {$A_{\text{OS}}^{{\it D K}}$}
    & {$A_{\text{SS}}^{{\it D \pi}}$}
    & {$A_{\text{OS}}^{{\it D \pi}}$}
    & {$R_{\text{SS}}^{{\it D K}/{\it D \pi}}$}
    & {$R_{\text{OS}}^{{\it D K}/{\it D \pi}}$}
    & {$R_{\text{SS/OS}}^{{\it D \pi}}$}
    \\
    
    \hline
    $A_{\text{SS}}^{{\it D K}}$     &    0.1973 & 0.1761 & -0.0492 & 0.2989 & 0.0067 & 0.2828 \\  
    $A_{\text{OS}}^{{\it D K}}$     &    &  0.0561 & -0.0247 & 0.0511 & 0.1520 & 0.0704 \\  
    $A_{\text{SS}}^{{\it D \pi}}$       &  &  &  0.0563 & 0.0345 & -0.0015 & 0.1262 \\  
    $A_{\text{OS}}^{{\it D \pi}}$       &  &  &  &  0.0085 & -0.0491 & 0.0589 \\  
    $R_{\text{SS}}^{{\it D K}/{\it D \pi}}$ &  &  &  &  &  0.5833 & 0.3132 \\  
    $R_{\text{OS}}^{{\it D K}/{\it D \pi}}$ &  &  &  &  &  &  -0.0586 \\  
    \hline
    \hline
  \end{tabular}
\end{table}

\acknowledgments
\input{acknowledgements}

\clearpage
\bibliographystyle{JHEP}
\bibliography{references}

\end{document}

%% file: authors.tex
\collaboration{The Belle and Belle II Collaboration}
  \author{I.~Adachi\,\orcidlink{0000-0003-2287-0173},} 
  \author{L.~Aggarwal\,\orcidlink{0000-0002-0909-7537},} 
  \author{H.~Aihara\,\orcidlink{0000-0002-1907-5964},} 
  \author{N.~Akopov\,\orcidlink{0000-0002-4425-2096},} 
  \author{A.~Aloisio\,\orcidlink{0000-0002-3883-6693},} 
  \author{N.~Anh~Ky\,\orcidlink{0000-0003-0471-197X},} 
  \author{D.~M.~Asner\,\orcidlink{0000-0002-1586-5790},} 
  \author{T.~Aushev\,\orcidlink{0000-0002-6347-7055},} 
  \author{V.~Aushev\,\orcidlink{0000-0002-8588-5308},} 
  \author{M.~Aversano\,\orcidlink{0000-0001-9980-0953},} 
% ...

%% file: acknowledgements.tex
This work, based on data collected using the Belle II detector, which was built and commissioned prior to March 2019, was supported by
Science Committee of the Republic of Armenia Grant No.~20TTCG-1C010;
Australian Research Council and research Grants
No.~DP200101792, 
No.~DP210101900, 
No.~DP210102831, 
No.~DE220100462, 
No.~LE210100098, 
and
No.~LE230100085; 
Austrian Federal Ministry of Education, Science and Research,
Austrian Science Fund
No.~P~31361-N36
and
No.~J4625-N,
and
Horizon 2020 ERC Starting Grant No.~947006 ``InterLeptons'';
Natural Sciences and Engineering Research Council of Canada, Compute Canada and CANARIE;
National Key R\&D Program of China under Contract No.~2022YFA1601903,
National Natural Science Foundation of China and research Grants
No.~11575017,
No.~11761141009,
No.~11705209,
No.~11975076,
No.~12135005,
No.~12150004,
No.~12161141008,
and
No.~12175041,
and Shandong Provincial Natural Science Foundation Project~ZR2022JQ02;
the Ministry of Education, Youth, and Sports of the Czech Republic under Contract No.~LTT17020 and
Charles University Grant No.~SVV 260448 and
the Czech Science Foundation Grant No.~22-18469S;
European Research Council, Seventh Framework PIEF-GA-2013-622527,
Horizon 2020 ERC-Advanced Grants No.~267104 and No.~884719,
Horizon 2020 ERC-Consolidator Grant No.~819127,
Horizon 2020 Marie Sklodowska-Curie Grant Agreement No.~700525 "NIOBE"
and
No.~101026516,
and
Horizon 2020 Marie Sklodowska-Curie RISE project JENNIFER2 Grant Agreement No.~822070 (European grants);
L'Institut National de Physique Nucl\'{e}aire et de Physique des Particules (IN2P3) du CNRS (France);
BMBF, DFG, HGF, MPG, and AvH Foundation (Germany);
Department of Atomic Energy under Project Identification No.~RTI 4002 and Department of Science and Technology (India);
Israel Science Foundation Grant No.~2476/17,
U.S.-Israel Binational Science Foundation Grant No.~2016113, and
Israel Ministry of Science Grant No.~3-16543;
Istituto Nazionale di Fisica Nucleare and the research grants BELLE2;
Japan Society for the Promotion of Science, Grant-in-Aid for Scientific Research Grants
No.~16H03968,
No.~16H03993,
No.~16H06492,
No.~16K05323,
No.~17H01133,
No.~17H05405,
No.~18K03621,
No.~18H03710,
No.~18H05226,
No.~19H00682, 
No.~22H00144,
No.~26220706,
and
No.~26400255,
the National Institute of Informatics, and Science Information NETwork 5 (SINET5), 
and
the Ministry of Education, Culture, Sports, Science, and Technology (MEXT) of Japan;  
National Research Foundation (NRF) of Korea Grants
No.~2016R1\-D1A1B\-02012900,
No.~2018R1\-A2B\-3003643,
No.~2018R1\-A6A1A\-06024970,
No.~2018R1\-D1A1B\-07047294,
No.~2019R1\-I1A3A\-01058933,
No.~2022R1\-A2C\-1003993,
and
No.~RS-2022-00197659,
Radiation Science Research Institute,
Foreign Large-size Research Facility Application Supporting project,
the Global Science Experimental Data Hub Center of the Korea Institute of Science and Technology Information
and
KREONET/GLORIAD;
Universiti Malaya RU grant, Akademi Sains Malaysia, and Ministry of Education Malaysia;
Frontiers of Science Program Contracts
No.~FOINS-296,
No.~CB-221329,
No.~CB-236394,
No.~CB-254409,
and
No.~CB-180023, and No.~SEP-CINVESTAV research Grant No.~237 (Mexico);
the Polish Ministry of Science and Higher Education and the National Science Center;
the Ministry of Science and Higher Education of the Russian Federation,
Agreement No.~14.W03.31.0026, and
the HSE University Basic Research Program, Moscow;
University of Tabuk research Grants
No.~S-0256-1438 and No.~S-0280-1439 (Saudi Arabia);
Slovenian Research Agency and research Grants
No.~J1-9124
and
No.~P1-0135;
Agencia Estatal de Investigacion, Spain
Grant No.~RYC2020-029875-I
and
Generalitat Valenciana, Spain
Grant No.~CIDEGENT/2018/020
Ministry of Science and Technology and research Grants
No.~MOST106-2112-M-002-005-MY3
and
No.~MOST107-2119-M-002-035-MY3,
and the Ministry of Education (Taiwan);
Thailand Center of Excellence in Physics;
TUBITAK ULAKBIM (Turkey);
National Research Foundation of Ukraine, project No.~2020.02/0257,
and
Ministry of Education and Science of Ukraine;
the U.S. National Science Foundation and research Grants
No.~PHY-1913789 
and
No.~PHY-2111604, 
and the U.S. Department of Energy and research Awards
No.~DE-AC06-76RLO1830, 
No.~DE-SC0007983, 
No.~DE-SC0009824, 
No.~DE-SC0009973, 
No.~DE-SC0010007, 
No.~DE-SC0010073, 
No.~DE-SC0010118, 
No.~DE-SC0010504, 
No.~DE-SC0011784, 
No.~DE-SC0012704, 
No.~DE-SC0019230, 
No.~DE-SC0021274, 
No.~DE-SC0022350, 
No.~DE-SC0023470; 
and
the Vietnam Academy of Science and Technology (VAST) under Grant No.~DL0000.05/21-23.

These acknowledgements are not to be interpreted as an endorsement of any statement made
by any of our institutes, funding agencies, governments, or their representatives.

We thank the SuperKEKB team for delivering high-luminosity collisions;
the KEK cryogenics group for the efficient operation of the detector solenoid magnet;
the KEK computer group and the NII for on-site computing support and SINET6 network support;
and the raw-data centers at BNL, DESY, GridKa, IN2P3, INFN, and the University of Victoria for offsite computing support.

%% file: main.bbl
\providecommand{\href}[2]{#2}\begingroup\raggedright\begin{thebibliography}{10}

\bibitem{Cabibbo:1963yz}
N.~Cabibbo, \emph{{Unitary Symmetry and Leptonic Decays}},
  \href{https://doi.org/10.1103/PhysRevLett.10.531}{\emph{Phys. Rev. Lett.}
  {\bfseries 10} (1963) 531}.

\bibitem{Kobayashi:1973fv}
M.~Kobayashi and T.~Maskawa, \emph{{CP Violation in the Renormalizable Theory
  of Weak Interaction}}, \href{https://doi.org/10.1143/PTP.49.652}{\emph{Prog.
  Theor. Phys.} {\bfseries 49} (1973) 652}.

\bibitem{Brod:2014bfa}
J.~Brod, A.~Lenz, G.~Tetlalmatzi-Xolocotzi and M.~Wiebusch, \emph{{New physics
  effects in tree-level decays and the precision in the determination of the
  quark mixing angle \ensuremath{\gamma}}},
  \href{https://doi.org/10.1103/PhysRevD.92.033002}{\emph{Phys. Rev. D}
  {\bfseries 92} (2015) 033002}
  [\href{https://arxiv.org/abs/1412.1446}{{\ttfamily 1412.1446}}].

\bibitem{Brod:2013sga}
J.~Brod and J.~Zupan, \emph{{The ultimate theoretical error on $\gamma$ from $B
  \to DK$ decays}}, \href{https://doi.org/10.1007/JHEP01(2014)051}{\emph{JHEP}
  {\bfseries 01} (2014) 051} [\href{https://arxiv.org/abs/1308.5663}{{\ttfamily
  1308.5663}}].

\bibitem{ParticleDataGroup:2022pth}
{\scshape Particle Data Group}, \emph{{Review of Particle Physics}},
  \href{https://doi.org/10.1093/ptep/ptac097}{\emph{Prog. Theor. Exp. Phys.}
  {\bfseries 2022} (2022) 083C01}.

\bibitem{BaBar:2013caj}
{\scshape BaBar collaboration}, \emph{{Observation of direct CP violation in
  the measurement of the Cabibbo-Kobayashi-Maskawa angle $\gamma$ with
  $B^\pm\to D^{(*)}K^{(*)\pm}$ decays}},
  \href{https://doi.org/10.1103/PhysRevD.87.052015}{\emph{Phys. Rev. D}
  {\bfseries 87} (2013) 052015}
  [\href{https://arxiv.org/abs/1301.1029}{{\ttfamily 1301.1029}}].

\bibitem{Belle:2021efh}
{\scshape Belle and Belle II collaborations}, \emph{{Combined analysis of Belle
  and Belle II data to determine the CKM angle $ \phi_{3} $ using $B^+ \to
  D(K_{S}^0 h^- h^+) h^+$ decays}},
  \href{https://doi.org/10.1007/JHEP02(2022)063}{\emph{JHEP} {\bfseries 02}
  (2022) 063} [\href{https://arxiv.org/abs/2110.12125}{{\ttfamily
  2110.12125}}].

\bibitem{LHCb:2022awq}
{\scshape LHCb collaboration}, \emph{{Simultaneous determination of the CKM
  angle $\gamma$ and parameters related to mixing and $CP$ violation in the
  charm sector}}, {\emph{LHCb-CONF-2022-003} (2022) }.

\bibitem{Grossman:2002aq}
Y.~Grossman, Z.~Ligeti and A.~Soffer, \emph{{Measuring $\gamma$ in B$^{\pm}$
  \textrightarrow{} K$^{\pm} (KK^*)_D$ decays}},
  \href{https://doi.org/10.1103/PhysRevD.67.071301}{\emph{Phys. Rev. D}
  {\bfseries 67} (2003) 071301}
  [\href{https://arxiv.org/abs/hep-ph/0210433}{{\ttfamily hep-ph/0210433}}].

\bibitem{CLEO:2012obf}
{\scshape CLEO collaboration}, \emph{{Studies of the decays $D^0 \rightarrow
  K_S^0K^-\pi^+$ and $D^0 \rightarrow K_S^0K^+\pi^-$}},
  \href{https://doi.org/10.1103/PhysRevD.85.092016}{\emph{Phys. Rev. D}
  {\bfseries 85} (2012) 092016}
  [\href{https://arxiv.org/abs/1203.3804}{{\ttfamily 1203.3804}}].

\bibitem{LHCb:2020vut}
{\scshape LHCb collaboration}, \emph{{Measurement of CP observables in B$^{±}$
  \textrightarrow{} DK$^{±}$ and B$^{±}$ \textrightarrow{}
  D\ensuremath{\pi}$^{±}$ with D \textrightarrow{} $
  {K}_{\mathrm{S}}^0{K}^{\pm }{\pi}^{\mp } $ decays}},
  \href{https://doi.org/10.1007/JHEP06(2020)058}{\emph{JHEP} {\bfseries 06}
  (2020) 058} [\href{https://arxiv.org/abs/2002.08858}{{\ttfamily
  2002.08858}}].

\bibitem{Belle:2000cnh}
{\scshape Belle collaboration}, \emph{{The Belle Detector}},
  \href{https://doi.org/10.1016/S0168-9002(01)02013-7}{\emph{Nucl. Instrum.
  Meth. A} {\bfseries 479} (2002) 117}.

\bibitem{Belle-II:2010dht}
{\scshape Belle II collaboration}, \emph{{Belle II Technical Design Report}},
  \href{https://arxiv.org/abs/1011.0352}{{\ttfamily 1011.0352}}.

\bibitem{Belle:2012iwr}
{\scshape Belle collaboration}, \emph{{Physics Achievements from the Belle
  Experiment}}, \href{https://doi.org/10.1093/ptep/pts072}{\emph{Prog. Theor.
  Exp. Phys.} {\bfseries 2012} (2012) 04D001}
  [\href{https://arxiv.org/abs/1212.5342}{{\ttfamily 1212.5342}}].

\bibitem{Kurokawa:2001nw}
S.~Kurokawa and E.~Kikutani, \emph{{Overview of the KEKB accelerators}},
  \href{https://doi.org/10.1016/S0168-9002(02)01771-0}{\emph{Nucl. Instrum.
  Meth. A} {\bfseries 499} (2003) 1}.

\bibitem{Abe:2013kxa}
T.~Abe et~al., \emph{{Achievements of KEKB}},
  \href{https://doi.org/10.1093/ptep/pts102}{\emph{Prog. Theor. Exp. Phys.}
  {\bfseries 2013} (2013) 03A001}.

\bibitem{Akai:2018mbz}
K.~Akai, K.~Furukawa and H.~Koiso, \emph{{SuperKEKB Collider}},
  \href{https://doi.org/10.1016/j.nima.2018.08.017}{\emph{Nucl. Instrum. Meth.
  A} {\bfseries 907} (2018) 188}
  [\href{https://arxiv.org/abs/1809.01958}{{\ttfamily 1809.01958}}].

\bibitem{Lange:2001uf}
D.J.~Lange, \emph{{The EvtGen particle decay simulation package}},
  \href{https://doi.org/10.1016/S0168-9002(01)00089-4}{\emph{Nucl. Instrum.
  Meth. A} {\bfseries 462} (2001) 152}.

\bibitem{Sjostrand:2007gs}
T.~Sj{\"o}strand, S.~Mrenna and P.Z.~Skands, \emph{{A Brief Introduction to
  PYTHIA 8.1}}, \href{https://doi.org/10.1016/j.cpc.2008.01.036}{\emph{Comput.
  Phys. Commun.} {\bfseries 178} (2008) 852}
  [\href{https://arxiv.org/abs/0710.3820}{{\ttfamily 0710.3820}}].

\bibitem{Jadach:1999vf}
S.~Jadach, B.F.L.~Ward and Z.~W\c{a}s, \emph{{The precision Monte Carlo event
  generator KK for two-fermion final states in $e^+e^-$ collisions}},
  \href{https://doi.org/10.1016/S0010-4655(00)00048-5}{\emph{Comput. Phys.
  Commun.} {\bfseries 130} (2000) 260}
  [\href{https://arxiv.org/abs/hep-ph/9912214}{{\ttfamily hep-ph/9912214}}].

\bibitem{Barberio:1993qi}
E.~Barberio and Z.~W\c{a}s, \emph{{PHOTOS: A Universal Monte Carlo for QED
  radiative corrections. Version 2.0}},
  \href{https://doi.org/10.1016/0010-4655(94)90074-4}{\emph{Comput. Phys.
  Commun.} {\bfseries 79} (1994) 291}.

\bibitem{Brun:1987ma}
R.~Brun, F.~Bruyant, M.~Maire, A.C.~McPherson and P.~Zanarini, \emph{{GEANT3}},
  {\emph{CERN Program Library Long Writeup W5013} (1987) unpublished}.

\bibitem{GEANT4:2002zbu}
{\scshape GEANT4 collaboration}, \emph{{GEANT4--a simulation toolkit}},
  \href{https://doi.org/10.1016/S0168-9002(03)01368-8}{\emph{Nucl. Instrum.
  Meth. A} {\bfseries 506} (2003) 250}.

\bibitem{Bernardini:1963dyf}
C.~Bernardini, G.F.~Corazza, G.~Di~Giugno, G.~Ghigo, R.~Querzoli, J.~Haissinski
  et~al., \emph{{Lifetime and beam size in a storage ring}},
  \href{https://doi.org/10.1103/PhysRevLett.10.407}{\emph{Phys. Rev. Lett.}
  {\bfseries 10} (1963) 407}.

\bibitem{Lewis:2018ayu}
P.M.~Lewis et~al., \emph{{First Measurements of Beam Backgrounds at
  SuperKEKB}}, \href{https://doi.org/10.1016/j.nima.2018.05.071}{\emph{Nucl.
  Instrum. Meth. A} {\bfseries 914} (2019) 69}
  [\href{https://arxiv.org/abs/1802.01366}{{\ttfamily 1802.01366}}].

\bibitem{Natochii:2023thp}
A.~Natochii et~al., \emph{{Measured and projected beam backgrounds in the Belle
  II experiment at the SuperKEKB collider}},
  \href{https://arxiv.org/abs/2302.01566}{{\ttfamily 2302.01566}}.

\bibitem{Gelb:2018agf}
M.~Gelb et~al., \emph{{B2BII: Data Conversion from Belle to Belle II}},
  \href{https://doi.org/10.1007/s41781-018-0016-x}{\emph{Comput. Softw. Big
  Sci.} {\bfseries 2} (2018) 9}
  [\href{https://arxiv.org/abs/1810.00019}{{\ttfamily 1810.00019}}].

\bibitem{Kuhr:2018lps}
T.~Kuhr, C.~Pulvermacher, M.~Ritter, T.~Hauth and N.~Braun, \emph{{The Belle II
  Core Software}},
  \href{https://doi.org/10.1007/s41781-018-0017-9}{\emph{Comput. Softw. Big
  Sci.} {\bfseries 3} (2019) 1}
  [\href{https://arxiv.org/abs/1809.04299}{{\ttfamily 1809.04299}}].

\bibitem{basf2-zenodo}
``{Belle~II Analysis Software Framework}
  (\href{https://doi.org/10.5281/zenodo.5574115}{basf2}).''

\bibitem{Belle:2018xst}
{\scshape Belle collaboration}, \emph{{Measurement of time-dependent $CP$
  asymmetries in $B^{0}\to K_S^0 \eta \gamma$ decays}},
  \href{https://doi.org/10.1103/PhysRevD.97.092003}{\emph{Phys. Rev. D}
  {\bfseries 97} (2018) 092003}
  [\href{https://arxiv.org/abs/1803.07774}{{\ttfamily 1803.07774}}].

\bibitem{Keck:2017gsv}
T.~Keck, \emph{{FastBDT: A Speed-Optimized Multivariate Classification
  Algorithm for the Belle II Experiment}},
  \href{https://doi.org/10.1007/s41781-017-0002-8}{\emph{Comput. Softw. Big
  Sci.} {\bfseries 1} (2017) 2}.

\bibitem{Fox:1978vu}
G.C.~Fox and S.~Wolfram, \emph{{Observables for the Analysis of Event Shapes in
  e$^{+}$ e$^{-}$ Annihilation and Other Processes}},
  \href{https://doi.org/10.1103/PhysRevLett.41.1581}{\emph{Phys. Rev. Lett.}
  {\bfseries 41} (1978) 1581}.

\bibitem{Belle:2003fgr}
{\scshape Belle collaboration}, \emph{{Evidence for $B^0 \to \pi^0 \pi^0$}},
  \href{https://doi.org/10.1103/PhysRevLett.91.261801}{\emph{Phys. Rev. Lett.}
  {\bfseries 91} (2003) 261801}
  [\href{https://arxiv.org/abs/hep-ex/0308040}{{\ttfamily hep-ex/0308040}}].

\bibitem{Farhi:1977sg}
E.~Farhi, \emph{{A QCD Test for Jets}},
  \href{https://doi.org/10.1103/PhysRevLett.39.1587}{\emph{Phys. Rev. Lett.}
  {\bfseries 39} (1977) 1587}.

\bibitem{Belle:2004uxp}
H.~Kakuno et~al., \emph{{Neutral B flavor tagging for the measurement of mixing
  induced CP violation at Belle}},
  \href{https://doi.org/10.1016/j.nima.2004.06.159}{\emph{Nucl. Instrum. Meth.
  A} {\bfseries 533} (2004) 516}
  [\href{https://arxiv.org/abs/hep-ex/0403022}{{\ttfamily hep-ex/0403022}}].

\bibitem{Belle-II:2021zvj}
{\scshape Belle II collaboration}, \emph{{B-flavor tagging at Belle II}},
  \href{https://doi.org/10.1140/epjc/s10052-022-10180-9}{\emph{Eur. Phys. J. C}
  {\bfseries 82} (2022) 283}
  [\href{https://arxiv.org/abs/2110.00790}{{\ttfamily 2110.00790}}].

\end{thebibliography}\endgroup
